\documentclass[twocolumn]{aastex631}

\received{June 19, 2025}
\revised{December 25, 2025}
\accepted{January 6, 2026}

\begin{document}

\title{Transverse Oscillations and Wave Propagation in the Magnetically Dominated M87 Jet}

\author[0000-0002-7322-6436]{Hyunwook Ro}
\affiliation{Korea Astronomy \& Space Science Institute, Daedeokdae-ro 776, Yuseong-gu, Daejeon 34055, Republic of Korea}
\affiliation{Department of Astronomy, Yonsei University, Yonsei-ro 50, Seodaemun-gu, Seoul 03722, Republic of Korea}
\email{hwro@kasi.re.kr}

\author[0000-0002-2709-7338]{Motoki Kino}
\affiliation{Kogakuin University of Technology \& Engineering, Academic Support Center, 2665-1 Nakano-machi, Hachioji, Tokyo 192-0015, Japan}
\affiliation{National Astronomical Observatory of Japan, 2-21-1 Osawa, Mitaka, Tokyo 181-8588, Japan}

\author[0000-0001-6906-772X]{Kazuhiro Hada}
\affiliation{Graduate School of Science, Nagoya City University, Yamanohata 1, Mizuho-cho, Mizuho-ku, Nagoya, Aichi 467-8501, Japan}
\affiliation{Mizusawa VLBI Observatory, National Astronomical Observatory of Japan, 2-12 Hoshigaoka, Mizusawa, Oshu, Iwate 023-0861, Japan}

\author[0000-0002-8131-6730]{Yosuke Mizuno}
\affiliation{Tsung-Dao Lee Institute, Shanghai Jiao Tong University, Shanghai, 201210, People's Republic of China}
\affiliation{School of Physics \& Astronomy, Shanghai Jiao Tong University, Shanghai, 200240, People's Republic of China}
\affiliation{Key Laboratory for Particle Physics, Astrophysics and Cosmology, Shanghai Key Laboratory for Particle Physics and Cosmology, Shanghai Jiao Tong University, Shanghai, 200240, People's Republic of China}

\author[0000-0001-6311-4345]{Yuzhu Cui}
\affiliation{Institute of Astrophysics, Central China Normal University, Wuhan 430079, People's Republic of China}

\author[0009-0007-8554-4507]{Kunwoo Yi}
\affiliation{School of Space Research, Kyung Hee University, 1732, Deogyeong-daero, Giheung-gu, Yongin-si, Gyeonggi-do 17104, Republic of Korea}

\author[0000-0001-8527-0496]{Tomohisa Kawashima}
\affiliation{National Institute of Technology, Ichinoseki College, Takanashi, Hagisho, Ichinoseki, Iwate, 021-8511, Japan}

\author[0000-0001-6558-9053]{Jongho Park}
\affiliation{School of Space Research, Kyung Hee University, 1732, Deogyeong-daero, Giheung-gu, Yongin-si, Gyeonggi-do 17104, Republic of Korea}
\affiliation{G-LAMP NEXUS Institute, Kyung Hee University, Yongin, 17104, Republic of Korea}
\affiliation{Institute of Astronomy and Astrophysics, Academia Sinica, P.O. Box 23-141, Taipei 10617, Taiwan, R. O. C}

\author[0000-0002-4148-8378]{Bong Won Sohn}
\affiliation{Korea Astronomy \& Space Science Institute, Daedeokdae-ro 776, Yuseong-gu, Daejeon 34055, Republic of Korea}
\affiliation{Department of Astronomy, Yonsei University, Yonsei-ro 50, Seodaemun-gu, Seoul 03722, Republic of Korea}
\affiliation{University of Science and Technology, Gajeong-ro 217, Yuseong-gu, Daejeon 34113, Republic of Korea}

\begin{abstract}
We present an in-depth analysis of transverse oscillations in the M87 jet, as identified in our previous study \citep{ro23}, which reported oscillatory patterns with a characteristic period of $\sim$1 year in the edge-brightened jet structure extending up to 12\,mas from the core.
This work is based on high-cadence KaVA 22\,GHz observations conducted from December 2013 to June 2016.
By analyzing the transverse velocity profiles and the spatial evolution of the oscillations, we find that the oscillations propagate downstream along the jet, with a wavelength of $\sim9-10$\,mas.
A single-mode sinusoidal wave model applied to the ridge lines successfully reproduces the observed transverse oscillations and yields superluminal wave speeds of $\sim2.7-2.9\,c$, consistent with the bulk jet velocity in this region.
These findings suggest that the transverse oscillations may be interpreted either as transverse MHD waves—possibly excited by jet precession, nutation, or quasi-periodic magnetic flux eruptions near the central engine—or as manifestations of jet instabilities, such as current-driven instabilities (CDIs).
Further investigation is required to distinguish between these scenarios and to clarify the dominant physical mechanism.
\end{abstract}


\section{Introduction}\label{chap:intro}

M87 is a giant elliptical galaxy at a distance of 16.8 Mpc~\citep{blakeslee09}, and it is classified as a Fanaroff-Riley Type I (FR I) radio galaxy. 
It hosts a supermassive black hole (SMBH) with a mass of $M_{\text{BH}}=6.5 \pm 0.7 \times10^9$\,$M_{\odot}$~\citep{ehtc19} and produces a prominent jet that extends several kiloparsecs from the galaxy.
At this distance, 1 milli-arcsecond (mas) corresponds to $\approx$ 0.081\,parsecs (pc), or about 130 Schwarzschild radii ($r_s=2GM_{\text{BH}}/c^2 $), and an apparent speed of 1\,mas/yr is equivalent to $\sim 0.264\,c$~\citep{walker2018}.
Thanks to its proximity and large black hole size, M87 serves as a unique cosmic laboratory for exploring accretion and jet-launching mechanisms in active galactic nuclei (AGNs). 
In particular, very long baseline interferometry (VLBI) observations have played a key role in unveiling the jet’s multi-scale structure and dynamics, providing the most critical observational data for understanding jet formation mechanisms.
For a comprehensive review and recent highlights of VLBI studies of the M87 jet, see \citet{hada2024}.

Recent VLBI monitoring has revealed complex transverse motions (i.e., displacements perpendicular to the jet axis) in the parsec-scale jet of M87.
These variations occur on multiple timescales and point to rich underlying dynamics, yet their physical origin remains poorly understood.
Notably, \citet{walker2018}, using 43\,GHz Very Long Baseline Array (VLBA) data from 1999 to 2016, reported a quasi-periodic lateral shift with a period of approximately $8 - 10$ years. 
They attributed this to Kelvin–Helmholtz instabilities propagating outward at a speed significantly slower than the jet’s bulk flow. 
This phenomenon was later supported and extended by \citet{cui23}, who incorporated additional 22 and 43\,GHz monitoring data from the East Asia VLBI Network (EAVN) up to 2022. They found a jet position angle variation with a period of roughly 11 years and interpreted it as evidence for Lense–Thirring precession of a compact, tilted accretion disk around a spinning SMBH.

In contrast, \citet{ro23} reported a short-term structural evolution of the M87 jet using high-cadence monitoring with the KVN and VERA Array (KaVA; \citealt{niinuma14}), a combined array of the Korean VLBI Network (KVN) and the VLBI Exploration of Radio Astrometry (VERA) in Japan.
Based on 24 epochs of KaVA 22\,GHz observations from December 2013 to June 2016, we analyzed the ridge lines of the M87 jet within 12\,mas of the radio core. 
This monitoring, with an average interval of $\sim0.1$ year, revealed fast transverse oscillations along both ridges of the jet. 
The oscillation period is nearly uniform across all distances, averaging 0.94 years with variations of about $\pm$0.12 years, and the amplitude is on the order of $\sim0.1$\,mas.
Moreover, the phases of the oscillations in the northern and southern limbs are nearly identical. 
Given the small amplitude, these oscillations are unlikely to be caused by Earth’s parallax and are instead intrinsic structural variations of the jet.

Taken together, recent findings reveal that the M87 jet exhibits transverse variations on both long and short timescales, likely driven by distinct physical mechanisms.
In this study, we extend the analysis of the $\sim$1-year transverse oscillations previously identified by \citet{ro23}, examining how they evolve over distance and time, and exploring their possible physical origin.
These coherent, small-scale motions may serve as a new diagnostic of the jet’s internal dynamics, potentially offering insights into processes such as energy dissipation and particle acceleration in the jet \citep[e.g., see][for a review]{perucho19}.

The paper is structured as follows: In Section \ref{sec:2}, we describe KaVA 22\,GHz monitoring data of the M87 jet and explain how to obtain the ridge lines of the jet. 
In Section \ref{sec:3}, we present the transverse velocity profiles of the M87 jet and analyze the wavelength of the oscillation.
In Section \ref{sec:4}, we show that the oscillations can be interpreted as waves propagating along the jet and apply the wave model to estimate the propagation speed.
In Section \ref{sec:5}, we discuss possible interpretations of the waves on the M87 jet.

\section{Observations and data analysis}\label{sec:2}

\subsection{KaVA observations}\label{sec:2.1}

We utilize 24 epochs of high-resolution KaVA 22\,GHz observations obtained between December 5, 2013, and June 13, 2016—identical to the dataset used in \citet{ro23}. 
These data represent the initial observations of the ongoing EAVN M87 monitoring program.
Since additional data may be affected by long-term structural evolution \citep[e.g.,][]{walker2018, cui23}, we focus solely on these early epochs to study short-term oscillations; a more extensive dataset will be analyzed in future work.

We performed a standard a-priori data calibration process using the NRAO’s Astronomical Image Processing System \citep[AIPS;][]{greisen2003}, and the images were subsequently produced via CLEAN and self-calibration using Difmap \citep{shepherd1997}.
Detailed data reduction procedures are described in previous KaVA/EAVN M87 monitoring studies \citep{niinuma14, hada17, park19, cui21, ro23, ro23_spix, cui23}.
The main observational parameters are summarized in Table \ref{tab:observation}.
The total observation period spans $\approx2.52$ years, with an average interval of $\sim0.1$ year, making the data well suited for tracking short-term oscillations $\lesssim1$ year and for investigating structural changes on similar spatial scales, as the $(u, v)-$coverage is nearly identical across epochs (see Figure 1 in \citet{park19} for a typical KaVA at 22\,GHz $(u, v)-$coverage).

\begin{deluxetable}{cccc}
\tablenum{1}
\tablecaption{Summary of KaVA 22\,GHz observations of the M87 jet from December 2013 to June 2016.\label{tab:observation}}
\tablewidth{0pt}
\tablehead{
\colhead{Exp. Code} & \colhead{Obs. Date} & \colhead{Beam size} & \colhead{Stations}
}
\decimalcolnumbers
\startdata
r13339b & 2013 Dec 5  & 1.31~$\times$~1.14, $-$18.0 & KaVA \\
r13360a & 2013 Dec 26 & 1.61~$\times$~1.23, $-$42.7 & KaVA \\
r14015a & 2014 Jan 15 & 1.83~$\times$~1.51, $-$54.3 & KaVA \\
r14061b & 2014 Mar 2  & 1.50~$\times$~1.21, 3.0     & KaVA \\
r14074b & 2014 Mar 15 & 1.34~$\times$~1.14, $-$15.9 & KaVA \\
r14093a & 2014 Apr 3  & 1.32~$\times$~1.22, 14.5    & KaVA \\
r14106b & 2014 Apr 16 & 1.27~$\times$~1.10, $-$17.7 & KaVA \\
r14123a & 2014 May 3  & 1.47~$\times$~1.26, 11.8    & KaVA \\
r14153a & 2014 Jun 2  & 1.08~$\times$~0.91, $-$18.3 & KaVA $-$IRK \\
r14165a & 2014 Jun 14 & 1.34~$\times$~1.10, $-$11.5 & KaVA $-$IRK \\
r14244a & 2014 Sep 1  & 1.35~$\times$~1.16, $-$7.9  & KaVA \\
r14257a & 2014 Sep 14 & 1.56~$\times$~1.15, $-$41.3 & KaVA \\
r14308c & 2014 Nov 4  & 1.41~$\times$~1.26, $-$32.2 & KaVA \\
r14349b & 2014 Dec 15 & 1.27~$\times$~1.10, 2.34    & KaVA \\
r15123a & 2015 May 3  & 1.22~$\times$~1.08, $-$4.1  & KaVA \\
r15136a & 2015 May 16 & 1.42~$\times$~1.15, $-$21.5 & KaVA \\
r16056c & 2016 Feb 25 & 1.31~$\times$~1.15, $-$5.9  & KaVA \\
r16069b & 2016 Mar 9  & 1.38~$\times$~1.15, $-$2.3  & KaVA \\
r16081b & 2016 Mar 21 & 1.55~$\times$~1.26, $-$14.9 & KaVA \\
r16099a & 2016 Apr 8  & 1.36~$\times$~1.12, 1.9     & KaVA \\
r16112b & 2016 Apr 21 & 1.28~$\times$~1.16, $-$3.7  & KaVA \\
r16124a & 2016 May 3  & 1.30~$\times$~1.05, $-$11.9 & KaVA \\
r16144a & 2016 May 23 & 1.24~$\times$~1.08, $-$12.6 & KaVA \\
r16165a & 2016 Jun 13 & 1.26~$\times$~1.14, 5.3     & KaVA \\
\enddata
\tablecomments{(1) Experiment code; (2) Observation date; (3) Synthesized beam size in mas $\times$ mas and position angle in degrees (natural weighting); (4) Participating stations. “KaVA” indicates all seven stations. On June 2 and 14, 2014, the VERA-Iriki station (IRK) was not included.}
\end{deluxetable}

\begin{figure*}
    \begin{minipage}{0.6\textwidth}
        \includegraphics[width=\textwidth]{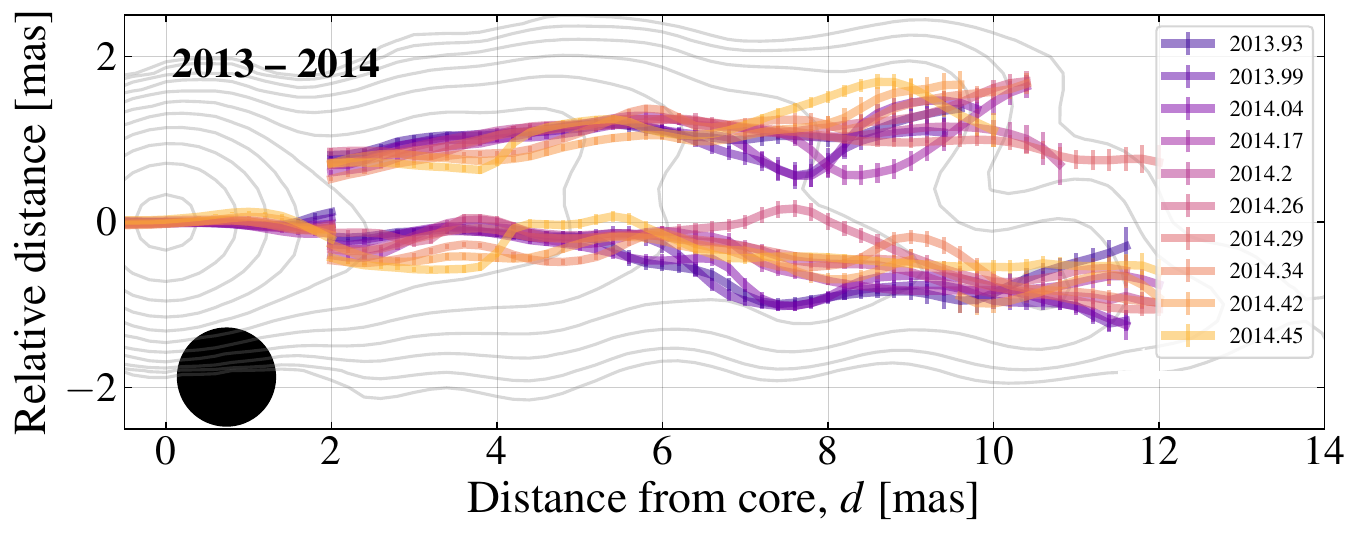}
        \includegraphics[width=\textwidth]{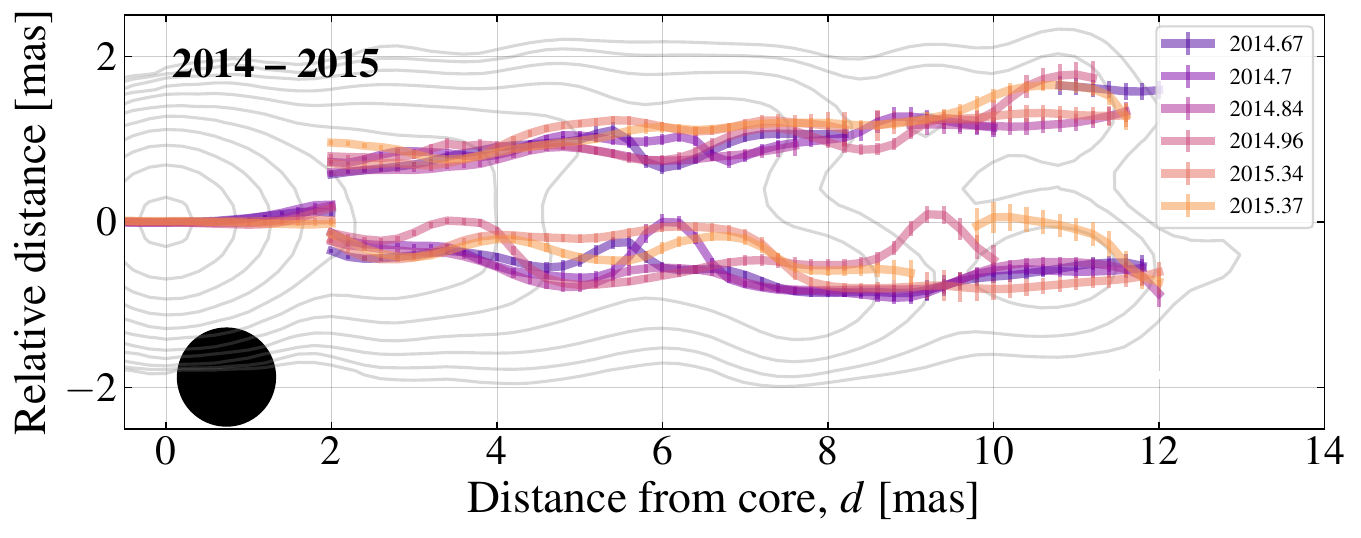}
        \includegraphics[width=\textwidth]{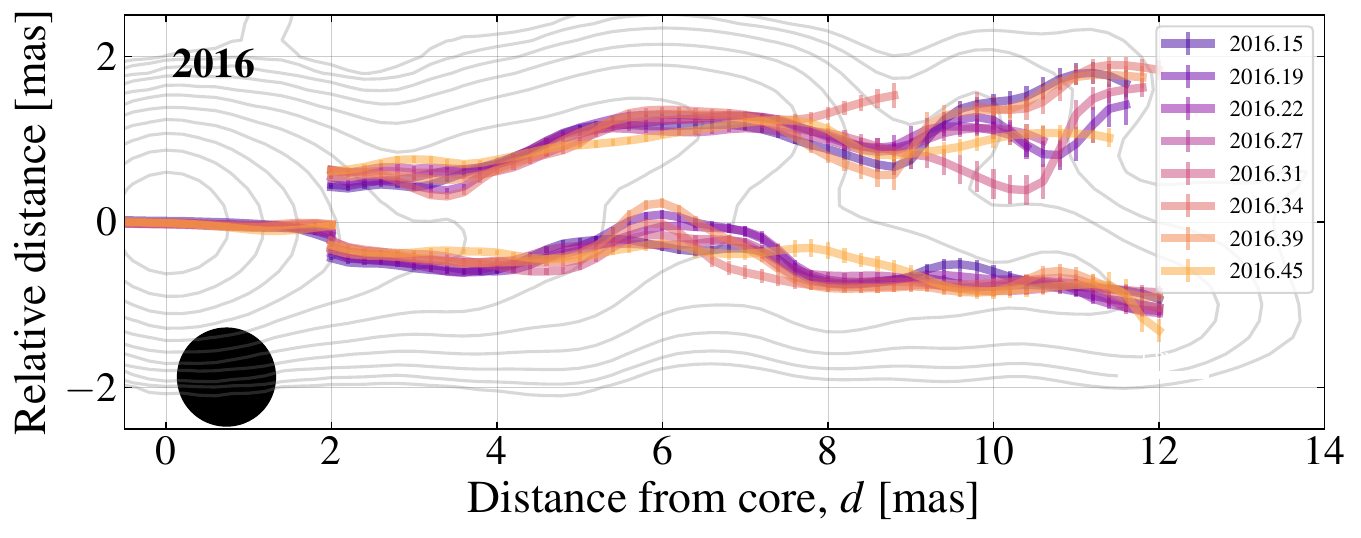}
    \end{minipage}
    \hfill
    \begin{minipage}{0.4\textwidth}
        \includegraphics[width=\textwidth]{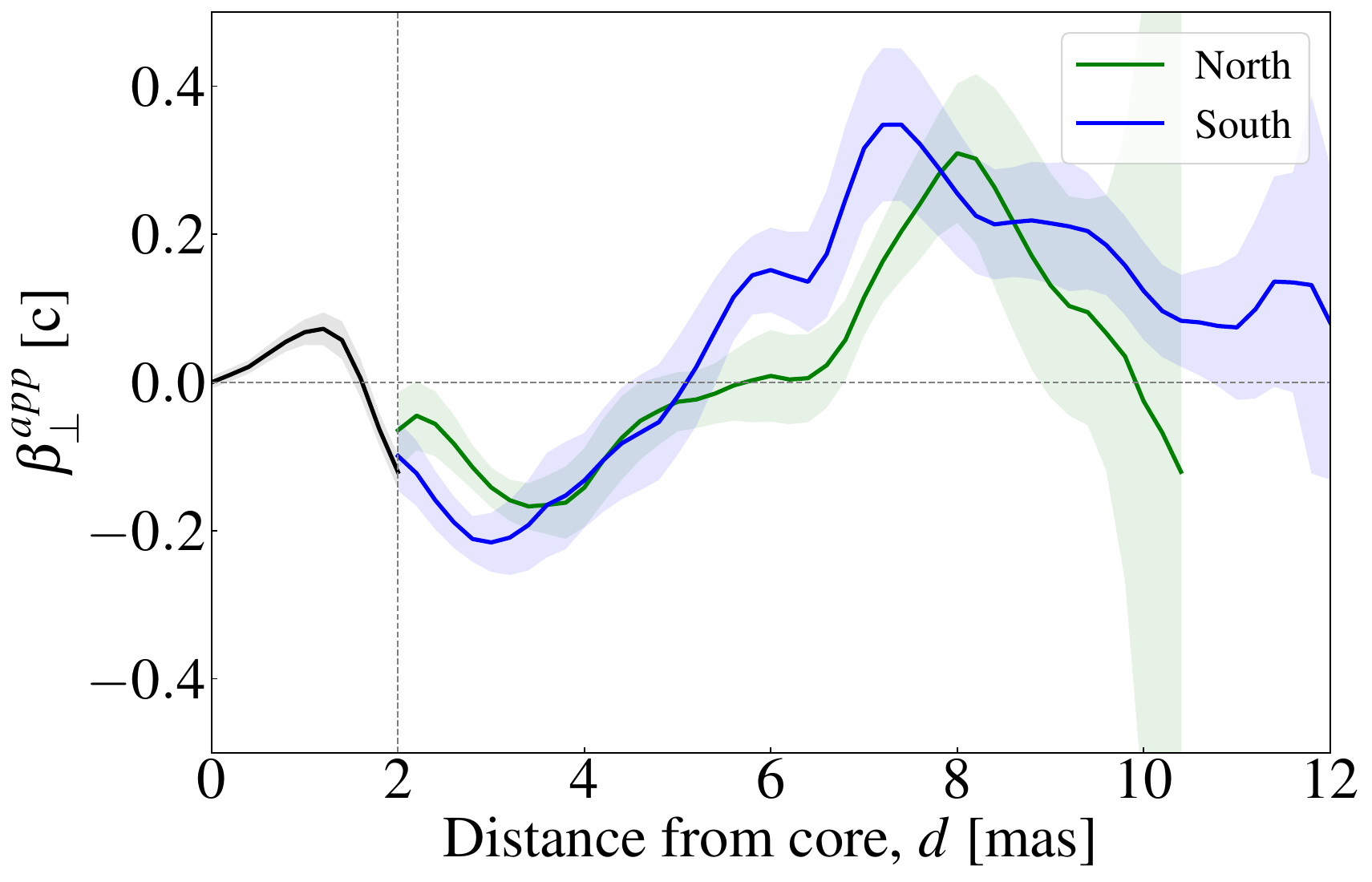}
        \includegraphics[width=\textwidth]{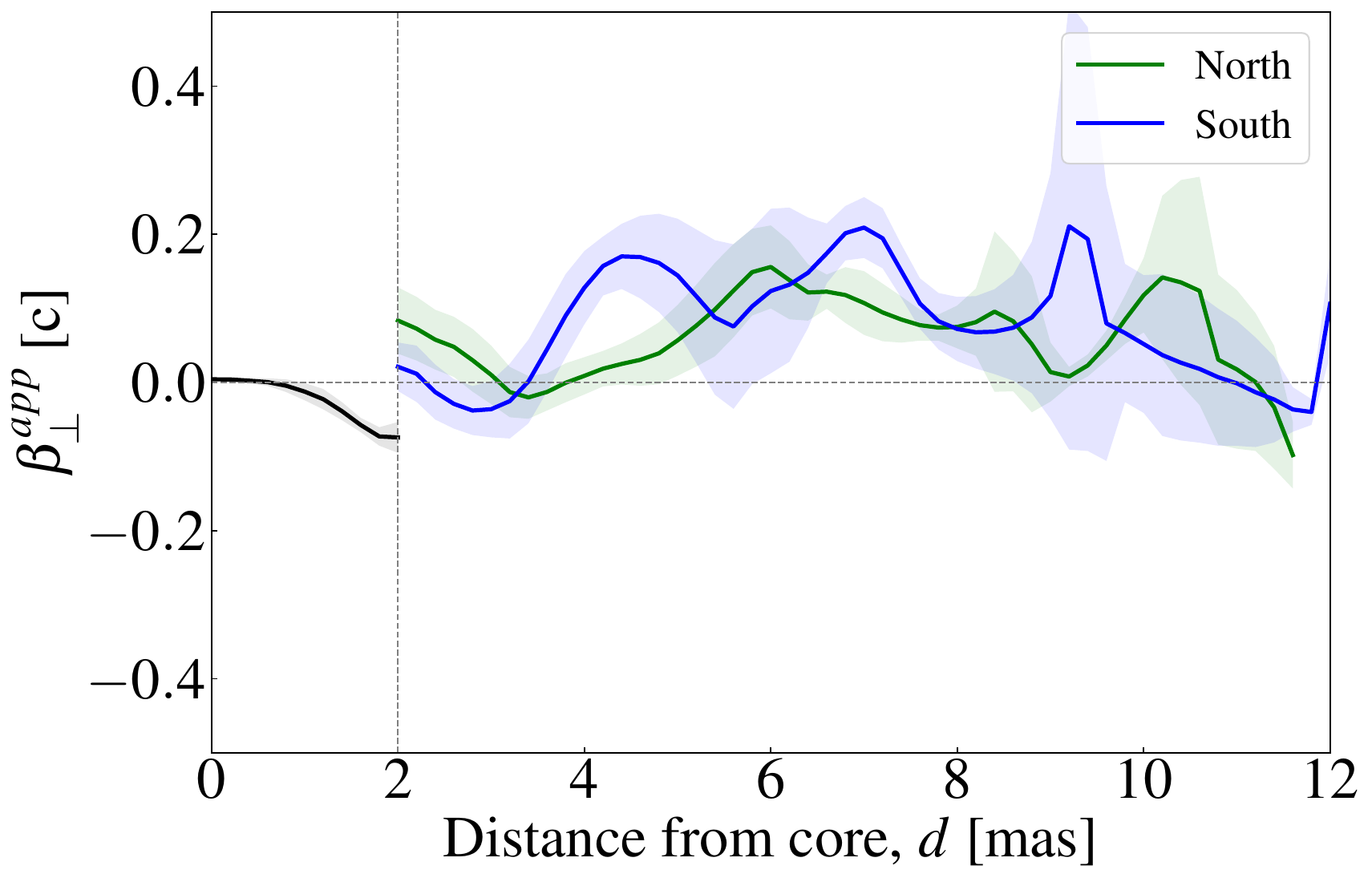}
        \includegraphics[width=\textwidth]{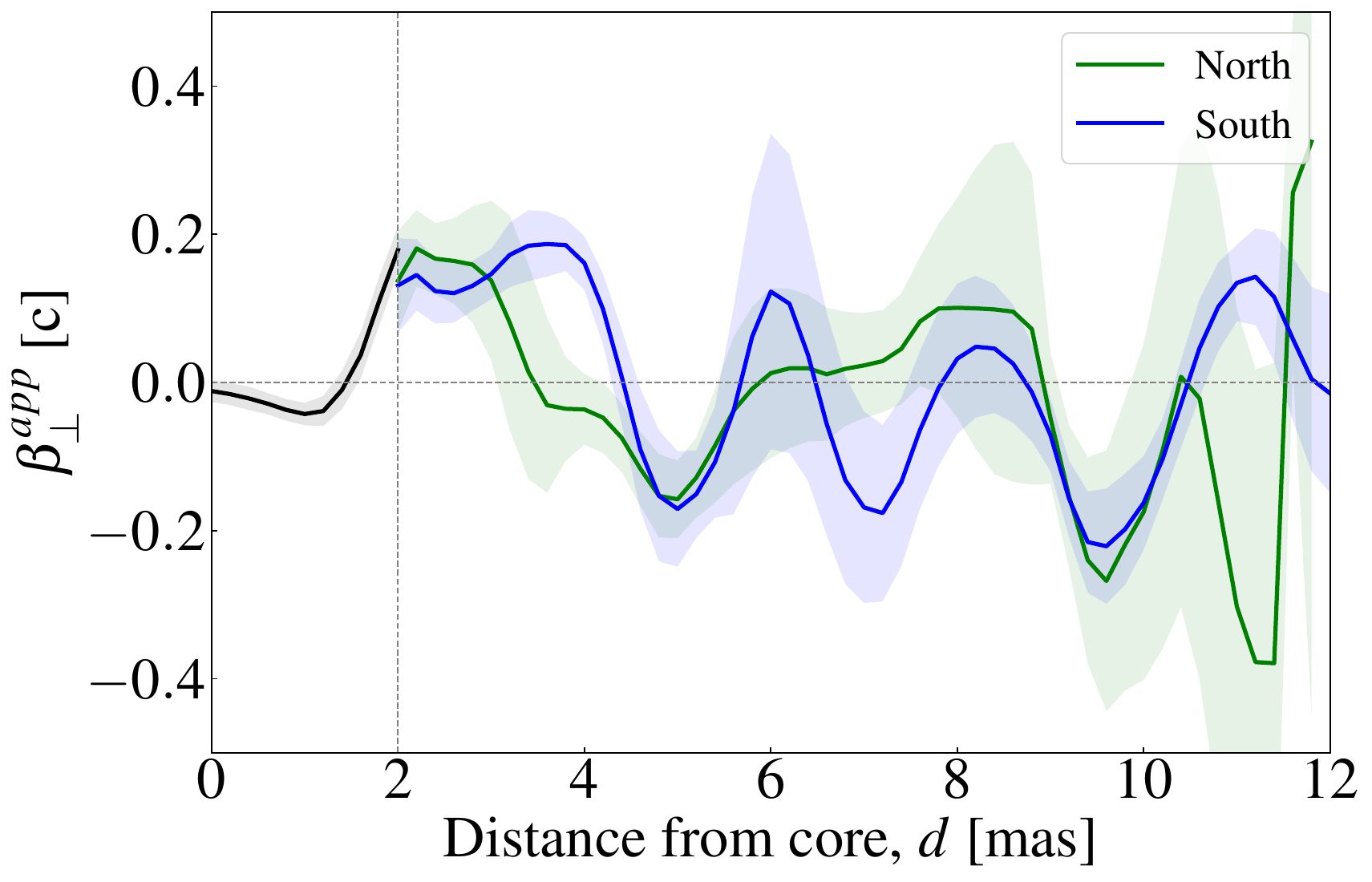}
    \end{minipage}
    \caption{Left: Distribution of ridge lines for the M87 jet at 22\,GHz from KaVA observations.
    The top, middle, and bottom panels correspond to observation periods from December 2013 to June 2014, September 2014 to May 2015, and February 2016 to June 2016, respectively.
    Individual observations are color-coded from purple to yellow to indicate the time evolution within each period, and the contours represent total intensity maps stacked over each period.
    The black circles in the lower left corner show the size of the restoring beam. 
    All images have been rotated by $18$° clockwise so that the jet axis is aligned horizontally.
    Right: Transverse velocity profiles of the M87 jet corresponding to the time ranges in the left panels. 
    The black line represents the velocity distribution derived from a single ridge, while the green and blue lines represent the distributions for the northern and southern ridges, respectively. 
    The shaded area denotes the $1\,\sigma$ error.}
    \label{fig:ridge_line_distribution_and_transverse_velocity_fields}
\end{figure*}

\subsection{Identification of ridge lines}\label{sec:2.2}
To study the transverse oscillations of the M87 jet, we need to identify its ridge lines.
To extract the ridge lines from the images, we performed the following procedure \citep{ro23}:
First, we restored all maps to a common circular beam of 1.2\,mas $\times$ 1.2\,mas and rotated the images by 18° clockwise so that the jet axis aligned horizontally\footnote{Here, the “jet axis” refers to the nominal position angle of 288°, which is commonly adopted as a reference direction for the M87 jet, despite known time variability in its actual direction \citep[e.g.,][]{cui23}.}.
We then extracted one-dimensional brightness profiles perpendicular to the jet axis in successive 0.2\,mas steps, starting at the core.
Each brightness profile was fitted with either a single or double Gaussian, and the ridge lines were constructed by connecting the peaks of these Gaussian fits along the jet.
We derived the ridge lines along the jet up to 12 mas from the core. 
In the inner 2 mas, the jet exhibits a single ridge structure.
We note that this single ridge is likely a result of the limited angular resolution blending the underlying limb-brightened structure \citep[e.g.,][]{walker2018, kim18, lu2023}. Thus, the derived positions should be interpreted as the flux-weighted centroids of the unresolved flow rather than the tracing of a distinct physical feature.
Beyond this distance ($>2$ mas), two distinct ridges become apparent \citep{ro23}.
When generating the ridge lines, we excluded regions where the peak intensity was less than three times the image rms noise (i.e., SNR $<$ 3) or where the position changed too abruptly.
This filtering process resulted in the exclusion of approximately 13\% of the data points, primarily due to the SNR criterion.
Detailed procedures, including error estimation and ridge line plots for individual epochs, are provided in \citet{ro23}.

\section{Transverse velocity profiles} of the M87 jet\label{sec:3}

The left panels of Figure~\ref{fig:ridge_line_distribution_and_transverse_velocity_fields} show the ridge line evolution of the M87 jet within 12\,mas from the core, based on observations from three periods: December 2013--June 2014 (top), September 2014--May 2015 (middle), and February--June 2016 (bottom), selected to capture densely sampled intervals while avoiding long observational gaps.
The ridge lines from individual epochs are shown in different colors, overlaid on average maps created by stacking total intensity images for each period.
The positional uncertainty of the ridge lines increases with distance from the core, and the ridge lines on the northern limb tend to be more scattered or truncated compared to the southern limb due to lower intensity.

Notably, a comparison among different epochs reveals small transverse displacement changes ($\lesssim1$\,mas), indicating subtle motion of the jet ridge lines over time.
To investigate how this small-scale transverse motion varies with distance, we performed a linear regression on the transverse displacement of the ridge lines for each observation period and derived the corresponding time-averaged transverse velocity of the jet.
These velocities were then converted using a factor of 1\,mas/yr $= 0.264\,c$ \cite[e.g.][]{walker2018}, yielding the apparent transverse velocity profiles ($\beta^{\text{app}}_{\perp}$) of the M87 jet for each period.

The right panels of Figure \ref{fig:ridge_line_distribution_and_transverse_velocity_fields} show the transverse velocity profiles ($\beta^{\text{app}}_{\perp}$) of the M87 jet for the three observation periods corresponding to those in the left panels. 
The black line represents the velocity profile obtained from the single ridge, while the green and blue lines correspond to the northern and southern ridge lines, respectively.

\begin{figure}
    \centering
    \includegraphics[width=0.45\textwidth]{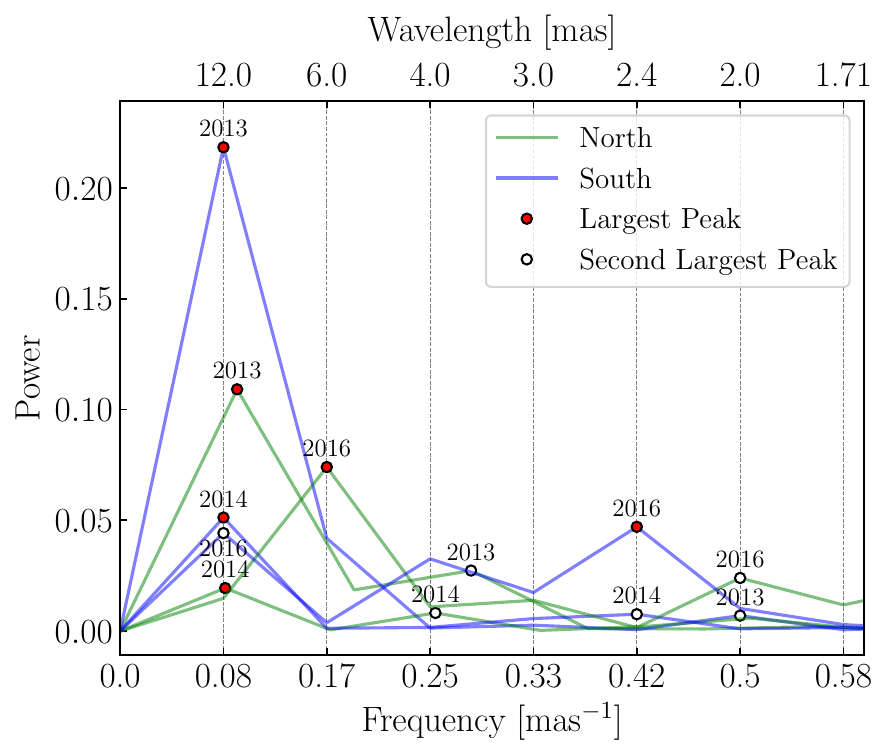}
        \caption{Power spectrum of the transverse velocity profiles of the M87 jet. The lower axis shows the spatial frequency, and the upper axis indicates the corresponding wavelength. The green and blue lines represent the power spectra for the northern and southern ridges, respectively. The red and white circle marks the largest and second-largest peaks in the power spectra with each peak annotated by the start year of the corresponding observation period.}
    \label{fig:power_spectrum_vtrans}
\end{figure}

Interestingly, we found that the time-averaged transverse velocity profiles on the two ridges exhibit a similar, sinusoidal-like pattern, with the velocity direction reversing with distance.
During the December 2013--June 2014 epoch, the velocity changes from positive to negative at approximately 1.5\,mas, becomes positive again around 5\,mas, and reverts to negative beyond roughly 10\,mas. 
Furthermore, the maximum apparent transverse velocity increases gradually with distance, reaching about 0.4\,$c$.
From September 2014 to May 2015, reversals in the transverse motion are observed at approximately 3.5\,mas and again beyond 9\,mas. 
In the February--June 2016 epoch, the velocity profile is more scattered, which makes these reversals harder to identify; however, the overall trend still indicates reversals near 1.5\,mas and beyond 3.5\,mas.
Despite being less pronounced in 2016, the sinusoidal shape observed in the earlier epochs allows us to estimate a characteristic wavelength of roughly $\gtrsim8.5$\,mas.

A comparison of the transverse velocity profiles across the three observation periods reveals clear temporal variations in the apparent transverse velocity of the M87 jet.
For example, at $d=1$\,mas (where $d$ is the distance from the core), the transverse velocity decreases from $\beta^{\text{app}}_{\perp}\sim +0.1\,c$ in 2013 to $\beta^{\text{app}}_{\perp}\sim -0.05\,c$ in 2016. 
In contrast, at $d=3$\,mas, the velocity increases from roughly $\beta^{\text{app}}_{\perp}\sim -0.2\,c$ in 2013 to $\beta^{\text{app}}_{\perp}\sim +0.2\,c$ in 2016.
Further downstream, at $d=8$\,mas the velocity is around $\beta^{\text{app}}_{\perp}\sim +0.4\,c$ in 2013, but drops to nearly zero or becomes negative in 2016.
These significant changes in the velocity profile over the $\sim$2.5-year monitoring period suggest a possible connection with the fast oscillations reported by \citet{ro23}.

In Figure~\ref{fig:power_spectrum_vtrans}, we compute the power spectrum of the transverse velocity profiles using both ridges across the three observing periods, in order to quantify the dominant spatial scales. 
We analyzed the velocity profile over a distance of 12\,mas\footnote{The two ridges are unresolved within 2\,mas (Figure~\ref{fig:ridge_line_distribution_and_transverse_velocity_fields}), but their nearly identical velocity patterns beyond this point suggest a shared transverse velocity profile in the inner region, which we accordingly assumed in our analysis.}, yielding a minimum spatial frequency resolution of $\sim$0.08\,mas$^{-1}$ \citep{vanderplas18}.
Except for the 2016 data, the power spectra show the largest peaks (red circles) at the lowest spatial frequency, corresponding to a wavelength of 12\,mas, with slight skewness toward shorter wavelengths.
This indicates that the velocity profiles are primarily composed of components with wavelengths $\gtrsim$8\,mas (see Appendix~\ref{sec:aa} for details).
In 2016, the southern ridge shows its second-largest peak (white circle) at 12\,mas, while the northern ridge’s dominant peak appears at a slightly shorter wavelength, $\sim$6\,mas.
Several additional peaks in the 2--4\,mas range are also detected—most prominently in 2016, but also present to some extent in other epochs (Figure \ref{fig:ridge_line_distribution_and_transverse_velocity_fields}).

The observed spatial and temporal variations in the transverse velocity profile suggest that the transverse oscillations exhibit an organized structure that evolves with distance, resembling the propagation of a wave along the jet.
However, the velocity profile alone is insufficient to determine whether the one-year oscillation truly corresponds to a propagating wave.
Moreover, while the power spectrum analysis provides an approximate wavelength, it does not constrain key wave properties such as propagation speed or phase evolution.
In the next chapter, we address these questions by analyzing the full temporal evolution of ridge displacements at each distance.

\section{Superluminal wave propagation in the M87 jet}\label{sec:4}
\subsection{Observational Evidence for Wave Propagation}\label{sec:4.1}

To characterize the spatial evolution of the transverse oscillations, we fit a sinusoidal function to the ridge displacement at every 0.2\,mas step along the jet.

The fitting was performed using the Markov chain Monte Carlo \citep[MCMC;][]{emcee2013, hogg18} method, yielding best-fit values for the period, amplitude, and phase at each location. The fitting function is defined as:
\begin{equation}
    y(t) = A \sin\left(\frac{2\pi}{P}t - \phi\right) + y_0,\label{eq:sinusoidal}
\end{equation} where $y(t)$ is the transverse displacement as a function of time $t$, and $A$, $P$, and $\phi$ represent the amplitude, period, and phase, respectively. The parameter $y_0$ denotes the displacement offset from the jet axis. 
The results at several selected locations were previously reported in \citet{ro23}.

\begin{figure*}
    \centering
    \includegraphics[width=1\textwidth]{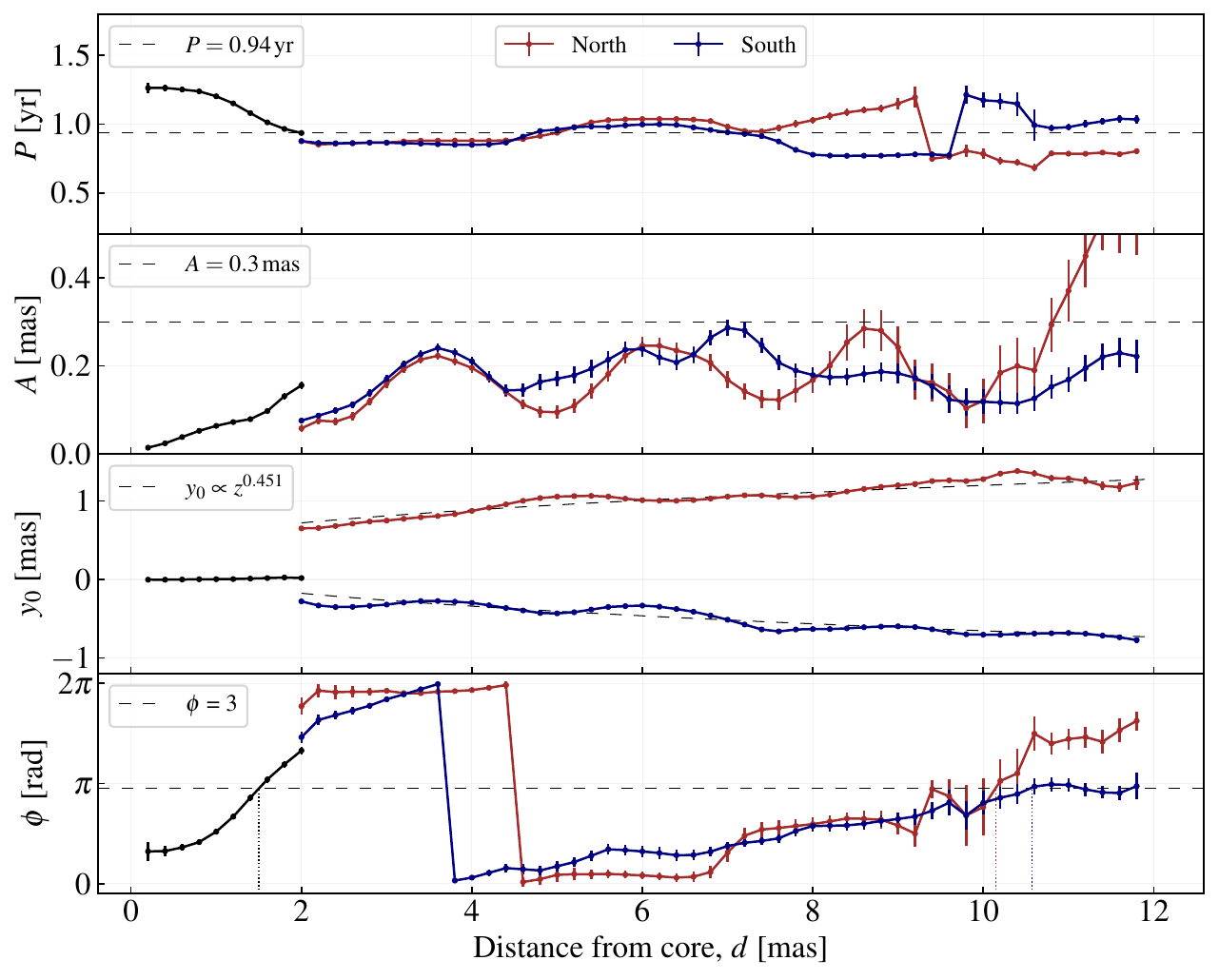}
        \caption{Summary of the sinusoidal fit to the transverse motion of the M87 jet as a function of distance from the core (Equation \ref{eq:sinusoidal}). {From top to bottom, period ($P$), amplitude ($A$), offset ($y_0$), and phase ($\phi$) of the oscillation}. The black line results from single ridges and the red and blue lines are the results of northern and southern ridges, respectively.
        {The dashed line of each panel indicates the average period ($P = 0.94$ yr), the maximum amplitude of the oscillations ($A = 0.3$\,mas), the best-fit jet width profile ($y_0\propto z^{0.451}$), and $\phi = 3$ radians, respectively.
        The phase of the oscillation increases continuously as a function of distance, implying that the oscillations {are propagating along} the jet.}}
    \label{fig:fitting_parameters_distribution}
\end{figure*}

Figure~\ref{fig:fitting_parameters_distribution} summarizes the full set of best-fit parameters.
The fitted periods, shown in the top panel, remain nearly constant along the jet, the average period of 0.94 years with a standard deviation of 0.12 years, consistent with \citet{ro23}.
The amplitude increases gradually with distance, as shown in the second panel, peaking at approximately 0.3\,mas around 7\,mas, and then remaining roughly constant or slightly decreasing farther downstream\footnote{An abrupt change in the amplitude is seen in the northern ridge after 10\,mas. However, this may result from systematic errors due to the lower intensity and fewer detected ridges in that limb (Figure \ref{fig:ridge_line_distribution_and_transverse_velocity_fields}, left panels). In the southern limb, where many ridges are detected beyond 10\,mas, no abrupt increase in the amplitude is observed.}.
The third panel shows the fitted offset $y_0$, which traces the underlying jet structure and reveals a near-parabolic expansion with $y_0 \propto d^{0.451\pm0.004}$, similar to the trend reported in previous studies \citep[e.g.,][]{asada12, hada13, nakamura18}.

The most prominent trend appears in the phase evolution (bottom panel of Figure~\ref{fig:fitting_parameters_distribution}), which increases continuously with distance.
This clearly indicates that the oscillation propagates downstream along the jet.
The phase advances by 2$\pi$ radians between $d \sim 1.5$ and $\sim$10.1 mas in the north, and $\sim$10.6 mas in the south, corresponding to wavelengths of approximately 8.7\,mas and 9.1\,mas, respectively.
These values are in good agreement with the dominant spatial scale identified in the power spectrum of the transverse velocity profiles (Section \ref{sec:3}).

Taken together, the nearly constant period and steadily increasing phase are consistent with a transverse oscillation propagating downstream along the jet as a wave.
The close agreement between the two limbs—evident in both the fitted oscillation properties (Figure~\ref{fig:fitting_parameters_distribution}) and the velocity profile analysis (Figure~\ref{fig:ridge_line_distribution_and_transverse_velocity_fields})—further supports that the waves propagating along the two limbs are nearly coherent.

\subsection{Single-Mode Sinusoidal Wave Model}\label{sec:4.2}

In this section, we introduce a single-mode wave model to describe the observed transverse oscillations and estimate the wave propagation speed along the jet. 
The model assumes a sinusoidal wave with a single frequency and wavelength propagating along the jet axis, with the amplitude increasing linearly with distance.
We note that this linear growth is adopted as an empirical parameterization to describe the oscillation behavior within the observed window.
This wave is superimposed on the underlying jet structure $y_0(d)$, which is defined by the offset profile shown in Figure~\ref{fig:fitting_parameters_distribution}.
The functional form of the wave model is given by:

\begin{equation}
y(t,d) = A_{\mathrm{wave}} \left(\frac{d}{d_{\mathrm{f}}}\right)\sin \left(\frac{2\pi}{P_{\mathrm{wave}}}t - \frac{2\pi}{\lambda_{\mathrm{wave}}}d + \phi_{\mathrm{wave}}\right) + y_{0}(d)
\label{eq:wave_model}
\end{equation}
where $y(t, d)$ is the transverse displacement at time $t$ and distance $d$ from the core, and $d_{\mathrm{f}} = 12 \,\mathrm{mas}$ is normalization factor corresponding to the outermost distance of our observational window.
The four parameters—$A_{\mathrm{wave}}$, $P_{\mathrm{wave}}$, $\lambda_{\mathrm{wave}}$, and $\phi_{\mathrm{wave}}$—represent the amplitude at $d_{\mathrm{f}}$, period, wavelength, and phase of the wave, respectively.

\begin{figure*}
    \centering
    \includegraphics[width=0.49\textwidth]{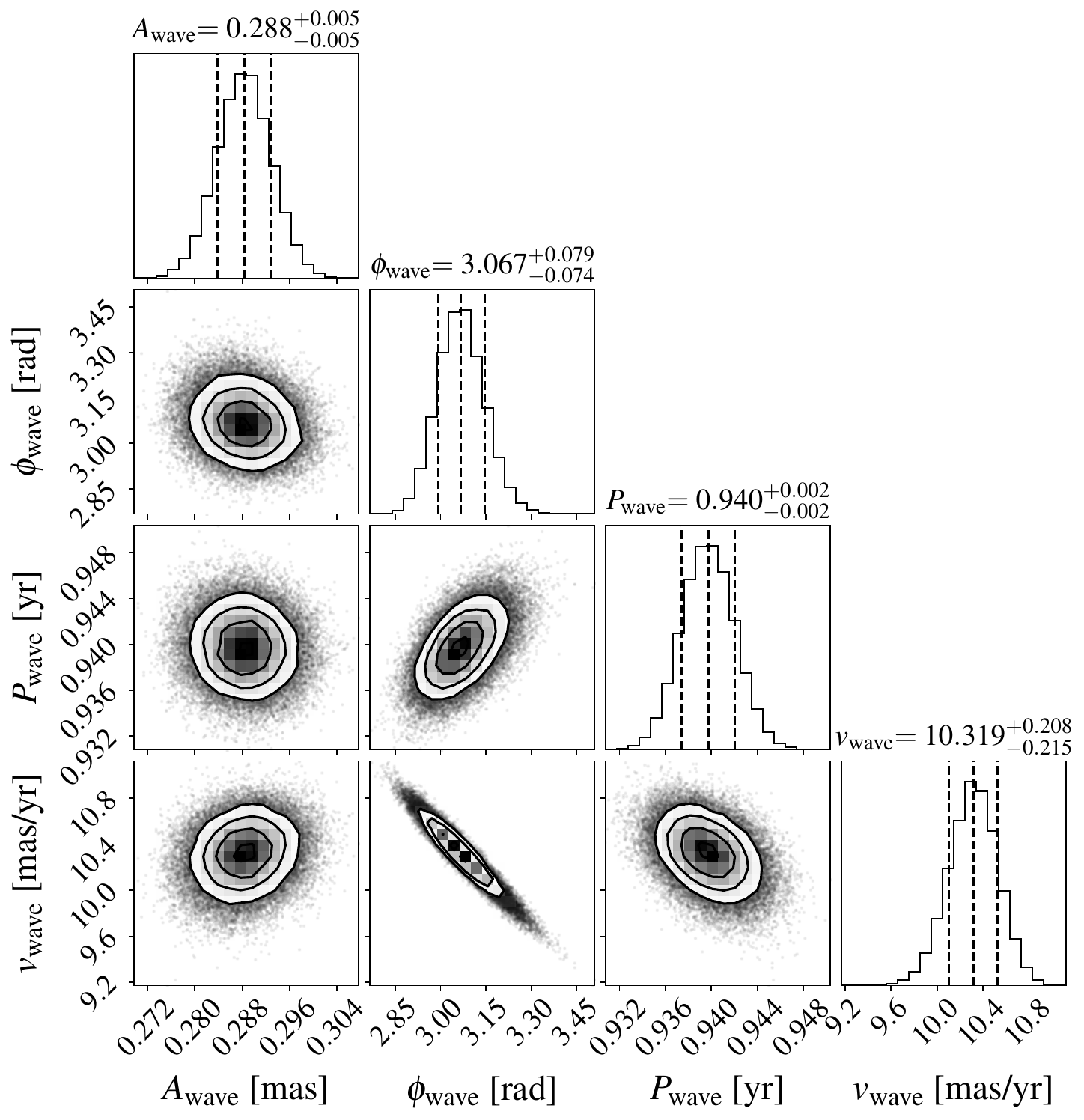}
    \includegraphics[width=0.49\textwidth]{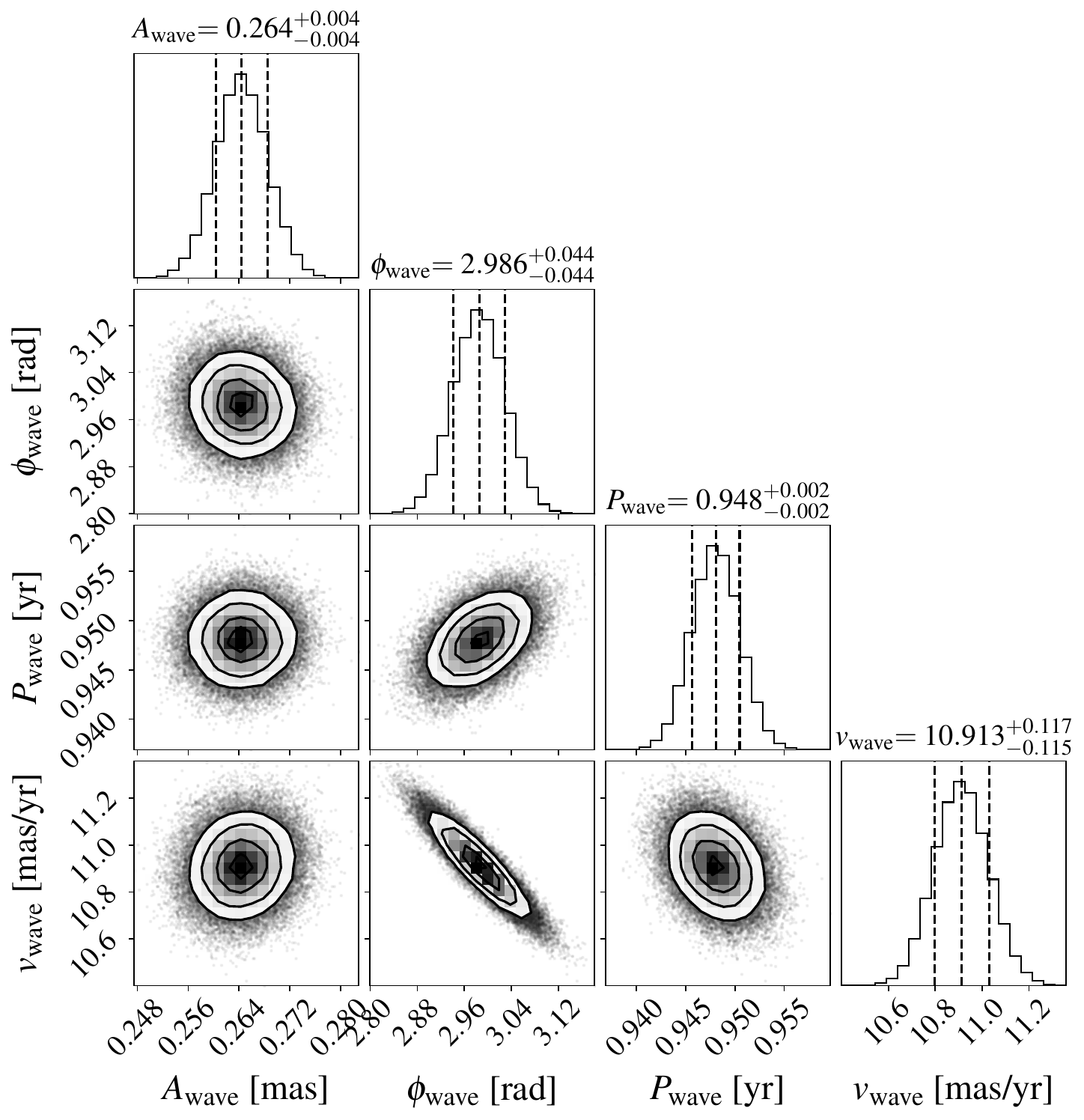}
        \caption{Corner plots showing the posterior distributions obtained from the MCMC fits of a single-mode sinusoidal wave model (Equation \ref{eq:wave_model}) to the ridge lines of the M87 jet’s northern limb (left) and southern limb (right). The fitted parameters are:
        (1) $A_\mathrm{wave}$, the wave amplitude evaluated at $d=12\,$mas,
        (2) $\phi_\mathrm{wave}$, the wave phase,
        (3) $P_\mathrm{wave}$, the wave period, and
        (4) $v_\mathrm{wave}$, the apparent wave propagation speed.
        Each panel along the diagonal shows the marginalized one‐dimensional distribution for that parameter, with vertical dashed lines marking the 16th, 50th, and 84th percentiles. 
        The off‐diagonal panels show the two‐dimensional posterior contours for the corresponding parameter pairs, where each contour encloses 68\%, 95\%, and 99.7\% of the samples. 
        The best‐fit values and uncertainties are summarized in Table \ref{tab:wave_mcmc_result}.}
    \label{fig:corner_plot}
\end{figure*}

\begin{deluxetable*}{cccccc}
\tablenum{2}
\tablecaption{Summary of the best‐fit parameters from the single-mode sinusoidal wave model applied to the M87 jet’s ridge lines between 2 and 12\,mas. Uncertainties denote the 16th and 84th percentiles of the MCMC posterior distributions. See the note below for descriptions of each column.\label{tab:wave_mcmc_result}}
\tablewidth{0pt}
\tablehead{
\colhead{Ridge} & \colhead{$A_{\text{wave}}$} & \colhead{$P_{\text{wave}}$} &  \colhead{$\lambda_{\text{wave}}$} & \colhead{$\phi_{\text{wave}}$}  & \colhead{$\beta^{\text{app}}_{\text{wave}}$}     \\
&  \colhead{(mas)}   &  \colhead{(year)}   &  \colhead{(mas)}   &  \colhead{(rad)}   &  \colhead{}}  
\decimalcolnumbers
\startdata
North & $0.288^{+0.005}_{-0.005}$ & $0.940^{+0.002}_{-0.002}$ & $9.698^{+0.188}_{-0.195}$ & $3.067^{+0.079}_{-0.074}$ & $2.724^{+0.055}_{-0.057}$ \\
South & $0.264^{+0.004}_{-0.004}$ & $0.948^{+0.002}_{-0.002}$ & $10.345^{+0.107}_{-0.104}$ & $2.986^{+0.044}_{-0.044}$ &  $2.881^{+0.031}_{-0.030}$ \\
\enddata
\tablecomments{(1) Ridges (2) Wave amplitude evaluated at 12\,mas (3) Wave period (4) Wave wavelength (5) Wave phase (6) Apparent wave propagation speed}
\end{deluxetable*}

\begin{figure*}
    \begin{minipage}{0.6\textwidth}
        \includegraphics[width=\textwidth]{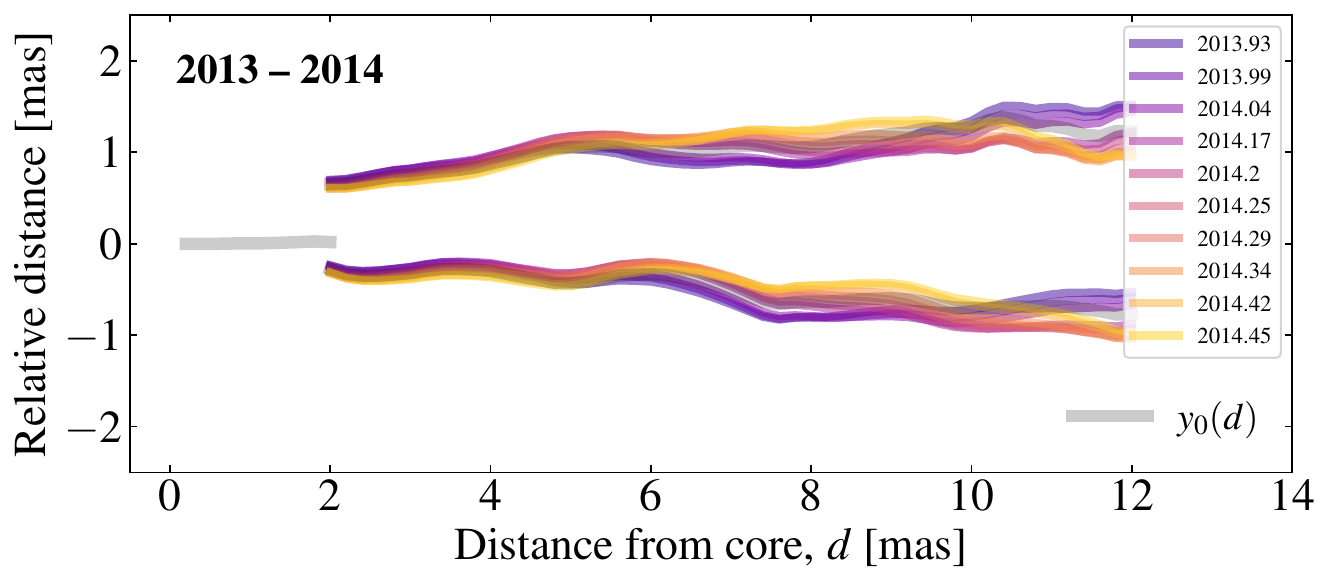}
        \includegraphics[width=\textwidth]{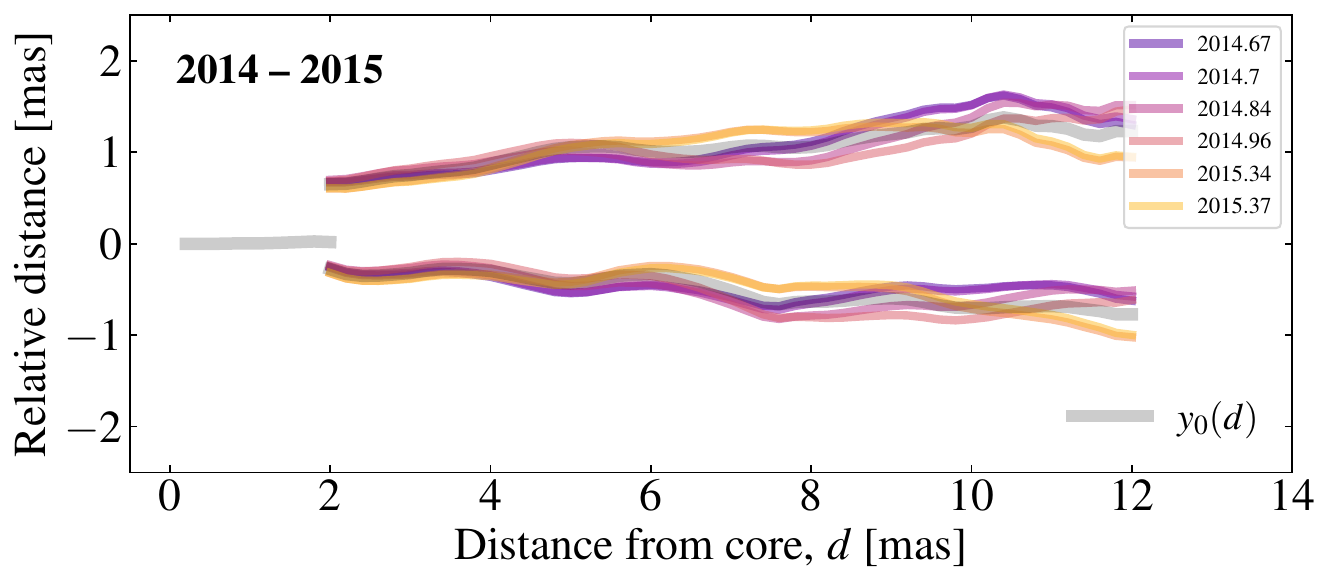}
        \includegraphics[width=\textwidth]{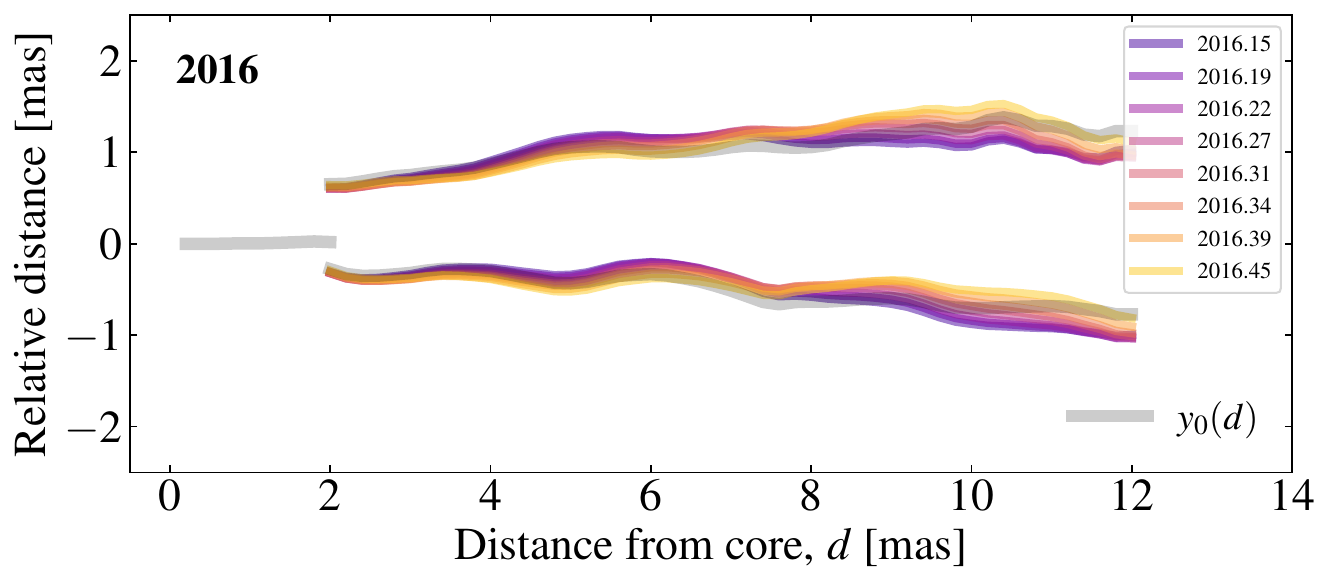}
    \end{minipage}
    \hfill
    \begin{minipage}{0.4\textwidth}
        \includegraphics[width=\textwidth]{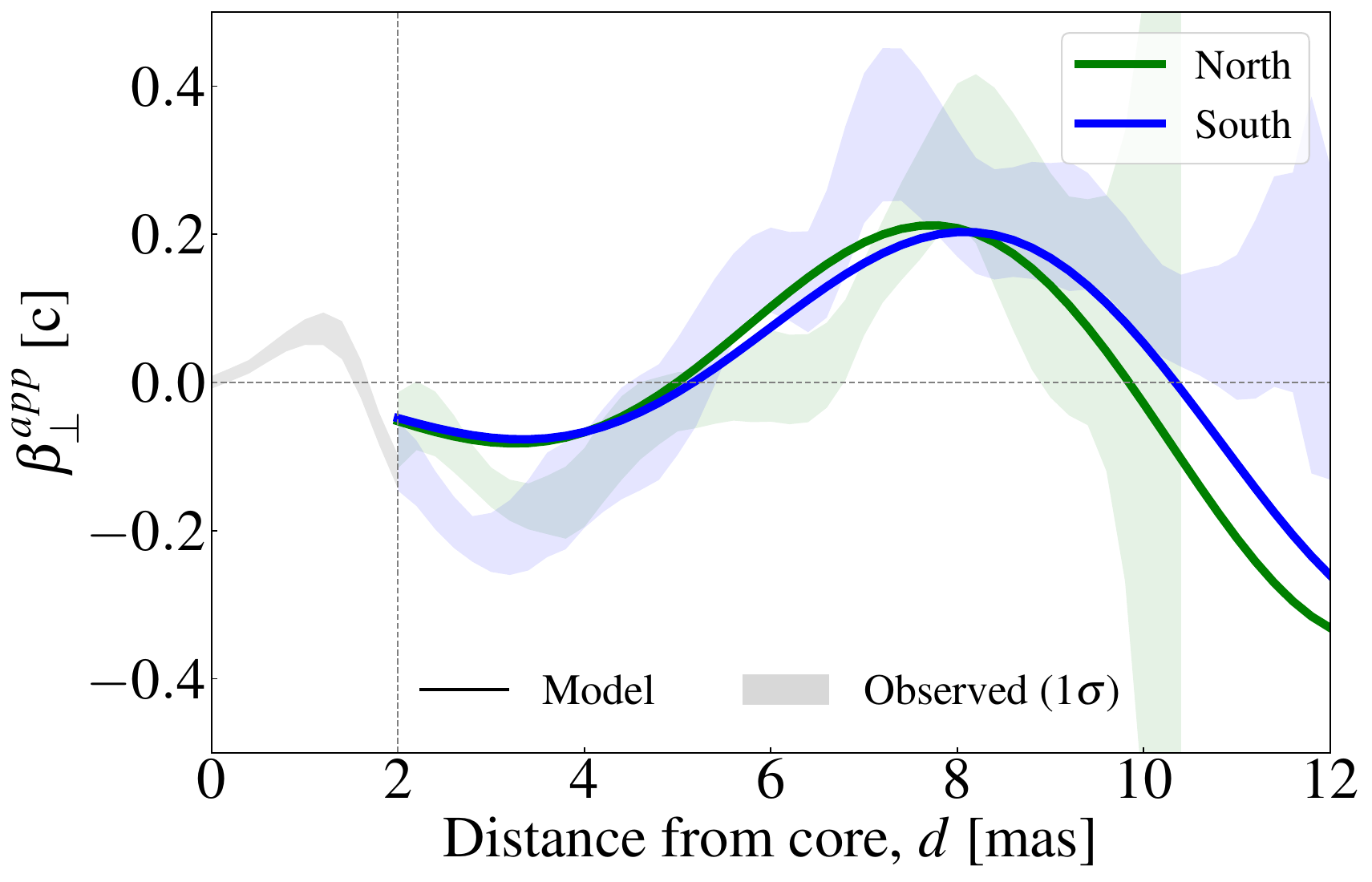}
        \includegraphics[width=\textwidth]{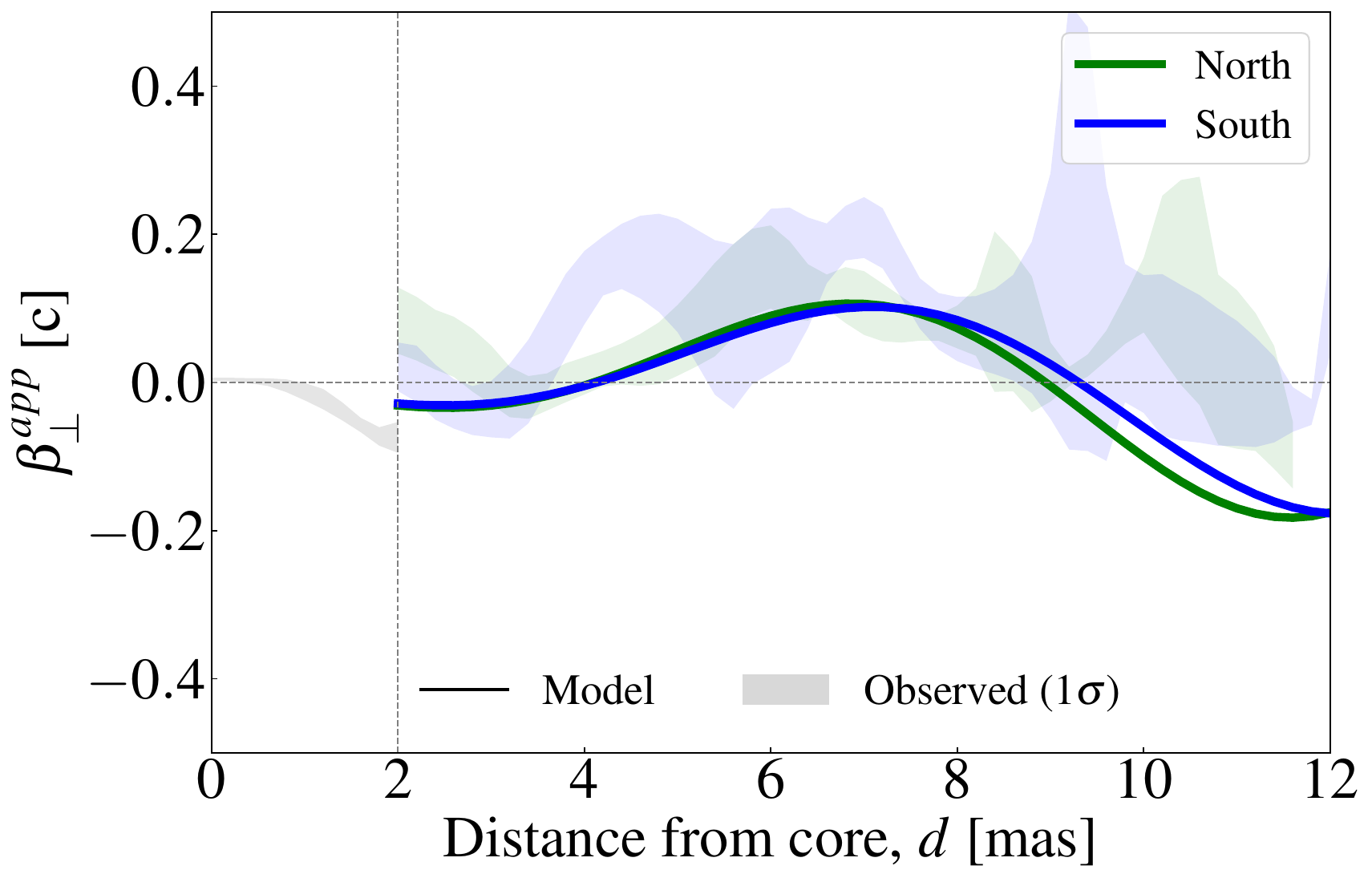}
        \includegraphics[width=\textwidth]{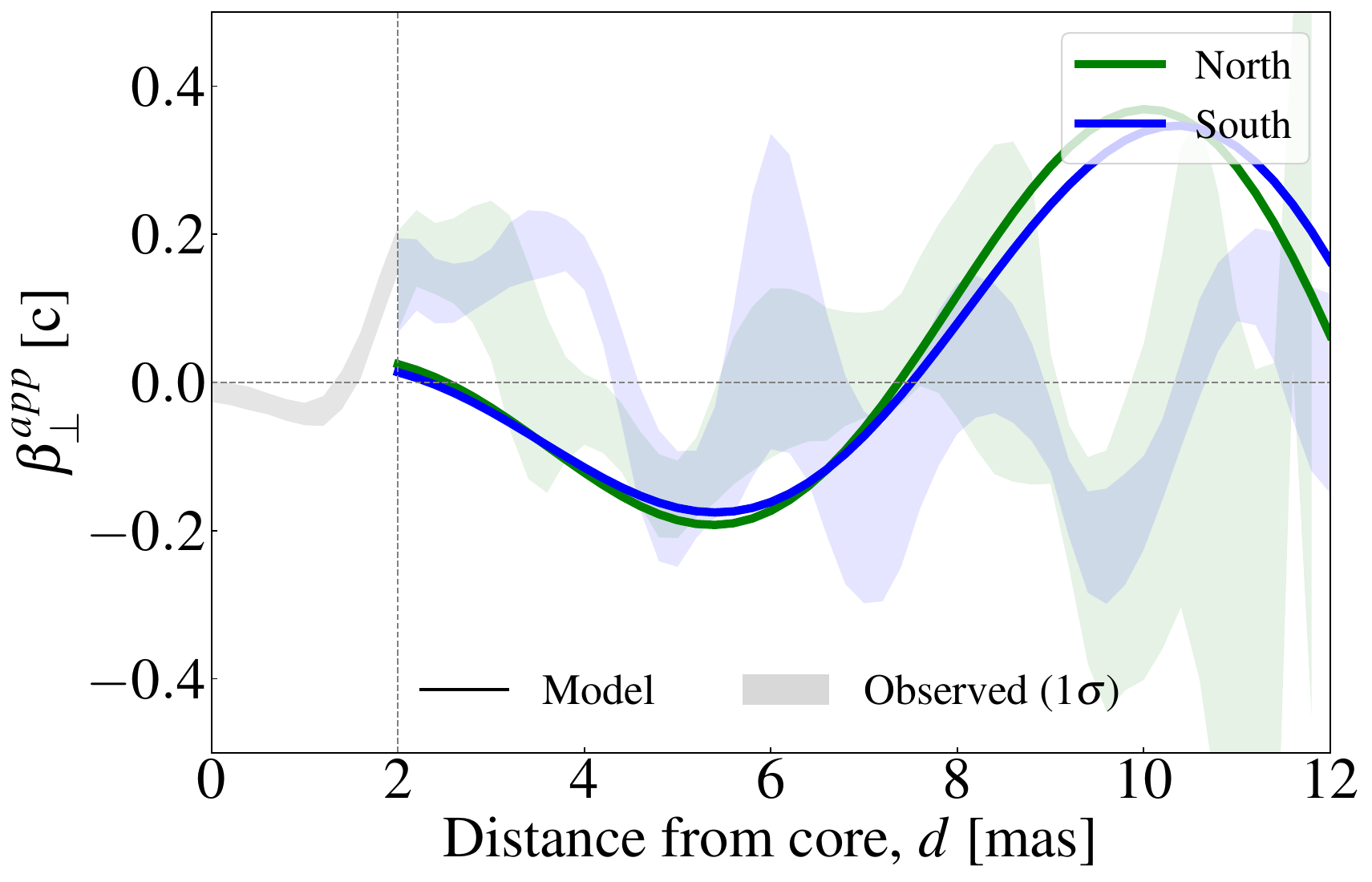}
    \end{minipage}    
        \caption{Comparison of the best‐fit single-mode wave model with the observed transverse velocity profiles of the M87 jet. Left panels: Model‐generated ridge lines plotted at the same epochs as the observations. The color gradient from purple to yellow indicates the time evolution, and the thick gray line denotes the underlying jet structure $y_0(d)$ (see Equation \ref{eq:wave_model}). Right panels: Transverse velocity profiles derived from the model (solid curves) over the corresponding time intervals, overlaid with the observed $1\,\sigma$ error intervals (shaded regions). In each right panel, the green and blue lines correspond to the northern and southern ridges, respectively.}
    \label{fig:bestfit_model_transv}
\end{figure*}

\begin{figure*}
    \centering
    
    \includegraphics[width=1.0\textwidth]{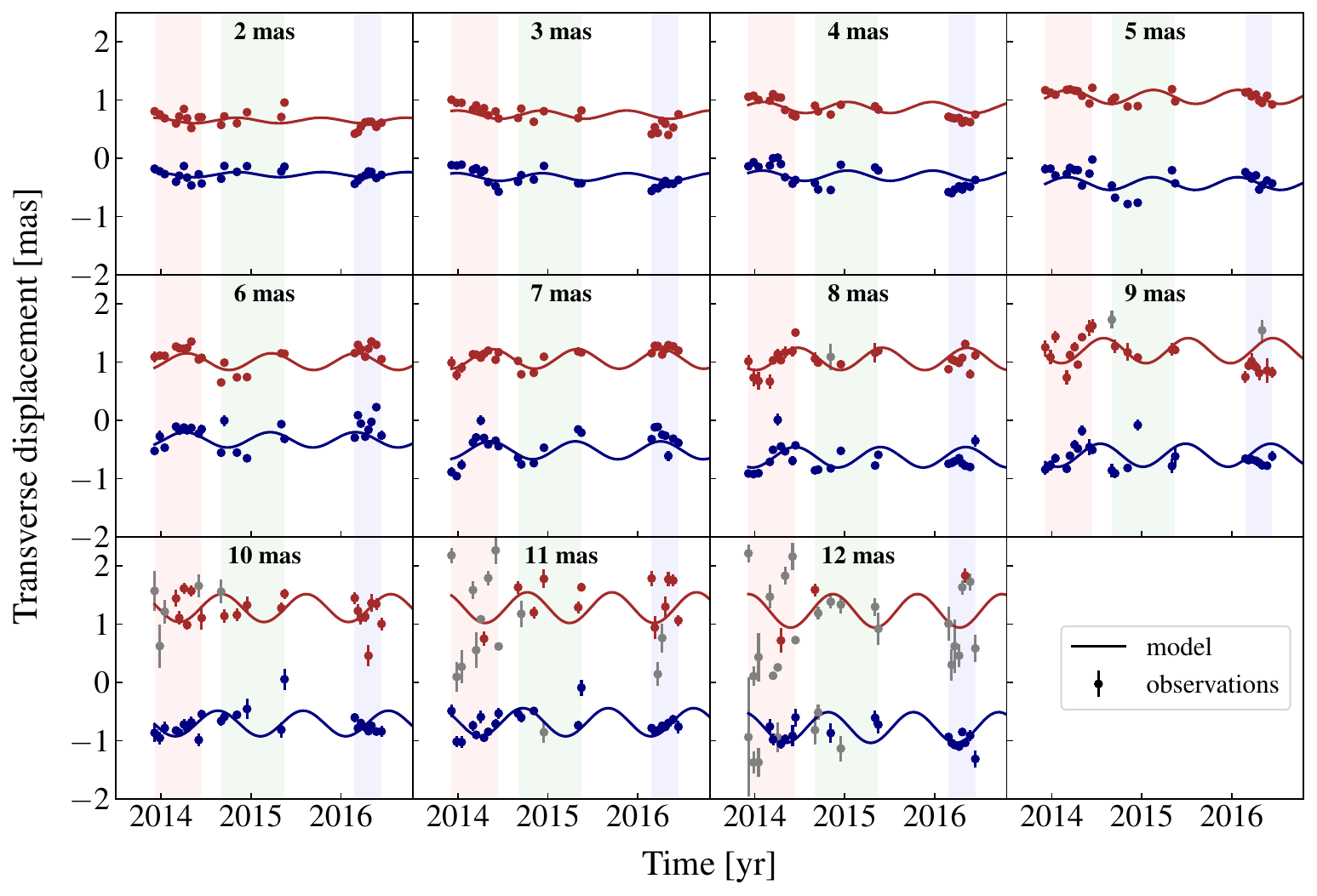}

        \caption{Comparison of observed and modeled transverse oscillations at distances of 2 $-$ 12 mas (in 1\,mas increments) from the core. Each panel corresponds to a specific distance, labeled at the top. The horizontal axis is time (years), and the vertical axis is the transverse displacement (mas) relative to the jet axis. The solid lines show the best‐fit wave model, while the points indicate the observed transverse displacements. Red and blue colors represent the northern and southern limbs, respectively, and grey colors mark excluded data points. The background shaded regions indicate the three periods used for the transverse velocity analysis in Figures \ref{fig:ridge_line_distribution_and_transverse_velocity_fields} and \ref{fig:bestfit_model_transv}: 2013--2014 (red), 2014--2015 (green), and 2016 (blue).}
    \label{fig:bestfit_model_oscillation}
\end{figure*}

We fitted the single-mode wave model to all ridge lines extracted from the monitoring dataset over the 2 -- 12\,mas region, including both the northern and southern limbs. 
The fitting was performed using the MCMC method with 100,000 samples. 
To directly estimate the apparent (projected) wave propagation speed, we sampled the wave velocity (\( v_{\mathrm{wave}} \)) instead of the wavelength, and computed the corresponding wavelength via \( \lambda_{\mathrm{wave}} = v_{\mathrm{wave}} P_{\mathrm{wave}} \).

Figure~\ref{fig:corner_plot} presents the posterior distributions of the fitted parameters, and the best-fit values are summarized in Table~\ref{tab:wave_mcmc_result}.
The derived wave periods for the northern and southern limbs are nearly identical, with $P_{\mathrm{wave,\,north}} = 0.940 \pm 0.002$ years and $P_{\mathrm{wave,\,south}} = 0.948 \pm 0.002$ years, consistent with the average period of the transverse oscillations shown in Figure \ref{fig:fitting_parameters_distribution}.
The amplitudes at {$d_{\mathrm{f}}=12$}\,mas are $A_{\rm{wave, north}} = 0.288\pm0.005$\,mas and $A_{\rm{wave, south}} = 0.264\pm0.004$\,mas.
The best-fit wavelengths are estimated as $\lambda_{\rm{wave, north}} = 9.698^{+0.188}_{-0.195}$\,mas, $\lambda_{\rm{wave, south}} = 10.345^{+0.107}_{-0.104}$\,mas, in good agreement with the characteristic wavelengths derived independently from the transverse velocity profile analysis (Section~\ref{sec:3}) and the direct sinusoidal fitting of the ridge lines (Section~\ref{sec:4.1}).
The fitted wave phases are nearly identical within uncertainties, with $\phi_{\rm{wave, north}} = 3.067^{+0.079}_{-0.074}$\,rad and $\phi_{\rm{wave, south}} = 2.986^{+0.044}_{-0.044}$\,rad, suggesting that the waves propagating along both limbs are coherent.

We estimate the wave propagation speeds from this analysis as $v_{\rm{wave, north}}=10.319^{+0.208}_{-0.215}\rm{\,mas\,yr^{-1}}$ and $v_{\rm{wave, south}}=10.913^{+0.117}_{-0.115}\rm{\,mas\,yr^{-1}}$ on the north and south ridges, respectively. 
Using a conversion factor of $0.264\,\rm{c} / \rm{mas\,yr^{-1}}$ \citep{walker2018}, we find that 
the apparent speeds of the waves are superluminal: $\beta^{\rm{app}}_{\rm{wave, north}} = 2.724^{+0.055}_{-0.057}$ and
$\beta^{\rm{app}}_{\rm{wave, south}} = 2.881^{+0.031}_{-0.030}$.
Interestingly, these speeds are slightly faster than the global bulk jet flow speed at the same distance
\citep[$\beta^{\mathrm{app}}_{\mathrm{flow}} \sim 0.5 - 1.7$;][]{park19},
but are comparable to the ``fast velocity component'' reported by \citet{mertens16} ($\beta^{\mathrm{app}}_{\mathrm{flow}} \sim 2 - 3$).

In Figure~\ref{fig:bestfit_model_transv}, we compare the best-fit single-mode wave model with the observations based on the transverse velocity profiles. 
The left panels show a series of model ridge lines generated for the three observation periods, with colors matching those in Figure~\ref{fig:ridge_line_distribution_and_transverse_velocity_fields}. 
The right panels display the corresponding time-averaged transverse velocity profiles ($\beta^{\text{app}}_{\perp}$), overlaid with the observed values shown as blurred dotted lines.
During the first two periods (December 2013 -- June 2014 and September 2014 -- May 2015), the model reproduces the overall trends in the observations with high fidelity. 
However, it fails to capture smaller-scale variations, which may correspond to short-wavelength components revealed by the power spectrum analysis (Figure~\ref{fig:power_spectrum_vtrans}).
In the 2016 data, where short-wavelength oscillations are more pronounced, the modeled $\beta^{\text{app}}_{\perp}$ exhibits increasingly significant deviations from the observed values, particularly beyond $\sim$8\,mas.

Figure \ref{fig:bestfit_model_oscillation} shows the transverse displacement of the ridge lines relative to the jet axis, as determined from the best-fit model (solid lines) and observations (circles) at 1\,mas intervals over a range of 2 -- 12\,mas.
The northern and southern limbs are plotted in red and blue, respectively, 
with grey dots indicating points excluded from the fitting (primarily due to SNR $<3$; see Section~\ref{sec:2.1}).
Overall, the single-mode wave model captures the essential features of the observed transverse oscillations; however, in the downstream region, noticeable discrepancies emerge.
For example, around 9\,mas, substantial offsets between the model and 2016 observations are apparent.
Furthermore, beyond 10\,mas, particularly in the northern limb, the data exhibit increased scatter and significant deviations from the model.
These discrepancies are partly due to systematic errors resulting from decreasing limb intensities at larger distances, which complicate the precise determination of the ridge lines. 
However, the complex oscillatory behavior observed downstream may also be attributable to intrinsic effects, which are further discussed in Chapter \ref{sec:5}.

In summary, the prominent features of the transverse oscillations in the M87 jet can be explained as the propagation of monochromatic waves with apparent superluminal speeds of approximately 2.7 to 2.9\,$c$. Nevertheless, we observe discrepancies between the model and the data at specific distances and periods, particularly in the downstream region. 
These differences suggest that the oscillations may not be perfectly periodic or that additional small-scale oscillations are present.

\section{Possible origins of the waves on the jet}\label{sec:5}

In this study, we refine our analysis of the fast transverse oscillations in the M87 jet reported by \citet{ro23} and demonstrate that these oscillations can be interpreted as superluminal waves propagating along the jet. 
Here we discuss several potential mechanisms for this phenomenon, which may not be mutually exclusive but could be interrelated. 

\subsection{Propagation of MHD wave launched from the vicinity of the black hole}\label{sec:5.1}
For highly magnetized jets, the transverse displacement of the magnetic field (and accompanying plasma) can give rise to transverse MHD waves (i.e., Alfv\'{e}n waves) due to the restoring force of magnetic tension \citep[e.g.,][]{meier2012}.
To date, only a few cases have reported evidence of Alfv\'{e}n wave propagation in AGN jets \citep{cohen15, cohen2017}.
In one notable example, \citet{cohen15} observed transverse patterns moving superluminally downstream in the jet of BL Lacertae (BL Lac), with apparent speeds ranging from $3.9\,c$ to $13.5\,c$,
which they interpreted as a transverse Alfv\'{e}n waves.

The waves in the M87 jet identified in our study appear to originate from a region different from that in BL Lac.
In BL Lac, the excitation of the Alfv\'{e}n wave seems to be triggered by position angle changes at the recollimation shock, located at scales of approximately $\sim2\times 10^5 r_s - 10^6 r_s$ ($\sim 3 - 15$ pc) from the core \citep{cohen14, cohen15}. 
In contrast, our results reveal that similar wave phenomena in M87 arise at scales of roughly $\sim 10^3 r_s$ ($\sim 0.6$ pc), a region much closer to the SMBH where jet acceleration and collimation actively occur.
Furthermore, the waves observed in BL Lac propagate at significantly higher apparent speeds, likely reflecting the different regions of origin.
Notably, no studies have reported Alfv\'{e}n wave propagation beyond a recollimation shock region (i.e., HST-1) in the M87 jet, and the presence of such high-speed waves in M87 remains unconfirmed.

If the observed waves are indeed Alfv\'{e}n waves, we can constrain the mass density of the jet ($\rho$) from their propagation speed, because the Alfv\'{e}n speed is determined by the magnetic field strength and mass density \citep[$V_A = {B}/({4\pi\rho})^{1/2}$;][]{meier2012, cohen14}.

Assuming a viewing angle of 17 degrees \citep{walker2018}, 
the intrinsic wave speed in the galaxy frame ($\beta_{\rm wave}$) can be derived from the apparent propagation speed ($\beta^{\rm app}_{\rm wave}$) using

\begin{equation}
    \beta_{\rm wave} = \frac{\beta^{\rm app}_{\rm wave}}{\sin\theta + \beta^{\rm app}_{\rm wave} \cos\theta}\label{eq:apparent_speed}
\end{equation}
If we take $\beta^{\rm app}_{\rm wave} \sim 2.7 - 2.9$ (Section \ref{sec:4.2}), then $\beta_{\rm{wave}} \sim 0.94 - 0.95$.
The wave speed in the jet frame ($\beta^{\rm jet}_{\rm wave}$) can then be calculated using a relativistic subtraction formula \citep[e.g.,][]{cohen15}:

\begin{equation}
    \beta^{\rm jet}_{\rm wave} = \frac{\beta_{\rm wave} - \beta_{\rm jet}}{1 - \beta_{\rm wave}\beta_{\rm jet}}\label{eq:rel_subtraction}
\end{equation}where $\beta_{\rm jet}$ is the jet flow speed, which is estimated to be $\sim 0.65 - 0.89$ in the $2-10$\,mas region \citep{park19}.
This yields $\beta^{\rm jet}_{\rm wave} \sim 0.31 - 0.78$ at $2-10$\,mas ($\sim 10^3 r_s$ in de-projected distance).
At this distance, the magnetic field strength is $B\sim0.1 - 1.0$ G \citep[][]{ro23_spix}.
Using these values, we estimate the possible range of the mass density $\rho$ as $\sim10^{-24} - 10^{-21} \rm{g/cm^3}$.
For comparison, the electron number density near the event horizon estimated from EHT observations is $n_e \sim 10^4 - 10^7 \rm{cm^{-3}}$ \citep{ehtc21, eht_mwl21}.
If we assume $n_e = n_p$ and a parabolic expansion of the jet, we obtain $\rho \sim10^{-24.5} - 10^{-21.5}\rm {g/cm^3}$ at $10^3 r_s$, which is in agreement with our estimation.

Alfv\'{e}n waves are analogous to transverse waves on a taut string, where disturbances at one end travel along its length.
This analogy implies that an upstream source is necessary to initiate these waves.
Given that wave propagation in M87 begins near the core, it is reasonable to attribute the excitation of these waves to processes occurring in the vicinity of the SMBH.

One possible source for generating MHD waves is the precession or nutation of the SMBH and accretion systems.
The one-year transverse oscillation identified in this study is much shorter in timescale than the $\sim$11-year variation in the position angle of M87’s innermost jet, which has been interpreted as Lense-Thirring precession \citep{cui23}.
Additionally, a structural variation on a similar long-term timescale has been reported to propagate at an apparent speed of $\beta^{\rm app}\approx0.81-0.89$ \citep{walker2018}, which is significantly slower than the speed associated with the one-year oscillation ($\beta^{\rm app}\sim2.7-2.9$).
These discrepancies suggest that the transverse oscillations in the M87 jet likely consist of at least two distinct components with different periods and speeds\footnote{In Appendix \ref{sec:ab}, we further explore the possibility of multiple wave propagation modes based on our data.}.
If the slower mode arises from precession, the faster one may correspond to nutation. 
Such a combination has been proposed in other astrophysical systems. 
For instance, long-term VLBI monitoring of the blazar OJ\,287 has been modeled with both jet precession ($\sim$27 years) and nutation ($\sim$1.6 years), yielding a period ratio of $\sim$17 \citep{britzen18, britzen23}, comparable to the ratio seen in M87 (11.2\,yr / 0.95\,yr $\sim$ 12).
A similar precession-to-nutation ratio is also observed in the X-ray binary SS\,433 \citep[16.2 days/6.5 days $\sim$25;][]{britzen18}. 
These analogies across different systems may support to the idea that M87 may exhibit both precession and nutation, potentially exciting multiple wave modes in the jet.

Another potential source of Alfv\'{e}n wave excitation near the SMBH is the ``magnetic flux eruption'' seen in magnetically arrested disks \citep[MADs;][]{narayan2003}.
GRMHD simulations have shown that quasi-periodic eruptions of magnetic flux in the innermost accretion flow can induce variability in the jet’s energy output and launch waves along the jet–wind interface \citep[e.g.,][]{tchekhovskoy11, mckinney12, ripperda2022, davelaar2023, chatterjee2023, lalakos2024, moriyama2024}.
Recent EHT polarization results suggest that M87’s accretion flow is in a MAD state \citep{ehtc21, eht2021_pol, ehtc23}, supporting the plausibility of this mechanism.
These simulations indicate characteristic timescales for flux eruptions of $10^3-10^4\,r_g/c$.
For M87, a timescale of $10^3\,r_g/c$ corresponds to approximately one year, suggesting that magnetic flux eruptions in MADs could naturally explain both the $\sim$1-year and $\sim$11-year periodicities observed in the M87 jet \citep[e.g.,][]{davelaar2020, lalakos2024}. 
Furthermore, additional oscillatory modes ranging from sub-week to decade-long periods may also arise \citep[][]{mckinney12, chatterjee2023}, raising the possibility of yet undetected variability components in the M87 jet\footnote{Recently, \citet{kino2025} discussed an alternative possibility of an SMBH binary in M87, which could also potentially trigger MHD waves. However, this scenario lies beyond the scope of this paper.}.

\subsection{Propagation of jet instabilities}\label{sec:5.2}

The propagation of distorted jet structures driven by instabilities is a plausible explanation for features observed in many AGN jets \citep[e.g.,][]{lobanov_zensus2001, lobanov03, hardee2005, hardee_eilek11, perucho2012, fromm13, cohen15, walker2018, vega-garcia2020, pasetto21, issaoun2022, jorstad2022, jorstad2023, nikonov23, fuentes23}.
Two main instability types are commonly considered \citep[see][]{hardee11, perucho2012_review, mizuno2016, perucho19}: the Kelvin–Helmholtz instability (KHI), which arises from velocity shear in kinetically dominated jets, and the current-driven instability (CDI), which develops in magnetically dominated regions with strong toroidal fields.
These instabilities can be triggered by perturbations near the SMBH and propagate downstream, deforming the jet structure \citep[e.g.,][]{lalakos2024}.
Our observations show coherent transverse motions in both limbs, which may be interpreted as either a KHI helical mode or a CDI kink mode.

Instabilities in the M87 jet have been studied across multiple scales and are often interpreted as manifestations of the KHI, with twisted filamentary structures in the kiloparsec-scale jet and spiral or helical patterns on smaller scales attributed to its development \citep[e.g.,][]{lobanov03, hardee_eilek11, walker2018, pasetto21, nikonov23}.
Such interpretations typically assume kinetically dominated flows with low magnetization.
However, MHD simulations suggest that KHI is suppressed in highly magnetized jets, especially when surrounded by a slower, magnetized sheath \citep[e.g.,][]{hardee2007, mizuno07, bodo2013}.
The region probed in this study lies within 12\,mas ($\sim5 \times 10^3\,r_s$) of the core—inside the acceleration and collimation zone (ACZ) \citep{kino15, kim18}, where magnetic-to-kinetic energy conversion is expected to occr \citep[e.g.,][]{boccardi2017, blandford2019, hada2024}.
Observations indicate that the ACZ in M87 extends out to $\sim10^6\,r_s$ \citep{asada12, hada13, nakamura18, park19}.
Since our observations focus on the innermost part of this region, the jet is likely to remain highly magnetized\footnote{\citet{park19} reported a relatively slow acceleration trend in the M87 jet compared to theoretical expectations, implying inefficient energy conversion and further supporting the strong magnetization.}.
Additionally, the rotation measure profiles further support the presence of a magnetized wind \citep{park19_fara}. 
These findings suggest that in the observed region, while it cannot be entirely ruled out, KHI is likely disfavored, and CDI may instead dominate.

The $\sim$1-year period transverse oscillations reported in this work follow a trend similar to that expected from the evolution of CDIs.
According to MHD simulations, CDIs in AGN jets evolve through a linear growth phase followed by a nonlinear phase \citep[e.g.,][]{mizuno09, mizuno11, mizuno14, o'neill2012, bodo2013, bodo2016, bodo2019, anjiri2014, bromberg2019}. 
In the linear stage, the fastest-growing kink mode—typically with a wavelength $\sim$10 times the jet width—generates a helical structure that induces sinusoidal transverse oscillations as the jet propagates downstream with the transverse velocity increasing exponentially until it peaks \citep[e.g.,][]{mizuno14, lalakos2024}. 
Once the system enters the nonlinear phase, the instability saturates through interactions with newly emerging modes, leading to a gradual decline in the transverse kinetic energy and the development of increasingly complex oscillation patterns \citep[e.g.,][]{mizuno09}.
In our observations, the oscillation amplitude increases gradually with distance, peaking around 7\,mas before becoming less pronounced (Figure~\ref{fig:fitting_parameters_distribution}). 
Similarly, the initially sinusoidal patterns become increasingly irregular downstream (Figure~\ref{fig:bestfit_model_oscillation}), consistent with the transition from the linear to nonlinear regime of CDIs.

We can estimate the transition from the linear to nonlinear CDI regime by comparing the instability’s growth timescale with the jet’s propagation time. 
The linear growth of the CDI kink mode typically lasts for 10 -- 100 Alfv\'{e}n crossing times ($\tau_A$), where $\tau_A$ is defined as the jet radius divided by the Alfv\'{e}n speed \citep[e.g.,][]{appl00, mizuno09, o'neill2012, anjiri2014, sobacchi2017, bromberg2019}.
Assuming a highly magnetized jet with Alfv\'{e}n speed $\approx c$ and a width of 1 -- 2\,mas (0.08 -- 0.16\,pc; Figure~\ref{fig:fitting_parameters_distribution}), we estimate $\tau_A \sim 0.26 - 0.52\, \mathrm{yr}$. 
The jet travel time over the 2 -- 10\,mas region is $\sim 6.9\,\mathrm{yr}$ (or $\sim2.4\,\mathrm{yr}$ for the fast component; see Appendix D of \citet{ro23_spix}), corresponding to $\sim 13 - 27\,\tau_A$ (or 4 -- 9\,$\tau_A$).

This suggests that the jet is likely still in the linear growth phase in the observed region. Further downstream, the accumulation of instability growth time (likely exceeding $\sim 100\,\tau_{A}$) combined with a weakening of the magnetic field is expected to facilitate a transition to KHI dominance, as supported by the detection of large-scale helical threads attributed to KHI extending up to $\sim 200\,$mas \citep{nikonov23}.

On the other hand, some studies suggest that nonlinear evolution could begin as early as $\sim$2\,mas \citep{sobacchi2017}\footnote{We note that differences in the assumed jet radius and black hole mass may also contribute to discrepancies in interpretation.}. 
Additionally, lower signal-to-noise ratios in downstream regions may increase uncertainties in ridge-line extraction, potentially contributing to the apparent complexity.
A full understanding of instability development in the M87 jet’s acceleration and collimation zone will require higher-sensitivity observations and further theoretical modeling.

It is noteworthy that the observed wave propagation speed ($\beta_{\mathrm{app}} \sim 2.7 - 2.9$) closely matches the fastest components of the bulk jet velocity profile reported in previous studies \citep[$\beta_{\mathrm{app}} \sim 2 - 3$;][]{mertens16, park19}.
According to these studies, the M87 jet exhibits velocity stratification, interpreted either as a faster jet stream interacting with a slower outer wind or as multiple streamlines with different acceleration profiles.
In this context, the observed one-year oscillations can be understood as CDIs propagating along the faster component within the stratified jet.
Such wave speeds are also in line with numerical simulations, which show that kink modes typically propagate at speeds comparable to the flow velocity \citep[e.g.,][]{mizuno11, mizuno14, singh16, lalakos2024}.

In summary, we propose that the transverse oscillations observed in the M87 jet may be explained by the propagation of Alfv\'{e}n waves or jet instabilities. 
The characteristics of the Alfv\'{e}n wave are determined by the excitation source near the central engine.
For example, if the source is periodic (e.g., due to jet precession or nutation), then the resulting transverse oscillations will exhibit periodic behavior.
In contrast, magnetic eruption events from MADs can produce quasi-periodic oscillations spanning a broad range of periods.
Meanwhile, such perturbations may also trigger instabilities that propagate along the jet.
In the magnetically dominated region near the jet base, CDIs are mainly considered.
Upstream, the oscillations appear as clear sinusoidal patterns dominated by the fastest-growing mode, while downstream, increasingly complex oscillations arise by interacting with different modes and different types of instabilities.

\section{Summary and Conclusions}\label{sec:6}

In this study, we perform an in-depth investigation of the transverse oscillations of the M87 jet with a period of $\sim1$ year, revealed by the high-cadence monitoring observations of KaVA at 22\,GHz from December 2013 to June 2016 \citep{ro23}.
The primary findings and major conclusions drawn from this study are summarized below:

\begin{enumerate}\itemsep3pt
    \item Firstly, we divided the entire observation period into three time ranges and derived the average transverse velocity profile during each period.
    We found that the transverse velocity profiles exhibit a sinusoidal pattern that changes direction with both time and distance, showing an apparent transverse velocity of up to $\beta^{\text{app}}_{\perp}\approx0.4$c. 
    Based on both the morphology of the velocity profile and the results of the power spectrum analysis, we deduce that the dominant component has a wavelength of $\gtrsim 8.5$\,mas, although smaller-scale features may also be present.

    \item Next, we performed a local sinusoidal fitting to the time evolution of the ridge-line displacements at individual radial distances. We examined the variation of the best-fit parameters as a function of distance and found that the phase of the oscillation steadily increases with distance and returns to its initial value after $\sim9$\,mas, while the amplitude gradually increases until $\sim7$\,mas, beyond which the growth becomes insignificant. 
    Together, these results suggest that the one-year transverse oscillations propagates downstream as a wave with a wavelength of $\sim9$\,mas.
    Additionally, we found nearly identical trends in the amplitude and phase of the oscillations on the northern and southern limbs, indicating that the waves on both limbs are coherent.    
    
    \item We applied a global single-mode wave model to the entire ridge line dataset and found that the waves propagate with superluminal speeds of approximately 2.7 -- 2.9$\,c$. 
    Overall, the model shows good agreement with the observed ridge line evolution; however, noticeable offsets between the model and observations were found at specific times and locations, particularly in the downstream region.
    These discrepancies suggest that, while the transverse oscillations are largely explained by the propagation of periodic waves, a non-periodic component may also contribute to the observed patterns.
    
    \item The observed transverse oscillations may be explained by either the propagation of MHD waves or the development of jet instabilities. In the wave scenario, Alfvén waves excited near the central engine—possibly triggered by jet precession/nutation or magnetic flux eruptions in a MAD—could account for the behavior. Alternatively, the oscillations may be driven by CDIs, which are favored given the jet’s likely highly magnetized, though some contribution from KHI cannot be excluded, especially farther downstream where the magnetic field may weaken.

\end{enumerate}

While this study focuses on the first three years of KaVA observations, we continue to monitor M87 through ongoing high-cadence VLBI programs using the EAVN \citep[East Asian VLBI Network;][]{eavn2022} and EATING VLBI \citep[From East Asia to Italy: Nearly Global VLBI;][]{giovannini23} arrays. 
These extended observations, with improved angular resolution and sensitivity, will allow us to probe finer-scale transverse structures and characterize longer-term variability \citep[e.g.,][see Appendix \ref{sec:ab}]{walker2018, cui23}.
In parallel, advances in imaging techniques, such as regularized maximum likelihood and Bayesian methods, will enable more precise reconstruction of ridge line variations \citep[e.g.,][]{chael2018, tiede2022, tazaki2023, park2024, foschi2024, kim_2024_resolve}. 
Together, these developments will facilitate a deeper understanding of the physical origin, multiplicity, and evolution of transverse oscillations in the M87 jet.

\begin{acknowledgments}
H.R. is supported by the National Research Council of Science \& Technology (NST) -- Korea Astronomy and Space Science Institute (KASI) Postdoctoral Fellowship Program for Young Scientists at KASI in South Korea.
This research was supported by funding from the Korea government (Korea AeroSpace Administration (KASA), grant number R25TA0065942000).
Y.M. is supported by the National Key R\&D Program of China (grant No. 2023YFE0101200), the National Natural Science Foundation of China (grant Nos. 12273022, 12511540053), and the Shanghai Municipality Orientation Program of Basic Research for International Scientists (grant no. 22JC1410600).
Y.C. is supported by the Natural Science Foundation of China (grant 12303021) and the China Postdoctoral Science Foundation under Grant Number 2024T170845. This work was supported by the National Research Foundation of Korea (NRF) grant funded by the Korea government (MSIT; RS-2024-00449206; RS-2025-02214038). This research has been supported by the POSCO Science Fellowship of POSCO TJ Park Foundation. 
This research was supported by Global-Learning \& Academic research institution for Master's \textperiodcentered~PhD students, and Postdocs (G-LAMP) Program of the National Research Foundation of Korea (NRF) grant funded by the Ministry of Education (RS-2025-25442355). 
This work was partially supported by the MEXT/JSPS KAKENHI (JP22H00157, JP23H00117). K.H. is supported by MEXT/JSPS KAKENHI (grants 25H00660, 22H00157, 21H04488), Mitsubishi Foundation (grant 202310034) and Daiko Foundation (grant J0SE807004). This work was supported in part by a University Research Support Grant from the National Astronomical Observatory of Japan (NAOJ), and by Grant-in-Aid for Outstanding Research Group Support Program in Nagoya City University Grant Number 2530002.
\end{acknowledgments}

\vspace{5mm}
\facilities{KaVA (KVN and VERA Array)}

\software{AIPS \citep{greisen2003},
          Difmap \citep{shepherd1997},
          astropy \citep{astropy2018},
          numpy \citep{numpy2020},
          matplotlib \citep{matplotlib2007},
          emcee \citep{emcee2013}
          }

\appendix

\section{Interpretation of the power spectrum of the transverse velocity profile}\label{sec:aa}

In Section \ref{sec:3}, we constructed power spectra of the velocity profiles of the M87 jet to estimate the oscillation wavelength (Figure \ref{fig:power_spectrum_vtrans}).
The velocity profile was sampled every 0.2\,mas.
For uniformly sampled data, the power spectrum captures all relevant frequency components present in the signal \citep{vanderplas18}.
However, due to the sensitivity limit of the observations, the velocity profile is available only up to 12\,mas from the core.
This is equivalent to applying a rectangular window of width 12\,mas to the data.
In the Fourier domain, this corresponds to a convolution with a sinc function of width $1/12\,\mathrm{mas}^{-1}$.
As a result, the frequency axis of the power spectrum is discretized with a resolution of approximately $0.08\,\mathrm{mas}^{-1}$ \citep{vanderplas18}.
This limited spectral resolution complicates precise wavelength estimation, especially when the wavelength is comparable to the total spatial extent of the data.

To better understand and interpret the power spectrum under these conditions, we generated artificial sinusoidal curves of total length 12\,mas with various wavelengths ($\lambda_{\mathrm{sine}}$), and examined their resulting power spectra. 
This allowed us to identify artifacts arising from limited frequency resolution. 

\begin{figure}
    \centering
    \includegraphics[width=0.45\textwidth]{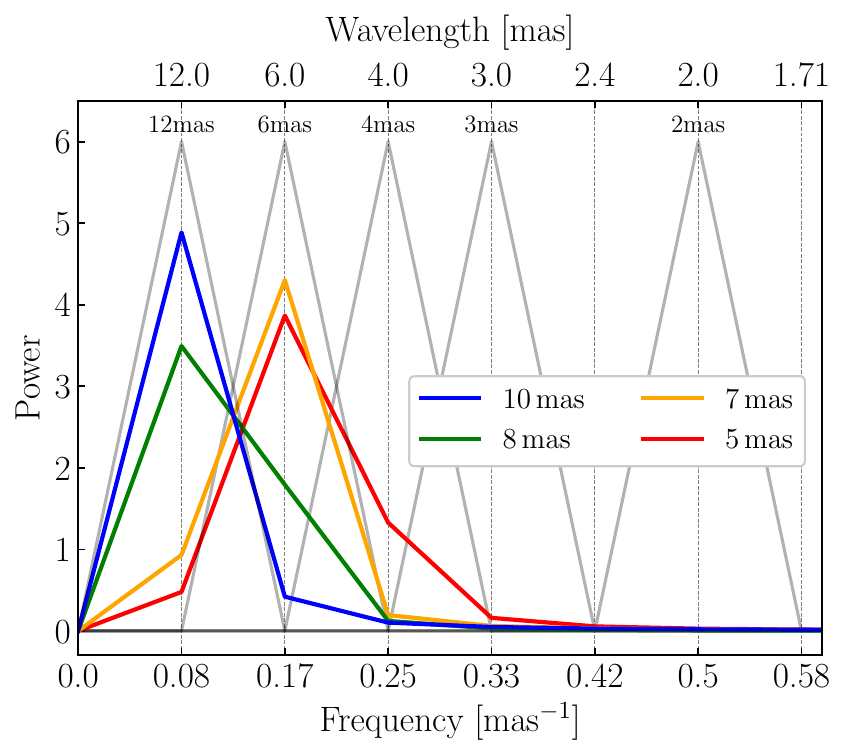}
        \caption{Power spectra of artificial sinusoidal curves with different wavelengths $\lambda_{\mathrm{sine}}$. The bottom axis shows the spatial frequency (mas$^{-1}$), and the top axis shows the corresponding wavelength (mas). Gray curves represent cases where $\lambda_{\mathrm{sine}}$ exactly matches the discrete frequency bins (2, 3, 4, 6, and 12\,mas), producing single peaks. Colored curves correspond to non-bin-centered values of $\lambda_{\mathrm{sine}}$ (5, 7, 8, and 10\,mas), which result in asymmetric spectral distributions due to limited frequency resolution.}
    \label{fig:power_ideal_sine}
\end{figure}

Figure \ref{fig:power_ideal_sine} summarizes the power spectra of sine curves with $\lambda_\mathrm{sine}$ ranging from 2 to 12\,mas. 
For wavelengths of $\lambda_{\mathrm{sine}}$ = 2, 3, 4, 6, and 12\,mas (grey lines), the power spectra exhibit a single peak at the corresponding frequency bin, with zero power at other frequencies.
However, when the wavelength does not match the discrete frequency bins, the power spectrum displays an asymmetric distribution with a peak at the nearest bin.
In such cases, the skewness of the distribution indicates whether the true wavelength is shorter or longer than the bin-centered value.
For example, $\lambda_{\mathrm{sine}}=5$ and 7\,mas (red and orange line, respectively) both show a peak at $\lambda_{\mathrm{sine}}=6$\,mas ($\sim 0.17\,\mathrm{mas}^{-1}$). 
The power spectrum for $\lambda_{\mathrm{sine}}=5$\,mas exhibits negative skewness (i.e., a tail toward shorter wavelengths), while that of $\lambda_{\mathrm{sine}}=7$\,mas shows positive skewness on the wavelength axis.
Similarly, for $\lambda_{\mathrm{sine}}=8$ and 10\,mas (green and blue lines), the peaks appear at 12\,mas, with a negative skewness.
The distribution becomes increasingly symmetric as $\lambda_{\mathrm{sine}}$ approaches the total window size of 12 mas.

In the power spectrum of the observed velocity profile (Figure \ref{fig:power_spectrum_vtrans}), most first peaks appear around 12\,mas, except in 2016. 
When compared with the synthetic test in Figure \ref{fig:power_ideal_sine}, this suggests that the transverse velocity profile of the M87 jet predominantly contains wavelength components of at least 8\,mas. 
Additionally, the velocity profile of the northern ridge in 2016 shows a peak at 6\,mas; since the skewness is not very pronounced, the wavelength is estimated to be close to 6\,mas.

\section{Hints of Multiple Wave Propagation in the M87 Jet}\label{sec:ab}

\begin{figure*}
    \centering
    \includegraphics[width=0.75\textwidth]{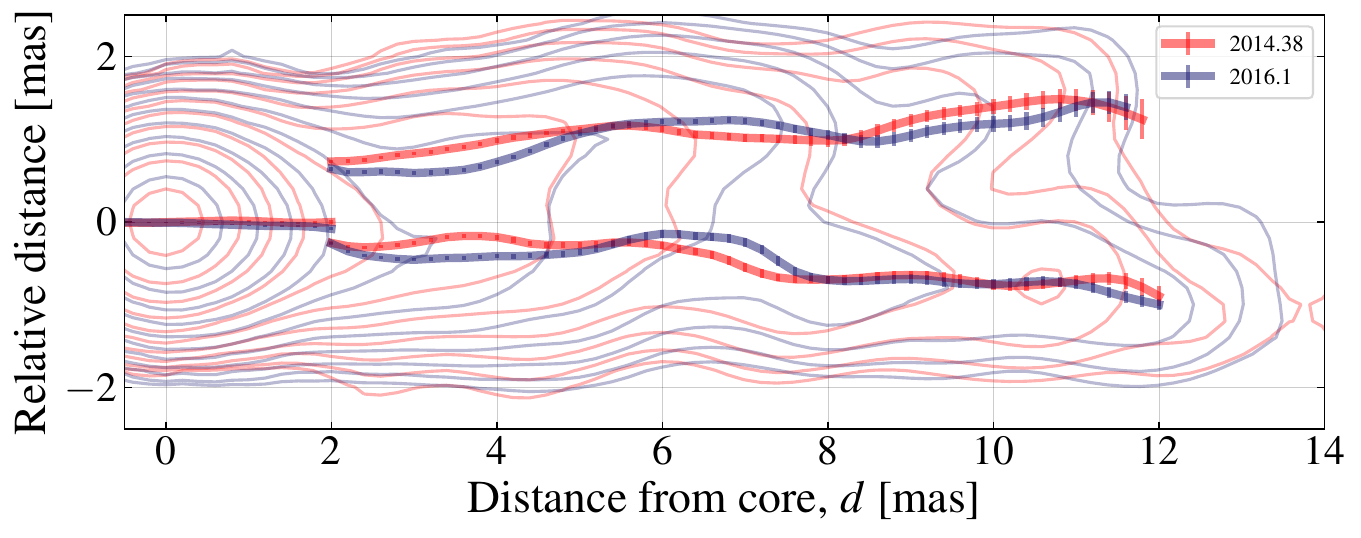}
    \caption{Ridge lines of the M87 jet based on images averaged over two time intervals: December 2013 to December 2014 (red) and May 2015 to June 2016 (blue). 
    Contours indicate the total intensity averaged over each period, and the legend shows the mean epoch of each time range.
    A downstream shift of the ridge line by approximately 2.5\,mas is observed over the two-year interval, suggesting the presence of a slower wave component propagating along the jet.
    }
    \label{fig:stacked_map_ridge_lines}
\end{figure*}

Figure \ref{fig:stacked_map_ridge_lines} shows the ridge lines derived from images averaged over two time intervals: red for the first half of the observations (December 2013 to December 2014), and blue for the second half (May 2015 to June 2016).
The stacked contours for each period are overlaid. 
Since each image represents an average over approximately one year, short-period oscillations are expected to be smoothed out, making it easier to detect longer-period structural trends.

A comparison between the two ridgelines reveals a downstream shift of the jet structure by approximately 2.5\,mas over a $\sim$2-year interval. 
This corresponds to a projected propagation speed of $\sim 1.25\,$mas/yr, or $\beta_{\text{app}} \approx 0.33$. 
This is about one-eighth of the apparent wave speed found in our main analysis ($\beta_{\text{app}}\sim2.7 - 2.9$; Table~\ref{tab:wave_mcmc_result}), but somewhat closer to the value reported by \citet{walker2018}, $\beta_{\text{app}} = 0.89 \pm 0.18$, though still significantly slower.

These results suggest that the M87 jet may host at least two distinct wave components: a fast mode and a slower one. 
The fact that the estimated speed of the slow wave is even lower than that reported by \citet{walker2018} raises the possibility of additional, slower modes being present. 
However, the inferred speed remains uncertain, as it is based on a simple two-epoch comparison of stacked ridge lines. 
Future high-cadence, long-term monitoring will be crucial for better characterizing the nature and multiplicity of wave components in the M87 jet.

\bibliography{main}{}

\begin{thebibliography}{}
\expandafter\ifx\csname natexlab\endcsname\relax\def\natexlab#1{#1}\fi
\providecommand{\url}[1]{\href{#1}{#1}}
\providecommand{\dodoi}[1]{doi:~\href{http://doi.org/#1}{\nolinkurl{#1}}}
\providecommand{\doeprint}[1]{\href{http://ascl.net/#1}{\nolinkurl{http://ascl.net/#1}}}
\providecommand{\doarXiv}[1]{\href{https://arxiv.org/abs/#1}{\nolinkurl{https://arxiv.org/abs/#1}}}

\bibitem[{{Akiyama} {et~al.}(2022){Akiyama}, {Algaba}, {An}, {Asada}, {Asanok}, {Byun}, {Chanapote}, {Chen}, {Chen}, {Cheng}, {Chibueze}, {Cho}, {Cho}, {Chung}, {Cui}, {Cui}, {Doi}, {Dong}, {Fujisawa}, {Gou}, {Guo}, {Hada}, {Hagiwara}, {Hirota}, {Hodgson}, {Honma}, {Imai}, {Jaroenjittichai}, {Jiang}, {Jiang}, {Jiang}, {Jike}, {Jung}, {Jung}, {Kawaguchi}, {Kim}, {Kim}, {Kim}, {Kim}, {Kim}, {Kim}, {Kino}, {Kobayashi}, {Koyama}, {Kramer}, {Lee}, {Lee}, {Lee}, {Lee}, {Li}, {Li}, {Li}, {Li}, {Liu}, {Liu}, {Lu}, {Motogi}, {Nakamura}, {Niinuma}, {Oh}, {Oh}, {Oh}, {Oh}, {Oyama}, {Park}, {Poshyachinda}, {Ro}, {Roh}, {Rujopakarn}, {Sakai}, {Sawada-Satoh}, {Shen}, {Shibata}, {Sohn}, {Soonthornthum}, {Sugiyama}, {Sun}, {Takamura}, {Tanabe}, {Tazaki}, {Trippe}, {Wajima}, {Wang}, {Wang}, {Wang}, {Wang}, {Xia}, {Xu}, {Yan}, {Yang}, {Yeom}, {Yi}, {Yi}, {Yonekura}, {Yoon}, {Yu}, {Yuan}, {Yun}, {Zhang}, {Zhang}, {Zhang}, {Zhao}, {Zhao}, {Zhong}, \& {East Asian VLBI Network Collaboration}}]{eavn2022}
{Akiyama}, K., {Algaba}, J.-C., {An}, T., {et~al.} 2022, Galaxies, 10, 113, \dodoi{10.3390/galaxies10060113}

\bibitem[{{Anjiri} {et~al.}(2014){Anjiri}, {Mignone}, {Bodo}, \& {Rossi}}]{anjiri2014}
{Anjiri}, M., {Mignone}, A., {Bodo}, G., \& {Rossi}, P. 2014, \mnras, 442, 2228, \dodoi{10.1093/mnras/stu1004}

\bibitem[{{Appl} {et~al.}(2000){Appl}, {Lery}, \& {Baty}}]{appl00}
{Appl}, S., {Lery}, T., \& {Baty}, H. 2000, \aap, 355, 818

\bibitem[{{Asada} \& {Nakamura}(2012)}]{asada12}
{Asada}, K., \& {Nakamura}, M. 2012, \apjl, 745, L28, \dodoi{10.1088/2041-8205/745/2/L28}

\bibitem[{{Astropy Collaboration} {et~al.}(2018){Astropy Collaboration}, {Price-Whelan}, {Sip{\H{o}}cz}, {G{\"u}nther}, {Lim}, {Crawford}, {Conseil}, {Shupe}, {Craig}, {Dencheva}, {Ginsburg}, {VanderPlas}, {Bradley}, {P{\'e}rez-Su{\'a}rez}, {de Val-Borro}, {Aldcroft}, {Cruz}, {Robitaille}, {Tollerud}, {Ardelean}, {Babej}, {Bach}, {Bachetti}, {Bakanov}, {Bamford}, {Barentsen}, {Barmby}, {Baumbach}, {Berry}, {Biscani}, {Boquien}, {Bostroem}, {Bouma}, {Brammer}, {Bray}, {Breytenbach}, {Buddelmeijer}, {Burke}, {Calderone}, {Cano Rodr{\'\i}guez}, {Cara}, {Cardoso}, {Cheedella}, {Copin}, {Corrales}, {Crichton}, {D'Avella}, {Deil}, {Depagne}, {Dietrich}, {Donath}, {Droettboom}, {Earl}, {Erben}, {Fabbro}, {Ferreira}, {Finethy}, {Fox}, {Garrison}, {Gibbons}, {Goldstein}, {Gommers}, {Greco}, {Greenfield}, {Groener}, {Grollier}, {Hagen}, {Hirst}, {Homeier}, {Horton}, {Hosseinzadeh}, {Hu}, {Hunkeler}, {Ivezi{\'c}}, {Jain}, {Jenness}, {Kanarek}, {Kendrew}, {Kern}, {Kerzendorf}, {Khvalko}, {King}, {Kirkby}, {Kulkarni}, {Kumar}, {Lee}, {Lenz}, {Littlefair}, {Ma}, {Macleod}, {Mastropietro}, {McCully}, {Montagnac}, {Morris}, {Mueller}, {Mumford}, {Muna}, {Murphy}, {Nelson}, {Nguyen}, {Ninan}, {N{\"o}the}, {Ogaz}, {Oh}, {Parejko}, {Parley}, {Pascual}, {Patil}, {Patil}, {Plunkett}, {Prochaska}, {Rastogi}, {Reddy Janga}, {Sabater}, {Sakurikar}, {Seifert}, {Sherbert}, {Sherwood-Taylor}, {Shih}, {Sick}, {Silbiger}, {Singanamalla}, {Singer}, {Sladen}, {Sooley}, {Sornarajah}, {Streicher}, {Teuben}, {Thomas}, {Tremblay}, {Turner}, {Terr{\'o}n}, {van Kerkwijk}, {de la Vega}, {Watkins}, {Weaver}, {Whitmore}, {Woillez}, {Zabalza}, \& {Astropy Contributors}}]{astropy2018}
{Astropy Collaboration}, {Price-Whelan}, A.~M., {Sip{\H{o}}cz}, B.~M., {et~al.} 2018, \aj, 156, 123, \dodoi{10.3847/1538-3881/aabc4f}

\bibitem[{{Blakeslee} {et~al.}(2009){Blakeslee}, {Jord{\'a}n}, {Mei}, {C{\^o}t{\'e}}, {Ferrarese}, {Infante}, {Peng}, {Tonry}, \& {West}}]{blakeslee09}
{Blakeslee}, J.~P., {Jord{\'a}n}, A., {Mei}, S., {et~al.} 2009, \apj, 694, 556, \dodoi{10.1088/0004-637X/694/1/556}

\bibitem[{{Blandford} {et~al.}(2019){Blandford}, {Meier}, \& {Readhead}}]{blandford2019}
{Blandford}, R., {Meier}, D., \& {Readhead}, A. 2019, \araa, 57, 467, \dodoi{10.1146/annurev-astro-081817-051948}

\bibitem[{{Boccardi} {et~al.}(2017){Boccardi}, {Krichbaum}, {Ros}, \& {Zensus}}]{boccardi2017}
{Boccardi}, B., {Krichbaum}, T.~P., {Ros}, E., \& {Zensus}, J.~A. 2017, \aapr, 25, 4, \dodoi{10.1007/s00159-017-0105-6}

\bibitem[{{Bodo} {et~al.}(2013){Bodo}, {Mamatsashvili}, {Rossi}, \& {Mignone}}]{bodo2013}
{Bodo}, G., {Mamatsashvili}, G., {Rossi}, P., \& {Mignone}, A. 2013, \mnras, 434, 3030, \dodoi{10.1093/mnras/stt1225}

\bibitem[{{Bodo} {et~al.}(2016){Bodo}, {Mamatsashvili}, {Rossi}, \& {Mignone}}]{bodo2016}
---. 2016, \mnras, 462, 3031, \dodoi{10.1093/mnras/stw1650}

\bibitem[{{Bodo} {et~al.}(2019){Bodo}, {Mamatsashvili}, {Rossi}, \& {Mignone}}]{bodo2019}
---. 2019, \mnras, 485, 2909, \dodoi{10.1093/mnras/stz591}

\bibitem[{{Britzen} {et~al.}(2023){Britzen}, {Zaja{\v{c}}ek}, {Gopal-Krishna}, {Fendt}, {Kun}, {Jaron}, {Sillanp{\"a}{\"a}}, \& {Eckart}}]{britzen23}
{Britzen}, S., {Zaja{\v{c}}ek}, M., {Gopal-Krishna}, {et~al.} 2023, \apj, 951, 106, \dodoi{10.3847/1538-4357/accbbc}

\bibitem[{{Britzen} {et~al.}(2018){Britzen}, {Fendt}, {Witzel}, {Qian}, {Pashchenko}, {Kurtanidze}, {Zajacek}, {Martinez}, {Karas}, {Aller}, {Aller}, {Eckart}, {Nilsson}, {Ar{\'e}valo}, {Cuadra}, {Subroweit}, \& {Witzel}}]{britzen18}
{Britzen}, S., {Fendt}, C., {Witzel}, G., {et~al.} 2018, \mnras, 478, 3199, \dodoi{10.1093/mnras/sty1026}

\bibitem[{{Bromberg} {et~al.}(2019){Bromberg}, {Singh}, {Davelaar}, \& {Philippov}}]{bromberg2019}
{Bromberg}, O., {Singh}, C.~B., {Davelaar}, J., \& {Philippov}, A.~A. 2019, \apj, 884, 39, \dodoi{10.3847/1538-4357/ab3fa5}

\bibitem[{{Chael} {et~al.}(2018){Chael}, {Johnson}, {Bouman}, {Blackburn}, {Akiyama}, \& {Narayan}}]{chael2018}
{Chael}, A.~A., {Johnson}, M.~D., {Bouman}, K.~L., {et~al.} 2018, \apj, 857, 23, \dodoi{10.3847/1538-4357/aab6a8}

\bibitem[{{Chatterjee} {et~al.}(2023){Chatterjee}, {Liska}, {Tchekhovskoy}, \& {Markoff}}]{chatterjee2023}
{Chatterjee}, K., {Liska}, M., {Tchekhovskoy}, A., \& {Markoff}, S. 2023, arXiv e-prints, arXiv:2311.00432, \dodoi{10.48550/arXiv.2311.00432}

\bibitem[{{Cohen}(2017)}]{cohen2017}
{Cohen}, M.~H. 2017, Galaxies, 5, 12, \dodoi{10.3390/galaxies5010012}

\bibitem[{{Cohen} {et~al.}(2014){Cohen}, {Meier}, {Arshakian}, {Homan}, {Hovatta}, {Kovalev}, {Lister}, {Pushkarev}, {Richards}, \& {Savolainen}}]{cohen14}
{Cohen}, M.~H., {Meier}, D.~L., {Arshakian}, T.~G., {et~al.} 2014, \apj, 787, 151, \dodoi{10.1088/0004-637X/787/2/151}

\bibitem[{{Cohen} {et~al.}(2015){Cohen}, {Meier}, {Arshakian}, {Clausen-Brown}, {Homan}, {Hovatta}, {Kovalev}, {Lister}, {Pushkarev}, {Richards}, \& {Savolainen}}]{cohen15}
---. 2015, \apj, 803, 3, \dodoi{10.1088/0004-637X/803/1/3}

\bibitem[{{Cui} {et~al.}(2023){Cui}, {Hada}, {Kawashima}, {Kino}, {Lin}, {Mizuno}, {Ro}, {Honma}, {Yi}, {Yu}, {Park}, {Jiang}, {Shen}, {Kravchenko}, {Algaba}, {Cheng}, {Cho}, {Giovannini}, {Giroletti}, {Jung}, {Lu}, {Niinuma}, {Oh}, {Ohsuga}, {Sawada-Satoh}, {Sohn}, {Takahashi}, {Takamura}, {Tazaki}, {Trippe}, {Wajima}, {Akiyama}, {An}, {Asada}, {Buttaccio}, {Byun}, {Cui}, {Hagiwara}, {Hirota}, {Hodgson}, {Kawaguchi}, {Kim}, {Lee}, {Lee}, {Lee}, {Maccaferri}, {Melis}, {Melnikov}, {Migoni}, {Oh}, {Sugiyama}, {Wang}, {Zhang}, {Chen}, {Hwang}, {Jung}, {Kim}, {Kim}, {Kobayashi}, {Li}, {Li}, {Li}, {Liu}, {Liu}, {Liu}, {Oh}, {Oyama}, {Roh}, {Wang}, {Wang}, {Wang}, {Xia}, {Yan}, {Yeom}, {Yonekura}, {Yuan}, {Zhang}, {Zhao}, \& {Zhong}}]{cui23}
{Cui}, Y., {Hada}, K., {Kawashima}, T., {et~al.} 2023, \nat, 621, 711, \dodoi{10.1038/s41586-023-06479-6}

\bibitem[{{Cui} {et~al.}(2021){Cui}, {Hada}, {Kino}, {Sohn}, {Park}, {Ro}, {Sawada-Satoh}, {Jiang}, {Cui}, {Honma}, {Shen}, {Tazaki}, {An}, {Cho}, {Zhao}, {Cheng}, {Niinuma}, {Wajima}, {Zhang}, {Kawaguchi}, {Algaba}, {Koyama}, {Hirota}, {Yonekura}, {Sakai}, {Xia}, {Jiang}, {Yu}, {Gou}, {Hwang}, {Jiang}, {Sun}, {Jung}, {Kim}, {Kim}, {Kobayashi}, {Lee}, {Lee}, {Zhang}, {Li}, {Xu}, {Li}, {Oh}, {Oh}, {Oh}, {Oyama}, {Roh}, {Shibata}, {Guo}, {Zhao}, {Zhong}, {Wang}, {Yang}, {Yan}, {Yeom}, {Li}, {Li}, {Yuan}, {Dong}, {Chen}, {Akiyama}, {Asada}, {Byun}, {Hagiwara}, {Hodgson}, {Jung}, {Kim}, {Lee}, {Yi}, {Liu}, {Liu}, {Lu}, {Nakamura}, {Trippe}, {Wang}, {Wang}, \& {Zhang}}]{cui21}
{Cui}, Y.-Z., {Hada}, K., {Kino}, M., {et~al.} 2021, Research in Astronomy and Astrophysics, 21, 205, \dodoi{10.1088/1674-4527/21/8/205}

\bibitem[{{Davelaar} {et~al.}(2020){Davelaar}, {Philippov}, {Bromberg}, \& {Singh}}]{davelaar2020}
{Davelaar}, J., {Philippov}, A.~A., {Bromberg}, O., \& {Singh}, C.~B. 2020, \apjl, 896, L31, \dodoi{10.3847/2041-8213/ab95a2}

\bibitem[{{Davelaar} {et~al.}(2023){Davelaar}, {Ripperda}, {Sironi}, {Philippov}, {Olivares}, {Porth}, {Berg}, {Bronzwaer}, {Chatterjee}, \& {Liska}}]{davelaar2023}
{Davelaar}, J., {Ripperda}, B., {Sironi}, L., {et~al.} 2023, \apjl, 959, L3, \dodoi{10.3847/2041-8213/ad0b79}

\bibitem[{{EHT MWL Science Working Group} {et~al.}(2021){EHT MWL Science Working Group}, {Algaba}, {Anczarski}, {Asada}, {Balokovi{\'c}}, {Chandra}, {Cui}, {Falcone}, {Giroletti}, {Goddi}, {Hada}, {Haggard}, {Jorstad}, {Kaur}, {Kawashima}, {Keating}, {Kim}, {Kino}, {Komossa}, {Kravchenko}, {Krichbaum}, {Lee}, {Lu}, {Lucchini}, {Markoff}, {Neilsen}, {Nowak}, {Park}, {Principe}, {Ramakrishnan}, {Reynolds}, {Sasada}, {Savchenko}, {Williamson}, {Event Horizon Telescope Collaboration}, {Akiyama}, {Alberdi}, {Alef}, {Anantua}, {Azulay}, {Baczko}, {Ball}, {Barrett}, {Bintley}, {Benson}, {Blackburn}, {Blundell}, {Boland}, {Bouman}, {Bower}, {Boyce}, {Bremer}, {Brinkerink}, {Brissenden}, {Britzen}, {Broderick}, {Broguiere}, {Bronzwaer}, {Byun}, {Carlstrom}, {Chael}, {Chan}, {Chatterjee}, {Chatterjee}, {Chen}, {Chen}, {Chesler}, {Cho}, {Christian}, {Conway}, {Cordes}, {Crawford}, {Crew}, {Cruz-Osorio}, {Davelaar}, {de Laurentis}, {Deane}, {Dempsey}, {Desvignes}, {Dexter}, {Doeleman}, {Eatough}, {Falcke}, {Farah}, {Fish}, {Fomalont}, {Ford}, {Fraga-Encinas}, {Friberg}, {Fromm}, {Fuentes}, {Galison}, {Gammie}, {Garc{\'\i}a}, {Gentaz}, {Georgiev}, {Gold}, {G{\'o}mez}, {G{\'o}mez-Ruiz}, {Gu}, {Gurwell}, {Hecht}, {Hesper}, {Ho}, {Ho}, {Honma}, {Huang}, {Huang}, {Hughes}, {Ikeda}, {Inoue}, {Issaoun}, {James}, {Jannuzi}, {Janssen}, {Jeter}, {Jiang}, {Jim{\'e}nez-Rosales}, {Johnson}, {Jung}, {Karami}, {Karuppusamy}, {Kettenis}, {Kim}, {Kim}, {Kim}, {Koay}, {Kofuji}, {Koch}, {Koyama}, {Kramer}, {Kramer}, {Kuo}, {Lauer}, {Levis}, {Li}, {Li}, {Lindqvist}, {Lico}, {Lindahl}, {Liu}, {Liu}, {Liuzzo}, {Lo}, {Lobanov}, {Loinard}, {Lonsdale}, {MacDonald}, {Mao}, {Marchili}, {Marrone}, {Marscher}, {Mart{\'\i}-Vidal}, {Matsushita}, {Matthews}, {Medeiros}, {Menten}, {Mizuno}, {Mizuno}, {Moran}, {Moriyama}, {Moscibrodzka}, {M{\"u}ller}, {Musoke}, {Mej{\'\i}as}, {Nagai}, {Nagar}, {Nakamura}, {Narayan}, {Narayanan}, {Natarajan}, {Nathanail}, {Neri}, {Ni}, {Noutsos}, {Okino}, {Olivares}, {Ortiz-Le{\'o}n}, {Oyama}, {{\"O}zel}, {Palumbo}, {Patel}, {Pen}, {Pesce}, {Pi{\'e}tu}, {Plambeck}, {Popstefanija}, {Porth}, {P{\"o}tzl}, {Prather}, {Preciado-L{\'o}pez}, {Psaltis}, {Pu}, {Rao}, {Rawlings}, {Raymond}, {Rezzolla}, {Ricarte}, {Ripperda}, {Roelofs}, {Rogers}, {Ros}, {Rose}, {Roshanineshat}, {Rottmann}, {Roy}, {Ruszczyk}, {Rygl}, {S{\'a}nchez}, {S{\'a}nchez-Arguelles}, {Savolainen}, {Schloerb}, {Schuster}, {Shao}, {Shen}, {Small}, {Sohn}, {Soohoo}, {Sun}, {Tazaki}, {Tetarenko}, {Tiede}, {Tilanus}, {Titus}, {Toma}, {Torne}, {Trent}, {Traianou}, {Trippe}, {van Bemmel}, {van Langevelde}, {van Rossum}, {Wagner}, {Ward-Thompson}, {Wardle}, {Weintroub}, {Wex}, {Wharton}, {Wielgus}, {Wong}, {Wu}, {Yoon}, {Young}, {Young}, {Younsi}, {Yuan}, {Yuan}, {Zensus}, {Zhao}, {Zhao}, {Fermi Large Area Telescope Collaboration}, {Principe}, {Giroletti}, {D'Ammando}, {Orienti}, {H.~E.~S.~S. Collaboration}, {Abdalla}, {Adam}, {Aharonian}, {Benkhali}, {Ang{\"u}ner}, {Arcaro}, {Armand}, {Armstrong}, {Ashkar}, {Backes}, {Baghmanyan}, {Barbosa Martins}, {Barnacka}, {Barnard}, {Becherini}, {Berge}, {Bernl{\"o}hr}, {Bi}, {B{\"o}ttcher}, {Boisson}, {Bolmont}, {de Lavergne}, {Breuhaus}, {Brun}, {Brun}, {Bryan}, {B{\"u}chele}, {Bulik}, {Bylund}, {Caroff}, {Carosi}, {Casanova}, {Chand}, {Chen}, {Cotter}, {Cury{\l}o}, {Damascene Mbarubucyeye}, {Davids}, {Davies}, {Deil}, {Devin}, {Dewilt}, {Dirson}, {Djannati-Ata{\"\i}}, {Dmytriiev}, {Donath}, {Doroshenko}, {Duffy}, {Dyks}, {Egberts}, {Eichhorn}, {Einecke}, {Emery}, {Ernenwein}, {Feijen}, {Fegan}, {Fiasson}, {de Clairfontaine}, {Fontaine}, {Funk}, {F{\"u}{\ss}ling}, {Gabici}, {Gallant}, {Giavitto}, {Giunti}, {Glawion}, {Glicenstein}, {Gottschall}, {Grondin}, {Hahn}, {Haupt}, {Hermann}, {Hinton}, {Hofmann}, {Hoischen}, {Holch}, {Holler}, {H{\"o}rbe}, {Horns}, {Huber}, {Jamrozy}, {Jankowsky}, {Jankowsky}, {Jardin-Blicq}, {Joshi}, {Jung-Richardt}, {Kasai}, {Kastendieck}, {Katarzy{\'n}ski}, {Katz}, {Khangulyan}, {Kh{\'e}lifi}, {Klepser}, {Klu{\'z}niak}, {Komin}, {Konno}, {Kosack}, {Kostunin}, {Kreter}, {Lamanna}, {Lemi{\`e}re}, {Lemoine-Goumard}, {Lenain}, {Levy}, {Lohse}, {Lypova}, {Mackey}, {Majumdar}, {Malyshev}, {Malyshev}, {Marandon}, {Marchegiani}, {Marcowith}, {Mares}, {Mart{\'\i}-Devesa}, {Marx}, {Maurin}, {Meintjes}, {Meyer}, {Moderski}, {Mohamed}, {Mohrmann}, {Montanari}, {Moore}, {Morris}, {Moulin}, {Muller}, {Murach}, {Nakashima}, {Nayerhoda}, {de Naurois}, {Ndiyavala}, {Niederwanger}, {Niemiec}, {Oakes}, {O'Brien}, {Odaka}, {Ohm}, {Olivera-Nieto}, {de Ona Wilhelmi}, {Ostrowski}, {Panter}, {Panny}, {Parsons}, {Peron}, {Peyaud}, {Piel}, {Pita}, {Poireau}, {Noel}, {Prokhorov}, {Prokoph}, {P{\"u}hlhofer}, {Punch}, {Quirrenbach}, {Rauth}, {Reichherzer}, {Reimer}, {Reimer}, {Remy}, {Renaud}, {Rieger}, {Rinchiuso}, {Romoli}, {Rowell}, {Rudak}, {Ruiz-Velasco}, {Sahakian}, {Sailer}, {Sanchez}, {Santangelo}, {Sasaki}, {Scalici}, {Schutte}, {Schwanke}, {Schwemmer}, {Seglar-Arroyo}, {Senniappan}, {Seyffert}, {Shafi}, {Shiningayamwe}, {Simoni}, {Sinha}, {Sol}, {Specovius}, {Spencer}, {Spir-Jacob}, {Stawarz}, {Sun}, {Steenkamp}, {Stegmann}, {Steinmassl}, {Steppa}, {Takahashi}, {Tavernier}, {Taylor}, {Terrier}, {Tiziani}, {Tluczykont}, {Tomankova}, {Trichard}, {Tsirou}, {Tuffs}, {Uchiyama}, {van der Walt}, {van Eldik}, {van Rensburg}, {van Soelen}, {Vasileiadis}, {Veh}, {Venter}, {Vincent}, {Vink}, {V{\"o}lk}, {Vuillaume}, {Wadiasingh}, {Wagner}, {Watson}, {Werner}, {White}, {Wierzcholska}, {Wong}, {Yusafzai}, {Zacharias}, {Zanin}, {Zargaryan}, {Zdziarski}, {Zech}, {Zhu}, {Zorn}, {Zouari}, {{\.Z}ywucka}, {MAGIC Collaboration}, {Acciari}, {Ansoldi}, {Antonelli}, {Engels}, {Artero}, {Asano}, {Baack}, {Babi{\'c}}, {Baquero}, {de Almeida}, {Barrio}, {Becerra Gonz{\'a}lez}, {Bednarek}, {Bellizzi}, {Bernardini}, {Bernardos}, {Berti}, {Besenrieder}, {Bhattacharyya}, {Bigongiari}, {Biland}, {Blanch}, {Bonnoli}, {Bo{\v{s}}njak}, {Busetto}, {Carosi}, {Ceribella}, {Cerruti}, {Chai}, {Chilingarian}, {Cikota}, {Colak}, {Colombo}, {Contreras}, {Cortina}, {Covino}, {D'Amico}, {D'Elia}, {da Vela}, {Dazzi}, {de Angelis}, {de Lotto}, {Delfino}, {Delgado}, {Delgado Mendez}, {Depaoli}, {di Pierro}, {di Venere}, {Do Souto Espi{\~n}eira}, {Dominis Prester}, {Donini}, {Dorner}, {Doro}, {Elsaesser}, {Ramazani}, {Fattorini}, {Ferrara}, {Fonseca}, {Font}, {Fruck}, {Fukami}, {Garc{\'\i}a L{\'o}pez}, {Garczarczyk}, {Gasparyan}, {Gaug}, {Giglietto}, {Giordano}, {Gliwny}, {Godinovi{\'c}}, {Green}, {Green}, {Hadasch}, {Hahn}, {Heckmann}, {Herrera}, {Hoang}, {Hrupec}, {H{\"u}tten}, {Inada}, {Inoue}, {Ishio}, {Iwamura}, {Jim{\'e}nez}, {Jormanainen}, {Jouvin}, {Kajiwara}, {Karjalainen}, {Kerszberg}, {Kobayashi}, {Kubo}, {Kushida}, {Lamastra}, {Lelas}, {Leone}, {Lindfors}, {Lombardi}, {Longo}, {L{\'o}pez-Coto}, {L{\'o}pez-Moya}, {L{\'o}pez-Oramas}, {Loporchio}, {Machado de Oliveira Fraga}, {Maggio}, {Majumdar}, {Makariev}, {Mallamaci}, {Maneva}, {Manganaro}, {Mannheim}, {Maraschi}, {Mariotti}, {Mart{\'\i}nez}, {Mazin}, {Menchiari}, {Mender}, {Mi{\'c}anovi{\'c}}, {Miceli}, {Miener}, {Minev}, {Miranda}, {Mirzoyan}, {Molina}, {Moralejo}, {Morcuende}, {Moreno}, {Moretti}, {Neustroev}, {Nigro}, {Nilsson}, {Nishijima}, {Noda}, {Nozaki}, {Ohtani}, {Oka}, {Otero-Santos}, {Paiano}, {Palatiello}, {Paneque}, {Paoletti}, {Paredes}, {Pavleti{\'c}}, {Pe{\~n}il}, {Perennes}, {Persic}, {Moroni}, {Prandini}, {Priyadarshi}, {Puljak}, {Rhode}, {Rib{\'o}}, {Rico}, {Righi}, {Rugliancich}, {Saha}, {Sahakyan}, {Saito}, {Sakurai}, {Satalecka}, {Saturni}, {Schleicher}, {Schmidt}, {Schweizer}, {Sitarek}, {{\v{S}}nidari{\'c}}, {Sobczynska}, {Spolon}, {Stamerra}, {Strom}, {Strzys}, {Suda}, {Suri{\'c}}, {Takahashi}, {Tavecchio}, {Temnikov}, {Terzi{\'c}}, {Teshima}, {Tosti}, {Truzzi}, {Tutone}, {Ubach}, {van Scherpenberg}, {Vanzo}, {Vazquez Acosta}, {Ventura}, {Verguilov}, {Vigorito}, {Vitale}, {Vovk}, {Will}, {Wunderlich}, {Zari{\'c}}, {VERITAS Collaboration}, {Adams}, {Benbow}, {Brill}, {Capasso}, {Christiansen}, {Chromey}, {Daniel}, {Errando}, {Farrell}, {Feng}, {Finley}, {Fortson}, {Furniss}, {Gent}, {Giuri}, {Hassan}, {Hervet}, {Holder}, {Hughes}, {Humensky}, {Jin}, {Kaaret}, {Kertzman}, {Kieda}, {Kumar}, {Lang}, {Lundy}, {Maier}, {Moriarty}, {Mukherjee}, {Nieto}, {Nievas-Rosillo}, {O'Brien}, {Ong}, {Otte}, {Patel}, {Pfrang}, {Pohl}, {Prado}, {Pueschel}, {Quinn}, {Ragan}, {Reynolds}, {Ribeiro}, {Richards}, {Roache}, {Rulten}, {Ryan}, {Santander}, {Sembroski}, {Shang}, {Weinstein}, {Williams}, {Williamson}, {Eavn Collaboration}, {Hirota}, {Cui}, {Niinuma}, {Ro}, {Sakai}, {Sawada-Satoh}, {Wajima}, {Wang}, {Liu}, \& {Yonekura}}]{eht_mwl21}
{EHT MWL Science Working Group}, {Algaba}, J.~C., {Anczarski}, J., {et~al.} 2021, \apjl, 911, L11, \dodoi{10.3847/2041-8213/abef71}

\bibitem[{{Event Horizon Telescope Collaboration} {et~al.}(2019){Event Horizon Telescope Collaboration}, {Akiyama}, {Alberdi}, {Alef}, {Asada}, {Azulay}, {Baczko}, {Ball}, {Balokovi{\'c}}, {Barrett}, {Bintley}, {Blackburn}, {Boland}, {Bouman}, {Bower}, {Bremer}, {Brinkerink}, {Brissenden}, {Britzen}, {Broderick}, {Broguiere}, {Bronzwaer}, {Byun}, {Carlstrom}, {Chael}, {Chan}, {Chatterjee}, {Chatterjee}, {Chen}, {Chen}, {Cho}, {Christian}, {Conway}, {Cordes}, {Crew}, {Cui}, {Davelaar}, {De Laurentis}, {Deane}, {Dempsey}, {Desvignes}, {Dexter}, {Doeleman}, {Eatough}, {Falcke}, {Fish}, {Fomalont}, {Fraga-Encinas}, {Freeman}, {Friberg}, {Fromm}, {G{\'o}mez}, {Galison}, {Gammie}, {Garc{\'\i}a}, {Gentaz}, {Georgiev}, {Goddi}, {Gold}, {Gu}, {Gurwell}, {Hada}, {Hecht}, {Hesper}, {Ho}, {Ho}, {Honma}, {Huang}, {Huang}, {Hughes}, {Ikeda}, {Inoue}, {Issaoun}, {James}, {Jannuzi}, {Janssen}, {Jeter}, {Jiang}, {Johnson}, {Jorstad}, {Jung}, {Karami}, {Karuppusamy}, {Kawashima}, {Keating}, {Kettenis}, {Kim}, {Kim}, {Kim}, {Kino}, {Koay}, {Koch}, {Koyama}, {Kramer}, {Kramer}, {Krichbaum}, {Kuo}, {Lauer}, {Lee}, {Li}, {Li}, {Lindqvist}, {Liu}, {Liuzzo}, {Lo}, {Lobanov}, {Loinard}, {Lonsdale}, {Lu}, {MacDonald}, {Mao}, {Markoff}, {Marrone}, {Marscher}, {Mart{\'\i}-Vidal}, {Matsushita}, {Matthews}, {Medeiros}, {Menten}, {Mizuno}, {Mizuno}, {Moran}, {Moriyama}, {Moscibrodzka}, {M{\"u}ller}, {Nagai}, {Nagar}, {Nakamura}, {Narayan}, {Narayanan}, {Natarajan}, {Neri}, {Ni}, {Noutsos}, {Okino}, {Olivares}, {Ortiz-Le{\'o}n}, {Oyama}, {{\"O}zel}, {Palumbo}, {Patel}, {Pen}, {Pesce}, {Pi{\'e}tu}, {Plambeck}, {PopStefanija}, {Porth}, {Prather}, {Preciado-L{\'o}pez}, {Psaltis}, {Pu}, {Ramakrishnan}, {Rao}, {Rawlings}, {Raymond}, {Rezzolla}, {Ripperda}, {Roelofs}, {Rogers}, {Ros}, {Rose}, {Roshanineshat}, {Rottmann}, {Roy}, {Ruszczyk}, {Ryan}, {Rygl}, {S{\'a}nchez}, {S{\'a}nchez-Arguelles}, {Sasada}, {Savolainen}, {Schloerb}, {Schuster}, {Shao}, {Shen}, {Small}, {Sohn}, {SooHoo}, {Tazaki}, {Tiede}, {Tilanus}, {Titus}, {Toma}, {Torne}, {Trent}, {Trippe}, {Tsuda}, {van Bemmel}, {van Langevelde}, {van Rossum}, {Wagner}, {Wardle}, {Weintroub}, {Wex}, {Wharton}, {Wielgus}, {Wong}, {Wu}, {Young}, {Young}, {Younsi}, {Yuan}, {Yuan}, {Zensus}, {Zhao}, {Zhao}, {Zhu}, {Algaba}, {Allardi}, {Amestica}, {Anczarski}, {Bach}, {Baganoff}, {Beaudoin}, {Benson}, {Berthold}, {Blanchard}, {Blundell}, {Bustamente}, {Cappallo}, {Castillo-Dom{\'\i}nguez}, {Chang}, {Chang}, {Chang}, {Chen}, {Chilson}, {Chuter}, {C{\'o}rdova Rosado}, {Coulson}, {Crawford}, {Crowley}, {David}, {Derome}, {Dexter}, {Dornbusch}, {Dudevoir}, {Dzib}, {Eckart}, {Eckert}, {Erickson}, {Everett}, {Faber}, {Farah}, {Fath}, {Folkers}, {Forbes}, {Freund}, {G{\'o}mez-Ruiz}, {Gale}, {Gao}, {Geertsema}, {Graham}, {Greer}, {Grosslein}, {Gueth}, {Haggard}, {Halverson}, {Han}, {Han}, {Hao}, {Hasegawa}, {Henning}, {Hern{\'a}ndez-G{\'o}mez}, {Herrero-Illana}, {Heyminck}, {Hirota}, {Hoge}, {Huang}, {Impellizzeri}, {Jiang}, {Kamble}, {Keisler}, {Kimura}, {Kono}, {Kubo}, {Kuroda}, {Lacasse}, {Laing}, {Leitch}, {Li}, {Lin}, {Liu}, {Liu}, {Lu}, {Marson}, {Martin-Cocher}, {Massingill}, {Matulonis}, {McColl}, {McWhirter}, {Messias}, {Meyer-Zhao}, {Michalik}, {Monta{\~n}a}, {Montgomerie}, {Mora-Klein}, {Muders}, {Nadolski}, {Navarro}, {Neilsen}, {Nguyen}, {Nishioka}, {Norton}, {Nowak}, {Nystrom}, {Ogawa}, {Oshiro}, {Oyama}, {Parsons}, {Paine}, {Pe{\~n}alver}, {Phillips}, {Poirier}, {Pradel}, {Primiani}, {Raffin}, {Rahlin}, {Reiland}, {Risacher}, {Ruiz}, {S{\'a}ez-Mada{\'\i}n}, {Sassella}, {Schellart}, {Shaw}, {Silva}, {Shiokawa}, {Smith}, {Snow}, {Souccar}, {Sousa}, {Sridharan}, {Srinivasan}, {Stahm}, {Stark}, {Story}, {Timmer}, {Vertatschitsch}, {Walther}, {Wei}, {Whitehorn}, {Whitney}, {Woody}, {Wouterloot}, {Wright}, {Yamaguchi}, {Yu}, {Zeballos}, {Zhang}, \& {Ziurys}}]{ehtc19}
{Event Horizon Telescope Collaboration}, {Akiyama}, K., {Alberdi}, A., {et~al.} 2019, \apjl, 875, L1, \dodoi{10.3847/2041-8213/ab0ec7}

\bibitem[{{Event Horizon Telescope Collaboration} {et~al.}(2021{\natexlab{a}}){Event Horizon Telescope Collaboration}, {Akiyama}, {Algaba}, {Alberdi}, {Alef}, {Anantua}, {Asada}, {Azulay}, {Baczko}, {Ball}, {Balokovi{\'c}}, {Barrett}, {Benson}, {Bintley}, {Blackburn}, {Blundell}, {Boland}, {Bouman}, {Bower}, {Boyce}, {Bremer}, {Brinkerink}, {Brissenden}, {Britzen}, {Broderick}, {Broguiere}, {Bronzwaer}, {Byun}, {Carlstrom}, {Chael}, {Chan}, {Chatterjee}, {Chatterjee}, {Chen}, {Chen}, {Chesler}, {Cho}, {Christian}, {Conway}, {Cordes}, {Crawford}, {Crew}, {Cruz-Osorio}, {Cui}, {Davelaar}, {De Laurentis}, {Deane}, {Dempsey}, {Desvignes}, {Dexter}, {Doeleman}, {Eatough}, {Falcke}, {Farah}, {Fish}, {Fomalont}, {Ford}, {Fraga-Encinas}, {Friberg}, {Fromm}, {Fuentes}, {Galison}, {Gammie}, {Garc{\'\i}a}, {Gelles}, {Gentaz}, {Georgiev}, {Goddi}, {Gold}, {G{\'o}mez}, {G{\'o}mez-Ruiz}, {Gu}, {Gurwell}, {Hada}, {Haggard}, {Hecht}, {Hesper}, {Himwich}, {Ho}, {Ho}, {Honma}, {Huang}, {Huang}, {Hughes}, {Ikeda}, {Inoue}, {Issaoun}, {James}, {Jannuzi}, {Janssen}, {Jeter}, {Jiang}, {Jimenez-Rosales}, {Johnson}, {Jorstad}, {Jung}, {Karami}, {Karuppusamy}, {Kawashima}, {Keating}, {Kettenis}, {Kim}, {Kim}, {Kim}, {Kim}, {Kino}, {Koay}, {Kofuji}, {Koch}, {Koyama}, {Kramer}, {Kramer}, {Krichbaum}, {Kuo}, {Lauer}, {Lee}, {Levis}, {Li}, {Li}, {Lindqvist}, {Lico}, {Lindahl}, {Liu}, {Liu}, {Liuzzo}, {Lo}, {Lobanov}, {Loinard}, {Lonsdale}, {Lu}, {MacDonald}, {Mao}, {Marchili}, {Markoff}, {Marrone}, {Marscher}, {Mart{\'\i}-Vidal}, {Matsushita}, {Matthews}, {Medeiros}, {Menten}, {Mizuno}, {Mizuno}, {Moran}, {Moriyama}, {Moscibrodzka}, {M{\"u}ller}, {Musoke}, {Mus Mej{\'\i}as}, {Michalik}, {Nadolski}, {Nagai}, {Nagar}, {Nakamura}, {Narayan}, {Narayanan}, {Natarajan}, {Nathanail}, {Neilsen}, {Neri}, {Ni}, {Noutsos}, {Nowak}, {Okino}, {Olivares}, {Ortiz-Le{\'o}n}, {Oyama}, {{\"O}zel}, {Palumbo}, {Park}, {Patel}, {Pen}, {Pesce}, {Pi{\'e}tu}, {Plambeck}, {PopStefanija}, {Porth}, {P{\"o}tzl}, {Prather}, {Preciado-L{\'o}pez}, {Psaltis}, {Pu}, {Ramakrishnan}, {Rao}, {Rawlings}, {Raymond}, {Rezzolla}, {Ricarte}, {Ripperda}, {Roelofs}, {Rogers}, {Ros}, {Rose}, {Roshanineshat}, {Rottmann}, {Roy}, {Ruszczyk}, {Rygl}, {S{\'a}nchez}, {S{\'a}nchez-Arguelles}, {Sasada}, {Savolainen}, {Schloerb}, {Schuster}, {Shao}, {Shen}, {Small}, {Sohn}, {SooHoo}, {Sun}, {Tazaki}, {Tetarenko}, {Tiede}, {Tilanus}, {Titus}, {Toma}, {Torne}, {Trent}, {Traianou}, {Trippe}, {van Bemmel}, {van Langevelde}, {van Rossum}, {Wagner}, {Ward-Thompson}, {Wardle}, {Weintroub}, {Wex}, {Wharton}, {Wielgus}, {Wong}, {Wu}, {Yoon}, {Young}, {Young}, {Younsi}, {Yuan}, {Yuan}, {Zensus}, {Zhao}, \& {Zhao}}]{ehtc21}
{Event Horizon Telescope Collaboration}, {Akiyama}, K., {Algaba}, J.~C., {et~al.} 2021{\natexlab{a}}, \apjl, 910, L13, \dodoi{10.3847/2041-8213/abe4de}

\bibitem[{{Event Horizon Telescope Collaboration} {et~al.}(2021{\natexlab{b}}){Event Horizon Telescope Collaboration}, {Akiyama}, {Algaba}, {Alberdi}, {Alef}, {Anantua}, {Asada}, {Azulay}, {Baczko}, {Ball}, {Balokovi{\'c}}, {Barrett}, {Benson}, {Bintley}, {Blackburn}, {Blundell}, {Boland}, {Bouman}, {Bower}, {Boyce}, {Bremer}, {Brinkerink}, {Brissenden}, {Britzen}, {Broderick}, {Broguiere}, {Bronzwaer}, {Byun}, {Carlstrom}, {Chael}, {Chan}, {Chatterjee}, {Chatterjee}, {Chen}, {Chen}, {Chesler}, {Cho}, {Christian}, {Conway}, {Cordes}, {Crawford}, {Crew}, {Cruz-Osorio}, {Cui}, {Davelaar}, {De Laurentis}, {Deane}, {Dempsey}, {Desvignes}, {Dexter}, {Doeleman}, {Eatough}, {Falcke}, {Farah}, {Fish}, {Fomalont}, {Ford}, {Fraga-Encinas}, {Freeman}, {Friberg}, {Fromm}, {Fuentes}, {Galison}, {Gammie}, {Garc{\'\i}a}, {Gentaz}, {Georgiev}, {Goddi}, {Gold}, {G{\'o}mez}, {G{\'o}mez-Ruiz}, {Gu}, {Gurwell}, {Hada}, {Haggard}, {Hecht}, {Hesper}, {Ho}, {Ho}, {Honma}, {Huang}, {Huang}, {Hughes}, {Ikeda}, {Inoue}, {Issaoun}, {James}, {Jannuzi}, {Janssen}, {Jeter}, {Jiang}, {Jimenez-Rosales}, {Johnson}, {Jorstad}, {Jung}, {Karami}, {Karuppusamy}, {Kawashima}, {Keating}, {Kettenis}, {Kim}, {Kim}, {Kim}, {Kim}, {Kino}, {Koay}, {Kofuji}, {Koch}, {Koyama}, {Kramer}, {Kramer}, {Krichbaum}, {Kuo}, {Lauer}, {Lee}, {Levis}, {Li}, {Li}, {Lindqvist}, {Lico}, {Lindahl}, {Liu}, {Liu}, {Liuzzo}, {Lo}, {Lobanov}, {Loinard}, {Lonsdale}, {Lu}, {MacDonald}, {Mao}, {Marchili}, {Markoff}, {Marrone}, {Marscher}, {Mart{\'\i}-Vidal}, {Matsushita}, {Matthews}, {Medeiros}, {Menten}, {Mizuno}, {Mizuno}, {Moran}, {Moriyama}, {Moscibrodzka}, {M{\"u}ller}, {Musoke}, {Mej{\'\i}as}, {Michalik}, {Nadolski}, {Nagai}, {Nagar}, {Nakamura}, {Narayan}, {Narayanan}, {Natarajan}, {Nathanail}, {Neilsen}, {Neri}, {Ni}, {Noutsos}, {Nowak}, {Okino}, {Olivares}, {Ortiz-Le{\'o}n}, {Oyama}, {{\"O}zel}, {Palumbo}, {Park}, {Patel}, {Pen}, {Pesce}, {Pi{\'e}tu}, {Plambeck}, {PopStefanija}, {Porth}, {P{\"o}tzl}, {Prather}, {Preciado-L{\'o}pez}, {Psaltis}, {Pu}, {Ramakrishnan}, {Rao}, {Rawlings}, {Raymond}, {Rezzolla}, {Ricarte}, {Ripperda}, {Roelofs}, {Rogers}, {Ros}, {Rose}, {Roshanineshat}, {Rottmann}, {Roy}, {Ruszczyk}, {Rygl}, {S{\'a}nchez}, {S{\'a}nchez-Arguelles}, \& {Sasada}}]{eht2021_pol}
---. 2021{\natexlab{b}}, \apjl, 910, L12, \dodoi{10.3847/2041-8213/abe71d}

\bibitem[{{Event Horizon Telescope Collaboration} {et~al.}(2023){Event Horizon Telescope Collaboration}, {Akiyama}, {Alberdi}, {Alef}, {Algaba}, {Anantua}, {Asada}, {Azulay}, {Bach}, {Baczko}, {Ball}, {Balokovi{\'c}}, {Barrett}, {Baub{\"o}ck}, {Benson}, {Bintley}, {Blackburn}, {Blundell}, {Bouman}, {Bower}, {Boyce}, {Bremer}, {Brinkerink}, {Brissenden}, {Britzen}, {Broderick}, {Broguiere}, {Bronzwaer}, {Bustamante}, {Byun}, {Carlstrom}, {Ceccobello}, {Chael}, {Chan}, {Chang}, {Chatterjee}, {Chatterjee}, {Chen}, {Chen}, {Cheng}, {Cho}, {Christian}, {Conroy}, {Conway}, {Cordes}, {Crawford}, {Crew}, {Cruz-Osorio}, {Cui}, {Dahale}, {Davelaar}, {De Laurentis}, {Deane}, {Dempsey}, {Desvignes}, {Dexter}, {Dhruv}, {Doeleman}, {Dougal}, {Dzib}, {Eatough}, {Emami}, {Falcke}, {Farah}, {Fish}, {Fomalont}, {Ford}, {Foschi}, {Fraga-Encinas}, {Freeman}, {Friberg}, {Fromm}, {Fuentes}, {Galison}, {Gammie}, {Garc{\'\i}a}, {Gentaz}, {Georgiev}, {Goddi}, {Gold}, {G{\'o}mez-Ruiz}, {G{\'o}mez}, {Gu}, {Gurwell}, {Hada}, {Haggard}, {Haworth}, {Hecht}, {Hesper}, {Heumann}, {Ho}, {Ho}, {Honma}, {Huang}, {Huang}, {Hughes}, {Ikeda}, {Impellizzeri}, {Inoue}, {Issaoun}, {James}, {Jannuzi}, {Janssen}, {Jeter}, {Jiang}, {Jim{\'e}nez-Rosales}, {Johnson}, {Jorstad}, {Joshi}, {Jung}, {Karami}, {Karuppusamy}, {Kawashima}, {Keating}, {Kettenis}, {Kim}, {Kim}, {Kim}, {Kim}, {Kino}, {Koay}, {Kocherlakota}, {Kofuji}, {Koch}, {Koyama}, {Kramer}, {Kramer}, {Kramer}, {Krichbaum}, {Kuo}, {La Bella}, {Lauer}, {Lee}, {Lee}, {Leung}, {Levis}, {Li}, {Lico}, {Lindahl}, {Lindqvist}, {Lisakov}, {Liu}, {Liu}, {Liuzzo}, {Lo}, {Lobanov}, {Loinard}, {Lonsdale}, {Lowitz}, {Lu}, {MacDonald}, {Mao}, {Marchili}, {Markoff}, {Marrone}, {Marscher}, {Mart{\'\i}-Vidal}, {Matsushita}, {Matthews}, {Medeiros}, {Menten}, {Michalik}, {Mizuno}, {Mizuno}, {Moran}, {Moriyama}, {Moscibrodzka}, {Mulaudzi}, {M{\"u}ller}, {M{\"u}ller}, {Mus}, {Musoke}, {Myserlis}, {Nadolski}, {Nagai}, {Nagar}, {Nakamura}, {Narayan}, {Narayanan}, {Natarajan}, {Nathanail}, {Fuentes}, {Neilsen}, {Neri}, {Ni}, {Noutsos}, {Nowak}, {Oh}, {Okino}, {Olivares}, {Ortiz-Le{\'o}n}, {Oyama}, {{\"O}zel}, {Palumbo}, {Paraschos}, {Park}, {Parsons}, {Patel}, {Pen}, \& {Pesce}}]{ehtc23}
{Event Horizon Telescope Collaboration}, {Akiyama}, K., {Alberdi}, A., {et~al.} 2023, \apjl, 957, L20, \dodoi{10.3847/2041-8213/acff70}

\bibitem[{{Foreman-Mackey} {et~al.}(2013){Foreman-Mackey}, {Hogg}, {Lang}, \& {Goodman}}]{emcee2013}
{Foreman-Mackey}, D., {Hogg}, D.~W., {Lang}, D., \& {Goodman}, J. 2013, \pasp, 125, 306, \dodoi{10.1086/670067}

\bibitem[{{Foschi} {et~al.}(2024){Foschi}, {G{\'o}mez}, {Fuentes}, {Cho}, {Marscher}, \& {Jorstad}}]{foschi2024}
{Foschi}, M., {G{\'o}mez}, J.~L., {Fuentes}, A., {et~al.} 2024, arXiv e-prints, arXiv:2412.09215, \dodoi{10.48550/arXiv.2412.09215}

\bibitem[{{Fromm} {et~al.}(2013){Fromm}, {Ros}, {Perucho}, {Savolainen}, {Mimica}, {Kadler}, {Lobanov}, \& {Zensus}}]{fromm13}
{Fromm}, C.~M., {Ros}, E., {Perucho}, M., {et~al.} 2013, \aap, 557, A105, \dodoi{10.1051/0004-6361/201321784}

\bibitem[{{Fuentes} {et~al.}(2023){Fuentes}, {G{\'o}mez}, {Mart{\'\i}}, {Perucho}, {Zhao}, {Lico}, {Lobanov}, {Bruni}, {Kovalev}, {Chael}, {Akiyama}, {Bouman}, {Sun}, {Cho}, {Traianou}, {Toscano}, {Dahale}, {Foschi}, {Gurvits}, {Jorstad}, {Kim}, {Marscher}, {Mizuno}, {Ros}, \& {Savolainen}}]{fuentes23}
{Fuentes}, A., {G{\'o}mez}, J.~L., {Mart{\'\i}}, J.~M., {et~al.} 2023, Nature Astronomy, 7, 1359, \dodoi{10.1038/s41550-023-02105-7}

\bibitem[{{Giovannini} {et~al.}(2023){Giovannini}, {Cui}, {Hada}, {Yi}, {Ro}, {Sohn}, {Takamura}, {Buttaccio}, {D'Ammando}, {Giroletti}, {Hagiwara}, {Kino}, {Kravchenko}, {Maccaferri}, {Melnikov}, {Niinuma}, {Orienti}, {Wajima}, {Akiyama}, {Doi}, {Byun}, {Hirota}, {Honma}, {Jung}, {Kobayashi}, {Koyama}, {Melis}, {Migoni}, {Murata}, {Nagai}, {Sawada-Satoh}, \& {Stagni}}]{giovannini23}
{Giovannini}, G., {Cui}, Y., {Hada}, K., {et~al.} 2023, Galaxies, 11, 49, \dodoi{10.3390/galaxies11020049}

\bibitem[{{Greisen}(2003)}]{greisen2003}
{Greisen}, E.~W. 2003, in Astrophysics and Space Science Library, Vol. 285, Information Handling in Astronomy - Historical Vistas, ed. A.~{Heck}, 109, \dodoi{10.1007/0-306-48080-8_7}

\bibitem[{{Hada} {et~al.}(2024){Hada}, {Asada}, {Nakamura}, \& {Kino}}]{hada2024}
{Hada}, K., {Asada}, K., {Nakamura}, M., \& {Kino}, M. 2024, \aapr, 32, 5, \dodoi{10.1007/s00159-024-00155-y}

\bibitem[{{Hada} {et~al.}(2013){Hada}, {Kino}, {Doi}, {Nagai}, {Honma}, {Hagiwara}, {Giroletti}, {Giovannini}, \& {Kawaguchi}}]{hada13}
{Hada}, K., {Kino}, M., {Doi}, A., {et~al.} 2013, \apj, 775, 70, \dodoi{10.1088/0004-637X/775/1/70}

\bibitem[{{Hada} {et~al.}(2017){Hada}, {Park}, {Kino}, {Niinuma}, {Sohn}, {Ro}, {Jung}, {Algaba}, {Zhao}, {Lee}, {Akiyama}, {Trippe}, {Wajima}, {Sawada-Satoh}, {Tazaki}, {Cho}, {Hodgson}, {Lee}, {Hagiwara}, {Honma}, {Koyama}, {Oh}, {Lee}, {Yoo}, {Kawaguchi}, {Roh}, {Oh}, {Yeom}, {Jung}, {Oh}, {Kim}, {Hwang}, {Byun}, {Cho}, {Kim}, {Kobayashi}, \& {Shibata}}]{hada17}
{Hada}, K., {Park}, J.~H., {Kino}, M., {et~al.} 2017, \pasj, 69, 71, \dodoi{10.1093/pasj/psx054}

\bibitem[{{Hardee}(2007)}]{hardee2007}
{Hardee}, P.~E. 2007, \apj, 664, 26, \dodoi{10.1086/518409}

\bibitem[{{Hardee}(2011)}]{hardee11}
{Hardee}, P.~E. 2011, in Jets at All Scales, ed. G.~E. {Romero}, R.~A. {Sunyaev}, \& T.~{Belloni}, Vol. 275, 41--49, \dodoi{10.1017/S1743921310015620}

\bibitem[{{Hardee} \& {Eilek}(2011)}]{hardee_eilek11}
{Hardee}, P.~E., \& {Eilek}, J.~A. 2011, \apj, 735, 61, \dodoi{10.1088/0004-637X/735/1/61}

\bibitem[{{Hardee} {et~al.}(2005){Hardee}, {Walker}, \& {G{\'o}mez}}]{hardee2005}
{Hardee}, P.~E., {Walker}, R.~C., \& {G{\'o}mez}, J.~L. 2005, \apj, 620, 646, \dodoi{10.1086/427083}

\bibitem[{{Harris} {et~al.}(2020){Harris}, {Millman}, {van der Walt}, {Gommers}, {Virtanen}, {Cournapeau}, {Wieser}, {Taylor}, {Berg}, {Smith}, {Kern}, {Picus}, {Hoyer}, {van Kerkwijk}, {Brett}, {Haldane}, {del R{\'\i}o}, {Wiebe}, {Peterson}, {G{\'e}rard-Marchant}, {Sheppard}, {Reddy}, {Weckesser}, {Abbasi}, {Gohlke}, \& {Oliphant}}]{numpy2020}
{Harris}, C.~R., {Millman}, K.~J., {van der Walt}, S.~J., {et~al.} 2020, \nat, 585, 357, \dodoi{10.1038/s41586-020-2649-2}

\bibitem[{{Hogg} \& {Foreman-Mackey}(2018)}]{hogg18}
{Hogg}, D.~W., \& {Foreman-Mackey}, D. 2018, \apjs, 236, 11, \dodoi{10.3847/1538-4365/aab76e}

\bibitem[{{Hunter}(2007)}]{matplotlib2007}
{Hunter}, J.~D. 2007, Computing in Science and Engineering, 9, 90, \dodoi{10.1109/MCSE.2007.55}

\bibitem[{{Issaoun} {et~al.}(2022){Issaoun}, {Wielgus}, {Jorstad}, {Krichbaum}, {Blackburn}, {Janssen}, {Chan}, {Pesce}, {G{\'o}mez}, {Akiyama}, {Mo{\'s}cibrodzka}, {Mart{\'\i}-Vidal}, {Chael}, {Lico}, {Liu}, {Ramakrishnan}, {Lisakov}, {Fuentes}, {Zhao}, {Moriyama}, {Broderick}, {Tiede}, {MacDonald}, {Mizuno}, {Traianou}, {Loinard}, {Davelaar}, {Gurwell}, {Lu}, {Alberdi}, {Alef}, {Algaba}, {Anantua}, {Asada}, {Azulay}, {Bach}, {Baczko}, {Ball}, {Balokovi{\'c}}, {Barrett}, {Baub{\"o}ck}, {Benson}, {Bintley}, {Blundell}, {Boland}, {Bouman}, {Bower}, {Boyce}, {Bremer}, {Brinkerink}, {Brissenden}, {Britzen}, {Broguiere}, {Bronzwaer}, {Bustamante}, {Byun}, {Carlstrom}, {Ceccobello}, {Chatterjee}, {Chatterjee}, {Chen}, {Chen}, {Cho}, {Christian}, {Conroy}, {Conway}, {Cordes}, {Crawford}, {Crew}, {Cruz-Osorio}, {Cui}, {De Laurentis}, {Deane}, {Dempsey}, {Desvignes}, {Dexter}, {Doeleman}, {Dhruv}, {Dzib Quijano}, {Eatough}, {Emami}, {Falcke}, {Farah}, {Fish}, {Fomalont}, {Ford}, {Fraga-Encinas}, {Freeman}, {Friberg}, {Fromm}, {Galison}, {Gammie}, {Garc{\'\i}a}, {Gentaz}, {Georgiev}, {Goddi}, {Gold}, {G{\'o}mez-Ruiz}, {Gu}, {Hada}, {Haggard}, {Hecht}, {Hesper}, {Ho}, {Ho}, {Honma}, {Huang}, {Huang}, {Hughes}, {Ikeda}, {Impellizzeri}, {Inoue}, {James}, {Jannuzi}, {Jeter}, {Jiang}, {Jimenez-Rosales}, {Johnson}, {Joshi}, {Jung}, {Karami}, {Karuppusamy}, {Kawashima}, {Keating}, {Kettenis}, {Kim}, {Kim}, {Kim}, {Kim}, {Kino}, {Koay}, {Kocherlakota}, {Kofuji}, {Koch}, {Koyama}, {Kramer}, {Kramer}, {Kuo}, {La Bella}, {Lauer}, {Lee}, {Lee}, {Leung}, {Levis}, {Li}, {Lico}, {Lindahl}, {Lindqvist}, {Liu}, {Liuzzo}, {Lo}, {Lobanov}, {Lonsdale}, {Mao}, {Marchili}, {Markoff}, {Marrone}, {Marscher}, {Matsushita}, {Matthews}, {Medeiros}, {Menten}, {Michalik}, {Mizuno}, {Mizuno}, {Moran}, {M{\"u}ller}, {Mus}, {Musoke}, {Myserlis}, {Nadolski}, {Nagai}, {Nagar}, {Nakamura}, {Narayan}, {Narayanan}, {Natarajan}, {Nathanail}, {Neilsen}, {Neri}, {Ni}, {Noutsos}, {Nowak}, {Oh}, {Okino}, {Olivares}, {Ortiz-Le{\'o}n}, {Oyama}, {{\"O}zel}, {Palumbo}, {Paraschos}, {Park}, {Parsons}, {Patel}, {Pen}, {Pi{\'e}tu}, {Plambeck}, {PopStefanija}, {Porth}, {P{\"o}tzl}, {Prather}, {Preciado-L{\'o}pez}, {Psaltis}, {Pu}, {Rao}, {Rawlings}, {Raymond}, {Rezzolla}, {Ricarte}, {Ripperda}, {Roelofs}, {Rogers}, {Ros}, {Romero-Canizales}, {Roshanineshat}, {Rottmann}, {Roy}, {Ruiz}, {Ruszczyk}, {Rygl}, {S{\'a}nchez}, {S{\'a}nchez-Arguelles}, {Sanchez-Portal}, {Sasada}, {Satapathy}, {Savolainen}, {Schloerb}, {Schuster}, {Shao}, {Shen}, {Small}, {Sohn}, {SooHoo}, {Souccar}, {Sun}, {Tazaki}, {Tetarenko}, {Tiede}, {Tilanus}, {Titus}, {Torne}, {Trent}, {Trippe}, {van Bemmel}, {van Langevelde}, {van Rossum}, {Vos}, {Wagner}, {Ward-Thompson}, {Wardle}, {Weintroub}, {Wex}, {Wharton}, {Wiik}, {Witzel}, {Wondrak}, {Wong}, {Wu}, {Yamaguchi}, {Yoon}, {Young}, {Young}, {Younsi}, {Yuan}, {Yuan}, {Zensus}, {Zhang}, \& {Zhao}}]{issaoun2022}
{Issaoun}, S., {Wielgus}, M., {Jorstad}, S., {et~al.} 2022, \apj, 934, 145, \dodoi{10.3847/1538-4357/ac7a40}

\bibitem[{{Jorstad} {et~al.}(2023){Jorstad}, {Wielgus}, {Lico}, {Issaoun}, {Broderick}, {Pesce}, {Liu}, {Zhao}, {Krichbaum}, {Blackburn}, {Chan}, {Janssen}, {Ramakrishnan}, {Akiyama}, {Alberdi}, {Algaba}, {Bouman}, {Cho}, {Fuentes}, {G{\'o}mez}, {Gurwell}, {Johnson}, {Kim}, {Lu}, {Mart{\'\i}-Vidal}, {Moscibrodzka}, {P{\"o}tzl}, {Traianou}, {van Bemmel}, {Alef}, {Anantua}, {Asada}, {Azulay}, {Bach}, {Baczko}, {Ball}, {Balokovi{\'c}}, {Barrett}, {Baub{\"o}ck}, {Benson}, {Bintley}, {Blundell}, {Bower}, {Boyce}, {Bremer}, {Brinkerink}, {Brissenden}, {Britzen}, {Broguiere}, {Bronzwaer}, {Bustamante}, {Byun}, {Carlstrom}, {Ceccobello}, {Chael}, {Chatterjee}, {Chatterjee}, {Chen}, {Chen}, {Cheng}, {Christian}, {Conroy}, {Conway}, {Cordes}, {Crawford}, {Crew}, {Cruz-Osorio}, {Cui}, {Davelaar}, {De Laurentis}, {Deane}, {Dempsey}, {Desvignes}, {Dexter}, {Dhruv}, {Doeleman}, {Dougal}, {Dzib}, {Eatough}, {Emami}, {Falcke}, {Farah}, {Fish}, {Fomalont}, {Ford}, {Fraga-Encinas}, {Freeman}, {Friberg}, {Fromm}, {Galison}, {Gammie}, {Garc{\'\i}a}, {Gentaz}, {Georgiev}, {Goddi}, {Gold}, {G{\'o}mez-Ruiz}, {Gu}, {Hada}, {Haggard}, {Haworth}, {Hecht}, {Hesper}, {Heumann}, {Ho}, {Ho}, {Honma}, {Huang}, {Huang}, {Hughes}, {Ikeda}, {Impellizzeri}, {Inoue}, {James}, {Jannuzi}, {Jeter}, {Jiang}, {Jim{\'e}nez-Rosales}, {Joshi}, {Jung}, {Karami}, {Karuppusamy}, {Kawashima}, {Keating}, {Kettenis}, {Kim}, {Kim}, {Kim}, {Kino}, {Koay}, {Kocherlakota}, {Kofuji}, {Koyama}, {Kramer}, {Kramer}, {Kuo}, {La Bella}, {Lauer}, {Lee}, {Lee}, {Leung}, {Levis}, {Li}, {Lindahl}, {Lindqvist}, {Lisakov}, {Liu}, {Liuzzo}, {Lo}, {Lobanov}, {Loinard}, {Lonsdale}, {MacDonald}, {Mao}, {Marchili}, {Markoff}, {Marrone}, {Marscher}, {Matsushita}, {Matthews}, {Medeiros}, {Menten}, {Michalik}, {Mizuno}, {Mizuno}, {Moran}, {Moriyama}, {M{\"u}ller}, {Mus}, {Musoke}, {Myserlis}, {Nadolski}, {Nagai}, {Nagar}, {Nakamura}, {Narayan}, {Narayanan}, {Natarajan}, {Nathanail}, {Fuentes}, {Neilsen}, {Neri}, {Ni}, {Noutsos}, {Nowak}, {Oh}, {Okino}, {Olivares}, {Ortiz-Le{\'o}n}, {Oyama}, {{\"O}zel}, {Palumbo}, {Paraschos}, {Park}, {Parsons}, {Patel}, {Pen}, {Pi{\'e}tu}, {Plambeck}, {PopStefanija}, {Porth}, {Prather}, {Preciado-L{\'o}pez}, {Psaltis}, {Pu}, {Rao}, {Rawlings}, {Raymond}, {Rezzolla}, {Ricarte}, {Ripperda}, {Roelofs}, {Rogers}, {Ros}, {Romero-Ca{\~n}izales}, {Roshanineshat}, {Rottmann}, {Roy}, {Ruiz}, {Ruszczyk}, {Rygl}, {S{\'a}nchez}, {S{\'a}nchez-Arg{\"u}elles}, {S{\'a}nchez-Portal}, {Sasada}, {Satapathy}, {Savolainen}, {Schloerb}, {Schonfeld}, {Schuster}, {Shao}, {Shen}, {Small}, {Sohn}, {SooHoo}, {Souccar}, {Sun}, {Tazaki}, {Tetarenko}, {Tiede}, {Tilanus}, {Titus}, {Torne}, {Trent}, {Trippe}, {Turk}, {van Langevelde}, {van Rossum}, {Vos}, {Wagner}, {Ward-Thompson}, {Wardle}, {Weintroub}, {Wex}, {Wharton}, {Wiik}, {Witzel}, {Wondrak}, {Wong}, {Wu}, {Yamaguchi}, {Yoon}, {Young}, {Young}, {Younsi}, {Yuan}, {Yuan}, {Zensus}, {Zhang}, \& {Zhao}}]{jorstad2023}
{Jorstad}, S., {Wielgus}, M., {Lico}, R., {et~al.} 2023, \apj, 943, 170, \dodoi{10.3847/1538-4357/acaea8}

\bibitem[{{Jorstad} {et~al.}(2022){Jorstad}, {Marscher}, {Raiteri}, {Villata}, {Weaver}, {Zhang}, {Dong}, {G{\'o}mez}, {Perel}, {Savchenko}, {Larionov}, {Carosati}, {Chen}, {Kurtanidze}, {Marchini}, {Matsumoto}, {Mortari}, {Aceti}, {Acosta-Pulido}, {Andreeva}, {Apolonio}, {Arena}, {Arkharov}, {Bachev}, {Banfi}, {Bonnoli}, {Borman}, {Bozhilov}, {Carnerero}, {Damljanovic}, {Ehgamberdiev}, {Els{\"a}sser}, {Frasca}, {Gabellini}, {Grishina}, {Gupta}, {Hagen-Thorn}, {Hallum}, {Hart}, {Hasuda}, {Hemrich}, {Hsiao}, {Ibryamov}, {Irsmambetova}, {Ivanov}, {Joner}, {Kimeridze}, {Klimanov}, {Kn{\"o}tt}, {Kopatskaya}, {Kurtanidze}, {Kurtenkov}, {Kuutma}, {Larionova}, {Leonini}, {Lin}, {Lorey}, {Mannheim}, {Marino}, {Minev}, {Mirzaqulov}, {Morozova}, {Nikiforova}, {Nikolashvili}, {Ovcharov}, {Papini}, {Pursimo}, {Rahimov}, {Reinhart}, {Sakamoto}, {Salvaggio}, {Semkov}, {Shakhovskoy}, {Sigua}, {Steineke}, {Stojanovic}, {Strigachev}, {Troitskaya}, {Troitskiy}, {Tsai}, {Valcheva}, {Vasilyev}, {Vince}, {Waller}, {Zaharieva}, \& {Chatterjee}}]{jorstad2022}
{Jorstad}, S.~G., {Marscher}, A.~P., {Raiteri}, C.~M., {et~al.} 2022, \nat, 609, 265, \dodoi{10.1038/s41586-022-05038-9}

\bibitem[{{Kim} {et~al.}(2024){Kim}, {Nikonov}, {Roth}, {En{\ss}lin}, {Janssen}, {Arras}, {M{\"u}ller}, \& {Lobanov}}]{kim_2024_resolve}
{Kim}, J.-S., {Nikonov}, A.~S., {Roth}, J., {et~al.} 2024, \aap, 690, A129, \dodoi{10.1051/0004-6361/202449663}

\bibitem[{{Kim} {et~al.}(2018){Kim}, {Krichbaum}, {Lu}, {Ros}, {Bach}, {Bremer}, {de Vicente}, {Lindqvist}, \& {Zensus}}]{kim18}
{Kim}, J.~Y., {Krichbaum}, T.~P., {Lu}, R.~S., {et~al.} 2018, \aap, 616, A188, \dodoi{10.1051/0004-6361/201832921}

\bibitem[{{Kino} {et~al.}(2025){Kino}, {Nagashima}, {Ro}, {Cui}, {Hada}, \& {Park}}]{kino2025}
{Kino}, M., {Nagashima}, M., {Ro}, H., {et~al.} 2025, \apj, 986, 49, \dodoi{10.3847/1538-4357/adceb6}

\bibitem[{{Kino} {et~al.}(2015){Kino}, {Takahara}, {Hada}, {Akiyama}, {Nagai}, \& {Sohn}}]{kino15}
{Kino}, M., {Takahara}, F., {Hada}, K., {et~al.} 2015, \apj, 803, 30, \dodoi{10.1088/0004-637X/803/1/30}

\bibitem[{{Lalakos} {et~al.}(2024){Lalakos}, {Tchekhovskoy}, {Bromberg}, {Gottlieb}, {Jacquemin-Ide}, {Liska}, \& {Zhang}}]{lalakos2024}
{Lalakos}, A., {Tchekhovskoy}, A., {Bromberg}, O., {et~al.} 2024, \apj, 964, 79, \dodoi{10.3847/1538-4357/ad0974}

\bibitem[{{Lobanov} {et~al.}(2003){Lobanov}, {Hardee}, \& {Eilek}}]{lobanov03}
{Lobanov}, A., {Hardee}, P., \& {Eilek}, J. 2003, \nar, 47, 629, \dodoi{10.1016/S1387-6473(03)00109-X}

\bibitem[{{Lobanov} \& {Zensus}(2001)}]{lobanov_zensus2001}
{Lobanov}, A.~P., \& {Zensus}, J.~A. 2001, Science, 294, 128, \dodoi{10.1126/science.1063239}

\bibitem[{{Lu} {et~al.}(2023){Lu}, {Asada}, {Krichbaum}, {Park}, {Tazaki}, {Pu}, {Nakamura}, {Lobanov}, {Hada}, {Akiyama}, {Kim}, {Marti-Vidal}, {G{\'o}mez}, {Kawashima}, {Yuan}, {Ros}, {Alef}, {Britzen}, {Bremer}, {Broderick}, {Doi}, {Giovannini}, {Giroletti}, {Ho}, {Honma}, {Hughes}, {Inoue}, {Jiang}, {Kino}, {Koyama}, {Lindqvist}, {Liu}, {Marscher}, {Matsushita}, {Nagai}, {Rottmann}, {Savolainen}, {Schuster}, {Shen}, {de Vicente}, {Walker}, {Yang}, {Zensus}, {Algaba}, {Allardi}, {Bach}, {Berthold}, {Bintley}, {Byun}, {Casadio}, {Chang}, {Chang}, {Chang}, {Chen}, {Chen}, {Chilson}, {Chuter}, {Conway}, {Crew}, {Dempsey}, {Dornbusch}, {Faber}, {Friberg}, {Garc{\'\i}a}, {Garrido}, {Han}, {Han}, {Hasegawa}, {Herrero-Illana}, {Huang}, {Huang}, {Impellizzeri}, {Jiang}, {Jinchi}, {Jung}, {Kallunki}, {Kirves}, {Kimura}, {Koay}, {Koch}, {Kramer}, {Kraus}, {Kubo}, {Kuo}, {Li}, {Lin}, {Liu}, {Liu}, {Lo}, {Lu}, {MacDonald}, {Martin-Cocher}, {Messias}, {Meyer-Zhao}, {Minter}, {Nair}, {Nishioka}, {Norton}, {Nystrom}, {Ogawa}, {Oshiro}, {Patel}, {Pen}, {Pidopryhora}, {Pradel}, {Raffin}, {Rao}, {Ruiz}, {Sanchez}, {Shaw}, {Snow}, {Sridharan}, {Srinivasan}, {Tercero}, {Torne}, {Traianou}, {Wagner}, {Walther}, {Wei}, {Yang}, \& {Yu}}]{lu2023}
{Lu}, R.-S., {Asada}, K., {Krichbaum}, T.~P., {et~al.} 2023, \nat, 616, 686, \dodoi{10.1038/s41586-023-05843-w}

\bibitem[{{McKinney} {et~al.}(2012){McKinney}, {Tchekhovskoy}, \& {Blandford}}]{mckinney12}
{McKinney}, J.~C., {Tchekhovskoy}, A., \& {Blandford}, R.~D. 2012, \mnras, 423, 3083, \dodoi{10.1111/j.1365-2966.2012.21074.x}

\bibitem[{{Meier}(2012)}]{meier2012}
{Meier}, D.~L. 2012, {Black Hole Astrophysics: The Engine Paradigm} (Berlin: Springer), \dodoi{10.1007/978-3-642-01936-4}

\bibitem[{{Mertens} {et~al.}(2016){Mertens}, {Lobanov}, {Walker}, \& {Hardee}}]{mertens16}
{Mertens}, F., {Lobanov}, A.~P., {Walker}, R.~C., \& {Hardee}, P.~E. 2016, \aap, 595, A54, \dodoi{10.1051/0004-6361/201628829}

\bibitem[{{Mizuno} {et~al.}(2016){Mizuno}, {G{\'o}mez}, {Nishikawa}, {Meli}, {Hardee}, {Rezzolla}, {Singh}, \& {de Gouveia Dal Pino}}]{mizuno2016}
{Mizuno}, Y., {G{\'o}mez}, J.~L., {Nishikawa}, K.-I., {et~al.} 2016, Galaxies, 4, 40, \dodoi{10.3390/galaxies4040040}

\bibitem[{{Mizuno} {et~al.}(2007){Mizuno}, {Hardee}, \& {Nishikawa}}]{mizuno07}
{Mizuno}, Y., {Hardee}, P., \& {Nishikawa}, K.-I. 2007, \apj, 662, 835, \dodoi{10.1086/518106}

\bibitem[{{Mizuno} {et~al.}(2011){Mizuno}, {Hardee}, \& {Nishikawa}}]{mizuno11}
{Mizuno}, Y., {Hardee}, P.~E., \& {Nishikawa}, K.-I. 2011, \apj, 734, 19, \dodoi{10.1088/0004-637X/734/1/19}

\bibitem[{{Mizuno} {et~al.}(2014){Mizuno}, {Hardee}, \& {Nishikawa}}]{mizuno14}
---. 2014, \apj, 784, 167, \dodoi{10.1088/0004-637X/784/2/167}

\bibitem[{{Mizuno} {et~al.}(2009){Mizuno}, {Lyubarsky}, {Nishikawa}, \& {Hardee}}]{mizuno09}
{Mizuno}, Y., {Lyubarsky}, Y., {Nishikawa}, K.-I., \& {Hardee}, P.~E. 2009, \apj, 700, 684, \dodoi{10.1088/0004-637X/700/1/684}

\bibitem[{{Moriyama} {et~al.}(2024){Moriyama}, {Cruz-Osorio}, {Mizuno}, {Fromm}, {Nathanail}, \& {Rezzolla}}]{moriyama2024}
{Moriyama}, K., {Cruz-Osorio}, A., {Mizuno}, Y., {et~al.} 2024, \apj, 960, 106, \dodoi{10.3847/1538-4357/ad07d4}

\bibitem[{{Nakamura} {et~al.}(2018){Nakamura}, {Asada}, {Hada}, {Pu}, {Noble}, {Tseng}, {Toma}, {Kino}, {Nagai}, {Takahashi}, {Algaba}, {Orienti}, {Akiyama}, {Doi}, {Giovannini}, {Giroletti}, {Honma}, {Koyama}, {Lico}, {Niinuma}, \& {Tazaki}}]{nakamura18}
{Nakamura}, M., {Asada}, K., {Hada}, K., {et~al.} 2018, \apj, 868, 146, \dodoi{10.3847/1538-4357/aaeb2d}

\bibitem[{{Narayan} {et~al.}(2003){Narayan}, {Igumenshchev}, \& {Abramowicz}}]{narayan2003}
{Narayan}, R., {Igumenshchev}, I.~V., \& {Abramowicz}, M.~A. 2003, \pasj, 55, L69, \dodoi{10.1093/pasj/55.6.L69}

\bibitem[{{Niinuma} {et~al.}(2014){Niinuma}, {Lee}, {Kino}, {Sohn}, {Akiyama}, {Zhao}, {Sawada-Satoh}, {Trippe}, {Hada}, {Jung}, {Hagiwara}, {Dodson}, {Koyama}, {Honma}, {Nagai}, {Chung}, {Doi}, {Fujisawa}, {Han}, {Kim}, {Lee}, {Lee}, {Miyazaki}, {Oyama}, {Sorai}, {Wajima}, {Bae}, {Byun}, {Cho}, {Choi}, {Chung}, {Chung}, {Han}, {Hirota}, {Hwang}, {Je}, {Jike}, {Jung}, {Jung}, {Kang}, {Kang}, {Kang}, {Kan-ya}, {Kanaguchi}, {Kawaguchi}, {Kim}, {Kim}, {Kim}, {Kim}, {Kim}, {Kim}, {Kim}, {Kobayashi}, {Kono}, {Kurayama}, {Lee}, {Lee}, {Lee}, {Minh}, {Matsumoto}, {Nakagawa}, {Oh}, {Oh}, {Park}, {Roh}, {Sasao}, {Shibata}, {Song}, {Tamura}, {Wi}, {Yeom}, \& {Yun}}]{niinuma14}
{Niinuma}, K., {Lee}, S.-S., {Kino}, M., {et~al.} 2014, \pasj, 66, 103, \dodoi{10.1093/pasj/psu104}

\bibitem[{{Nikonov} {et~al.}(2023){Nikonov}, {Kovalev}, {Kravchenko}, {Pashchenko}, \& {Lobanov}}]{nikonov23}
{Nikonov}, A.~S., {Kovalev}, Y.~Y., {Kravchenko}, E.~V., {Pashchenko}, I.~N., \& {Lobanov}, A.~P. 2023, \mnras, 526, 5949, \dodoi{10.1093/mnras/stad3061}

\bibitem[{{O'Neill} {et~al.}(2012){O'Neill}, {Beckwith}, \& {Begelman}}]{o'neill2012}
{O'Neill}, S.~M., {Beckwith}, K., \& {Begelman}, M.~C. 2012, \mnras, 422, 1436, \dodoi{10.1111/j.1365-2966.2012.20721.x}

\bibitem[{{Park} {et~al.}(2019{\natexlab{a}}){Park}, {Hada}, {Kino}, {Nakamura}, {Ro}, \& {Trippe}}]{park19_fara}
{Park}, J., {Hada}, K., {Kino}, M., {et~al.} 2019{\natexlab{a}}, \apj, 871, 257, \dodoi{10.3847/1538-4357/aaf9a9}

\bibitem[{{Park} {et~al.}(2019{\natexlab{b}}){Park}, {Hada}, {Kino}, {Nakamura}, {Hodgson}, {Ro}, {Cui}, {Asada}, {Algaba}, {Sawada-Satoh}, {Lee}, {Cho}, {Shen}, {Jiang}, {Trippe}, {Niinuma}, {Sohn}, {Jung}, {Zhao}, {Wajima}, {Tazaki}, {Honma}, {An}, {Akiyama}, {Byun}, {Kim}, {Zhang}, {Cheng}, {Kobayashi}, {Shibata}, {Lee}, {Roh}, {Oh}, {Yeom}, {Jung}, {Oh}, {Kim}, {Hwang}, \& {Hagiwara}}]{park19}
---. 2019{\natexlab{b}}, \apj, 887, 147, \dodoi{10.3847/1538-4357/ab5584}

\bibitem[{{Park} {et~al.}(2024){Park}, {Zhao}, {Nakamura}, {Mizuno}, {Pu}, {Asada}, {Takahashi}, {Toma}, {Kino}, {Cho}, {Hada}, {Edwards}, {Ro}, {Kam}, {Yi}, {Lee}, {Koyama}, {Byun}, {Phillips}, {Reynolds}, {Hodgson}, \& {Lee}}]{park2024}
{Park}, J., {Zhao}, G.-Y., {Nakamura}, M., {et~al.} 2024, arXiv e-prints, arXiv:2408.09069, \dodoi{10.48550/arXiv.2408.09069}

\bibitem[{{Pasetto} {et~al.}(2021){Pasetto}, {Carrasco-Gonz{\'a}lez}, {G{\'o}mez}, {Mart{\'\i}}, {Perucho}, {O'Sullivan}, {Anderson}, {D{\'\i}az-Gonz{\'a}lez}, {Fuentes}, \& {Wardle}}]{pasetto21}
{Pasetto}, A., {Carrasco-Gonz{\'a}lez}, C., {G{\'o}mez}, J.~L., {et~al.} 2021, \apjl, 923, L5, \dodoi{10.3847/2041-8213/ac3a88}

\bibitem[{{Perucho}(2012)}]{perucho2012_review}
{Perucho}, M. 2012, in International Journal of Modern Physics Conference Series, Vol.~8, International Journal of Modern Physics Conference Series, 241--252, \dodoi{10.1142/S2010194512004667}

\bibitem[{{Perucho}(2019)}]{perucho19}
{Perucho}, M. 2019, Galaxies, 7, 70, \dodoi{10.3390/galaxies7030070}

\bibitem[{{Perucho} {et~al.}(2012){Perucho}, {Kovalev}, {Lobanov}, {Hardee}, \& {Agudo}}]{perucho2012}
{Perucho}, M., {Kovalev}, Y.~Y., {Lobanov}, A.~P., {Hardee}, P.~E., \& {Agudo}, I. 2012, \apj, 749, 55, \dodoi{10.1088/0004-637X/749/1/55}

\bibitem[{{Ripperda} {et~al.}(2022){Ripperda}, {Liska}, {Chatterjee}, {Musoke}, {Philippov}, {Markoff}, {Tchekhovskoy}, \& {Younsi}}]{ripperda2022}
{Ripperda}, B., {Liska}, M., {Chatterjee}, K., {et~al.} 2022, \apjl, 924, L32, \dodoi{10.3847/2041-8213/ac46a1}

\bibitem[{{Ro} {et~al.}(2023{\natexlab{a}}){Ro}, {Yi}, {Cui}, {Kino}, {Hada}, {Kawashima}, {Mizuno}, {Sohn}, \& {Tazaki}}]{ro23}
{Ro}, H., {Yi}, K., {Cui}, Y., {et~al.} 2023{\natexlab{a}}, Galaxies, 11, 33, \dodoi{10.3390/galaxies11010033}

\bibitem[{{Ro} {et~al.}(2023{\natexlab{b}}){Ro}, {Kino}, {Sohn}, {Hada}, {Park}, {Nakamura}, {Cui}, {Yi}, {Chung}, {Hodgson}, {Kawashima}, {An}, {Trippe}, {Algaba}, {Kim}, {Sawada-Satoh}, {Wajima}, {Shen}, {Cheng}, {Cho}, {Jiang}, {Jung}, {Lee}, {Niinuma}, {Oh}, {Tazaki}, {Zhao}, {Akiyama}, {Honma}, {Lee}, {Lu}, {Zhang}, {Asada}, {Cui}, {Hagiwara}, {Hirota}, {Kawaguchi}, {Koyama}, {Lee}, {Oh}, {Sugiyama}, {Takamura}, {Wang}, {Hwang}, {Jung}, {Kim}, {Kim}, {Kobayashi}, {Oh}, {Oyama}, {Roh}, \& {Yeom}}]{ro23_spix}
{Ro}, H., {Kino}, M., {Sohn}, B.~W., {et~al.} 2023{\natexlab{b}}, \aap, 673, A159, \dodoi{10.1051/0004-6361/202142988}

\bibitem[{{Shepherd}(1997)}]{shepherd1997}
{Shepherd}, M.~C. 1997, in Astronomical Society of the Pacific Conference Series, Vol. 125, Astronomical Data Analysis Software and Systems VI, ed. G.~{Hunt} \& H.~{Payne}, 77

\bibitem[{{Singh} {et~al.}(2016){Singh}, {Mizuno}, \& {de Gouveia Dal Pino}}]{singh16}
{Singh}, C.~B., {Mizuno}, Y., \& {de Gouveia Dal Pino}, E.~M. 2016, \apj, 824, 48, \dodoi{10.3847/0004-637X/824/1/48}

\bibitem[{{Sobacchi} {et~al.}(2017){Sobacchi}, {Lyubarsky}, \& {Sormani}}]{sobacchi2017}
{Sobacchi}, E., {Lyubarsky}, Y.~E., \& {Sormani}, M.~C. 2017, \mnras, 468, 4635, \dodoi{10.1093/mnras/stx807}

\bibitem[{{Tazaki} {et~al.}(2023){Tazaki}, {Cui}, {Hada}, {Kino}, {Cho}, {Zhao}, {Akiyama}, {Mizuno}, {Ro}, {Honma}, {Lu}, {Shen}, {Cui}, \& {Yonekura}}]{tazaki2023}
{Tazaki}, F., {Cui}, Y., {Hada}, K., {et~al.} 2023, Galaxies, 11, 39, \dodoi{10.3390/galaxies11020039}

\bibitem[{{Tchekhovskoy} {et~al.}(2011){Tchekhovskoy}, {Narayan}, \& {McKinney}}]{tchekhovskoy11}
{Tchekhovskoy}, A., {Narayan}, R., \& {McKinney}, J.~C. 2011, \mnras, 418, L79, \dodoi{10.1111/j.1745-3933.2011.01147.x}

\bibitem[{{Tiede}(2022)}]{tiede2022}
{Tiede}, P. 2022, The Journal of Open Source Software, 7, 4457, \dodoi{10.21105/joss.04457}

\bibitem[{{VanderPlas}(2018)}]{vanderplas18}
{VanderPlas}, J.~T. 2018, \apjs, 236, 16, \dodoi{10.3847/1538-4365/aab766}

\bibitem[{{Vega-Garc{\'\i}a} {et~al.}(2020){Vega-Garc{\'\i}a}, {Lobanov}, {Perucho}, {Bruni}, {Ros}, {Anderson}, {Agudo}, {Davis}, {G{\'o}mez}, {Kovalev}, {Krichbaum}, {Lisakov}, {Savolainen}, {Schinzel}, \& {Zensus}}]{vega-garcia2020}
{Vega-Garc{\'\i}a}, L., {Lobanov}, A.~P., {Perucho}, M., {et~al.} 2020, \aap, 641, A40, \dodoi{10.1051/0004-6361/201935168}

\bibitem[{{Walker} {et~al.}(2018){Walker}, {Hardee}, {Davies}, {Ly}, \& {Junor}}]{walker2018}
{Walker}, R.~C., {Hardee}, P.~E., {Davies}, F.~B., {Ly}, C., \& {Junor}, W. 2018, \apj, 855, 128, \dodoi{10.3847/1538-4357/aaafcc}

\end{thebibliography}
\bibliographystyle{aasjournal}

\end{document}